\documentclass[longauth]{aa}

\usepackage[dvipsnames,table,xcdraw]{xcolor}
\usepackage{longtable}
\usepackage{booktabs} 
\usepackage{natbib, twoopt}
\usepackage[hyphenbreaks]{breakurl}
\usepackage[colorlinks=true, urlcolor=blue, linkcolor=blue, citecolor=blue, breaklinks]{hyperref}
\bibpunct{(}{)}{;}{a}{}{,}

\usepackage{graphics, graphicx}
\usepackage{subcaption}
\usepackage{tabularx}
\usepackage{multirow}
\usepackage{placeins}
\usepackage[labelfont=bf]{caption}
\usepackage{pdfpages}

\usepackage{txfonts}
\usepackage{soul}
\usepackage{float}

\usepackage{comment}

\usepackage[capitalise]{cleveref}
\crefname{section}{Sect.}{Sect.}

\newcommand{\Halpha}{\mathrm{H\alpha}}
\newcommand{\Hbeta}{\mathrm{H\beta}}
\newcommand{\NII}{[\ion{N}{II}]}
\newcommand{\OII}{[\ion{O}{II}]}
\newcommand{\OIII}{[\ion{O}{III}]}

\defcitealias{SP15}{SP15}

\begin{document}
    \title{GLACE survey: OSIRIS/GTC multi-object spectroscopy of the rich galaxy cluster ZwCl 0024.0+1652 at $z\sim0.4$}
    \subtitle{III. The mass--SFR relation and the quenching of cluster galaxies}

    \author{{Simon B. De Daniloff}
            \inst{\ref{inst:IRAM},\ref{inst:UGR}}
        \and
            {Ángel Bongiovanni} 
            \inst{\ref{inst:IRAM},\ref{inst:AsocAstro}}
        \and
            {Miguel Sánchez-Portal}
            \inst{\ref{inst:IRAM},\ref{inst:AsocAstro}}
        \and
            {Bernabé Cedrés}
            \inst{\ref{inst:IAC},\ref{inst:ULL},\ref{inst:AsocAstro}}
        \and
            {Carmen P. Padilla-Torres}
            \inst{\ref{inst:AsocAstro},\ref{inst:IAC},\ref{inst:ULL},\ref{inst:FundGG}}
        \and
            {Ana María Pérez-García}
            \inst{\ref{inst:ISDEFE},\ref{inst:AsocAstro}}
        \and
            {Ricardo Pérez-Martínez}
            \inst{\ref{inst:ISDEFE},\ref{inst:AsocAstro}}
        \and
            {Daniel Espada}
            \inst{\ref{inst:UGR},\ref{inst:IC1}}
        \and
            {Clara C. de la Casa}
            \inst{\ref{inst:IAA}}
        \and
            {Gloria Torres-Ríos}
            \inst{\ref{inst:UGR}}
        \and
            {Mauro Gónzalez-Otero}
            \inst{\ref{inst:IAA},\ref{inst:AsocAstro}}
        \and
            {José A. de Diego}
            \inst{\ref{inst:AstroUNAM}}
        \and
            {Mónica I. Rodríguez}
            \inst{\ref{inst:IRAM}}
        \and
            {Miguel Cerviño}
            \inst{\ref{inst:CAB}}
        \and
            {Maritza A. Lara-López}
            \inst{\ref{inst:UCM},\ref{inst:IFPC}}
        \and
            {Jordi Cepa}
            \inst{\ref{inst:IAC},\ref{inst:ULL},\ref{inst:AsocAstro}}
        \and
            {Ivan Valtchanov}
            \inst{\ref{inst:Telespazio},\ref{inst:AsocAstro}}
        \and
            {J. Ignacio González-Serrano}
            \inst{\ref{inst:IFCant},\ref{inst:AsocAstro}}
        \and
            {Irene Cruz-González}
            \inst{\ref{inst:AstroUNAM}}
        \and
            {Castalia Alenka Negrete}
            \inst{\ref{inst:AstroUNAM},\ref{inst:UNAM_SECIHTI}}
        \and
            {Zeleke Beyoro-Amado}
            \inst{\ref{inst:KUE}}
        \and
            {Manuel Castillo-Fraile}
            \inst{\ref{inst:IRAM}}
        \and
            {Brisa Mancillas}
            \inst{\ref{inst:IRAM}}
        \and
            {Mirjana Pović}
            \inst{\ref{inst:IAA},\ref{inst:SSGI},\ref{inst:MUST}}
           }
            
    \institute{Institut de Radioastonomie Millimétrique (IRAM), Av. Divina Pastora 7, Núcleo Central 18012, Granada, Spain\label{inst:IRAM}
        \and
               Universidad de Granada, Departamento de Física Teórica y del Cosmos, Campus Fuente Nueva, Edificio Mecenas, E-18071, Granada, Spain\label{inst:UGR}
        \and
               Asociación Astrofísica para la Promoción de la Investigación, Instrumentación y su Desarrollo, ASPID, 38205 La Laguna, Tenerife, Spain\label{inst:AsocAstro}
        \and
               Instituto de Astrofísica de Canarias, 38205 La Laguna, Tenerife, Spain\label{inst:IAC}
        \and
               Departamento de Astrofísica, Universidad de La Laguna (ULL), 38205 La Laguna, Tenerife, Spain\label{inst:ULL}
        \and
               Fundación Galileo Galilei-INAF Rambla José Ana Fernández Pérez, 7, 38712 Breña Baja, Tenerife, Spain\label{inst:FundGG}
        \and
               ISDEFE, Beatriz de Bobadilla 3, E-28040 Madrid, Spain\label{inst:ISDEFE}
        \and
               Instituto Carlos I de Física Teórica y Computacional, Universidad de Granada, E18071, Granada, Spain\label{inst:IC1}
        \and
               Instituto de Astrofísica de Andalucía (CSIC), Glorieta de la Astronomía s/n, 18008 Granada, Spain\label{inst:IAA}
        \and
               Instituto de Astronomía, Universidad Nacional Autónoma de México (UNAM), Mexico City, Mexico\label{inst:AstroUNAM}
        \and
               Centro de Astrobiología (CSIC/INTA), 28692 ESAC Campus, Villanueva de la Cañada, Madrid, Spain\label{inst:CAB}
        \and
               Departamento de Física de la Tierra y Astrofísica, Fac. de C.C. Físicas, Universidad Complutense de Madrid, E-28040 Madrid, Spain\label{inst:UCM}
        \and
               Instituto de Física de Partículas y del Cosmos, IPARCOS, Fac. C.C. Físicas, Universidad Complutense de Madrid, E-28040 Madrid, Spain\label{inst:IFPC}
        \and
               Telespazio UK for European Space Agency, European Space Astronomy Centre, Villanueva de la Cañada, Madrid, Spain\label{inst:Telespazio}
        \and
               Instituto de Física de Cantabria (CSIC-Universidad de Cantabria), E-39005, Santander, Spain\label{inst:IFCant}
        \and
               SECIHTI Research Fellow – Instituto de Astronomía, Universidad Nacional Autónoma de México (UNAM), Mexico City, Mexico\label{inst:UNAM_SECIHTI}
        \and
               Kotebe University of Education, Department of Physics, Addis Ababa, Ethiopia\label{inst:KUE}
        \and
               Space Science and Geospatial Institute (SSGI), Entoto Observatory and Research Center (EORC), Astronomy and Astrophysics Research Division, PO Box 33679, Addis Abbaba, Ethiopia\label{inst:SSGI}
        \and
               Physics Department, Mbarara University of Science and Technology (MUST), Mbarara, Uganda\label{inst:MUST}
        \\
               {\href{mailto:sbonnal@iram.es}{\email{sbonnal@iram.es}}}
              }

   \date{Received: --; accepted: --}

    \abstract
    {Galaxy clusters are among the largest and densest structures in the Universe. Their high density generally increases the suppression of star formation, known as quenching, altering galaxy properties.}
    {We study the quenching of emission-line galaxies (ELGs) in the rich cluster ZwCl 0024.0+1652 (Cl0024) at redshift $z\sim0.4$, aiming to determine if and how star formation is suppressed.}
    {Using multi-object spectroscopy from the GLACE survey, we extracted fluxes and redshifts of \OII$\lambda\lambda3727,3729$, $\Hbeta$, and \OIII$\lambda5007$ emission lines to derive star formation rates (SFRs) for 173 ELGs. We also performed spectral energy distribution fitting to obtain key evolutionary parameters such as stellar masses ($M_\star$) and the 4000 \AA\ break ($D4000$) index.}
    {We derived the $M_\star-{\rm SFR}$ relation for 98 star-forming galaxies (SFGs), finding 34.7\% exhibit suppressed SFRs in the cluster, compared to 11.0\% in the field. While the SFRs show no significant variation with local density, the fraction of SFGs is 1.55 times higher in the cluster outskirts than in intermediate-density regions. The specific SFR decreases strongly with $D4000$ for active SFGs but remains constant for suppressed galaxies. The fraction of suppressed galaxies in the infall region is 2.6 times higher than in the core, especially in the infalling structure B of the cluster. The cluster’s total mass does not appear to be a key factor in SFG quenching.}
    {Star formation in Cl0024 galaxies is suppressed by the dense cluster environment. This suppression is evident in SFG fractions and parameters tracing long-term evolution, indicating prolonged quenching. The SFGs preferentially reside in low-density regions, while suppressed galaxies dominate the infall region, supporting a `delayed-then-rapid' quenching scenario.}

    \titlerunning{GLACE survey: The mass--SFR relation and the quenching of cluster galaxies}
    \maketitle

\section{Introduction}
The mechanisms responsible for the evolution of galaxies are still debated, particularly regarding their dependence on the environment, whether on large or local scales. It is known that galaxies in cluster environments have globally lower star formation rates (SFRs) \citep[e.g.][]{Balogh2004,Peng2010,wetzel_galaxy_2012,PerezMillan2023}, are redder than galaxies outside cluster regions or field galaxies \citep[e.g.][]{Kodama2001}, and present larger fractions of early-type morphologies than the field \citep{Dressler1980,dens_morph_goto,postman_morphology-density_2005}. It is also thought that galaxies in clusters have been accreted from lower-density regions \citep{Gunn_Gott1972,Berrier2009,Haines2015}, but the transformation of star-forming galaxies (SFGs) to quiescence is not fully understood.

Several internal and external processes are known to affect star formation (SF) in galaxies \citep[see][for a recent review of the different processes]{de_lucia_cosmic_2025}. The result of processes that lead to a significant and sustained reduction of SF in galaxies is generally referred to as `quenching'. We distinguish secular quenching \citep{kormendy_secular_2004}, which is the gradual suppression of SF due to internal, long-term processes, from environmental quenching, which is caused by external processes. However, these two categories are not the only way to classify the quenching phenomena; differentiation based on the speed of the processes is also made in the literature.

It is also known that galaxies in different large-scale structures display different levels of SF activity, and it has recently been confirmed that galaxies evolve differently depending on their large-scale environments \citep{gomez_galaxy_2003,dominguez-gomez_galaxies_2023,torres-rios_effect_2024}. Environmental processes of galaxy-galaxy interactions, such as dynamical friction and galactic cannibalism \citep{OstrikerTremaine1975}, and galaxy-cluster interaction, such as ram pressure stripping \citep{Gunn_Gott1972} and strangulation \citep{Larson1980}, are suspected to play a large role in the quenching of galaxies, depending on the local density of the galaxies within the cluster \citep{Treu2003}. However, local-scale gradients of the properties of the galaxies are not as distinct and direct in groups and clusters across different redshifts. For instance, the SFRs of galaxies in groups and clusters do not show a significant dependence with regard to the local density of their environment, whether it is at low \cite[e.g.][]{McGee2011,LimaDias2021}, intermediate \cite[e.g.][]{LaganaUlmer2018}, or high redshift \cite[e.g.][]{Tadaki2012}, so the determination of the causes of this quenching represents a significant challenge.

Simulations supported by observations have helped to hypothesise some scenarios contributing to the truncation of SF in cluster galaxies \citep[see for example the review of][]{somerville_physical_2015}. The initial absence of detection of the ongoing truncation would suggest a `rapid quenching' scenario in which the SF quenches in less than 1 Gyr after the galaxy is accreted by the cluster \citep[e.g.][]{Balogh2004,Peng2010,Pallero2022}. This scenario appears to be preferred for galaxies with high stellar masses ($M_\star$). However, the detection of galaxies with suppressed SF within the clusters indicates that they can also go through `slow quenching' \citep[e.g.][]{vonderLinden2010,Haines2015,Jian2020}. In recent years, a more nuanced scenario has appeared, called the `delayed-then-rapid quenching'. In this context, the cluster environment has little or no effect on galaxies for 2-4 Gyr after the infall, and during that time, the galaxies suffer from slow quenching processes such as strangulation during a phase of `delay' but are then rapidly quenched, for example, via ram pressure stripping \citep[e.g.][]{Wetzel2013,Roberts2019,cleland_environmental_2021,Morgan2024}.

To study the activity and evolution of galaxies, different methods have been employed. For example, \cite{Oman2016,Finn2023} studied the fraction of galaxies with low SF and found that galaxies tend to quench rapidly and possibly during the infall towards the centre of the cluster. \cite{Jaffe2015,Rhee2017,Rhee2020} studied the position of galaxies in the projected phase-space of the cluster and determined that ram pressure stripping plays an important role, combined with pre-processing, in the quenching within cluster galaxies. Other studies have used fundamental relations between the properties of galaxies, such as the correlation between the $M_\star$ of a galaxy and its SFR \citep[e.g.][]{Brinchmann2003,Noeske2007,Paccagnella2016} or metallicity \citep[e.g.][]{Tremonti2004,LaraLopez2009a,LaraLopez2009b}, and found correlations between these diagrams and the level of SF activity within galaxies.

In this work, we study how the SF activity correlates with the local environment of the galaxy cluster ZwCl 0024.0+1652 (hereafter Cl0024), at redshift $z\sim0.4$, in order to understand how quenching occurs for galaxies in this particular case. We calculated the SFRs of 173 emission-line galaxies (ELGs) using multi-object spectroscopy (MOS) and specifically analysed their position with respect to the star-forming main sequence (SFMS) using spectroscopic and photometric data from current and previous studies. Once we determined their level of SF activity, we studied how quenching affects them depending on their position in the cluster by employing several techniques and comparison samples.

This paper is organised as follows. In \cref{sec:methodology} we describe the GaLAxy Cluster Evolution (GLACE) survey, present the cluster and the data used, and discuss the methodology employed to estimate SFRs. In \cref{sec:results} we show the results, including the products of the MOS data reduction, the $M_\star-{\rm SFR}$ relation, the connection with the local and large-scale environments, and the effects on the SF activity. In \cref{sec:discussion} we discuss these results in the context of previous literature. In \cref{sec:summary} we summarise the results and conclusions of this work, and lastly, in \cref{sec:examples_spectra,sec:data_corrections,sec:sed_fittings_parameters}, we present complementary information on the reduction of the data and obtention of parameters.

We assumed a $\Lambda$CDM cosmology with $H_{\rm 0}=70\ \rm km\ Mpc^{-1}\ s^{-1}$, $T_{\rm CMB}=2.725$ K, and $\Omega_{\rm m,0}=0.3$, and a Chabrier initial mass function  \citep[IMF;][]{Chabrier_IMF}.

\section{Methodology: GLACE survey, ZwCl 0024.0+1652 cluster, and MOS data}\label{sec:methodology}

\subsection{The GLACE survey}
The GLACE survey is a deep panoramic narrow-band survey of ELGs that aims to study the SF and active galactic nucleus (AGN) activity, morphology, and metallicity variations of galaxies in clusters at different redshifts (\citealt{Perez_Martinez2013, Pintos_Castro2013, SP15, Pintos_Castro2016, Zeleke2019, Zeleke2021, Cedres2024}). For this purpose, clusters in three ranges of redshifts were selected at $z$$\sim$0.40, 0.63, and 0.86 to study their properties using $\Halpha$, $\Hbeta$, \NII$\lambda6583$, \OII, and \OIII$\lambda5007$ emission lines. In \cite{SP15}, the GLACE project was presented together with its first blind survey using tunable filter (TF) observations from the OSIRIS instrument at the 10.4 m GTC telescope, which focused on the cluster Cl0024 located at $z\sim0.395$. The authors presented a catalogue of 174 ELGs with spectroscopic redshifts and analysed the properties of the $\Halpha/$\NII$\lambda6583$ emission lines, distinguishing the AGN population, and calculating the $\Halpha$ SFRs of the SFGs. The main conclusion was that these galaxies suffer a suppression of their SF and AGN activity in the innermost region of the cluster. The following papers of the GLACE consortium studied the SFRs of the galaxies, and their relationships with the cluster environment.

\cite{Zeleke2021} used TF data targeting the \OIII$\lambda\lambda4989,5007$ and $\Hbeta$ emission lines to derive complementary results concerning the SFR and its relation with the environmental parameters of Cl0024. They respectively found 59 galaxies with $\Hbeta$ and 35 with \OIII$\lambda5007$ emission lines, and studied the relations between the SFR, $M_\star$, and local density. They confirmed the previous findings and suggested an accelerated evolution of galaxies in clusters with respect to the field. However, the analyses carried out biases due to the large uncertainties inherently associated with the estimation of the emission line fluxes from the TF observations.

Therefore, $\Halpha/$\NII$\lambda6583$ TF data were newly analysed in \cite{Cedres2024}. A Monte Carlo inverse convolution method \citep{InvConvMethod_Nadolny} was used to obtain the emission line fluxes as well as the metallicities of the gas within the 174 sources and to study the mass-metallicity relation of galaxies. They found that massive galaxies and galaxies located in the innermost region of the cluster tend to have a higher gas metallicity, but not a metallicity gradient within the cluster. Also, more massive clusters, considered to exhibit more effective interactions between galaxies and larger regions of high density, seem to present a lower value of the turnover $M_\star$ in the mass-metallicity relation (that of saturation of the gas metallicity), consistent with a truncation of the SF.

The TF observations were the basis of a blind survey, which served to identify the ELGs of the cluster; it allowed for low-resolution pseudo-spectra of $R\sim500$ with a limited spectral range around the $\Halpha$ emission line. To complement these observations, we decided to observe the ELGs of Cl0024 using the MOS mode of the OSIRIS instrument at the GTC telescope, which have a wider spectral range and offer the possibility to cover several emission lines, as well as being more time-efficient after candidate selection (details can be found in \cref{sec:glace_mos_data}). We observed the $\Hbeta$, \OII\ and \OIII$\lambda5007$ emission lines of the previously introduced sample. This new dataset allows the galaxies in the cluster to be studied further through the emission lines used by \cite{Zeleke2021}, but with more reliable and precise results, as well as revision and completion of the previous analysis using the $\Halpha/$\NII$\lambda6583$ TF data.

\subsection{The ZwCl 0024.0+1652 cluster}\label{sec:Cl0024}
The Cl0024 cluster at $z\sim0.395$, is a well-studied cluster at intermediate redshift. One of the first mentions of the cluster appears in \cite{ButcherOemler1978}, where the authors suggested for the first time that clusters at intermediate redshift contain a larger fraction of blue galaxies than clusters at low redshift. It was suggested that this effect could be caused by interactions and the particular disturbed systems in the population of blue galaxies and starbursts of the cluster \citep{Lavery1992,Oemler1997}. The dark matter distribution of the cluster has also been well-studied \citep{Tyson1998,Kneib2003,Diaferio2005,jee_discovery_2007,Umetsu2009}, concluding to the possible existence of a `ring-like dark matter structure' and that the cluster might be in a post-collision state. The particular kinematics of the cluster was studied by \cite{Czoske2001,Czoske2002}, who found that the cluster is experiencing a high-speed collision with a smaller group of galaxies located in our line of sight. Furthermore, \cite{Treu2003,Moran2005} used the Hubble Space Telescope (HST) to create a photometric catalogue of Cl0024 with 22\,000 entries. \cite{leethochawalit_evolution_2018} used this same catalogue and reported an evolution of the $M_\star-Z$ relation with respect to redshift using the quiescent population of galaxies in the cluster; and \cite{Costa2024} found a high percentage of early-type galaxies in the outskirts of the cluster, where pre-processing might be happening. Additionally, \cite{kodama2004} studied the luminosity function of the galaxies in the cluster using narrow-band $\Halpha$ from Subaru, and \cite{Geach2008,Geach2009} used infrared $24\ \rm \mu m$ and CO observations to study ultra-luminous infrared galaxies and determine their mass assembly time to be of $\sim150\ {\rm Myr}$. Finally, \cite{Medezinski2011} included Cl0024 in their sample of clusters to measure the weak lensing of massive clusters.

Cl0024 is a kinematically active cluster, with a principal structure (A) assembling onto the cluster core from the north-west with an orientation almost on the plane of the sky and an infalling group (B) at high velocity nearly along the line of sight to the cluster centre. A third structure (C) was also observed at $z\sim0.42$ by \cite{Czoske2001} and \cite{SP15}. This cluster is estimated to have a virial mass $M_{200} = (4.1\pm0.2) \times 10^{14}\ \mathrm{M_\odot}\ h^{-1}$ and a virial radius $R_{200} = (1.21\pm0.1)\ \mathrm{Mpc}\ h^{-1}$ (\citealt{SP15}, hereafter \citetalias{SP15})\footnote{We define $M_{200}$ as the mass enclosed within a sphere of radius $R_{200}$ whose mean density is 200 times the critical density of the Universe. We use the term `virial mass' as a convention referring to $M_{200}$, but we do not imply that the cluster is necessarily virialised.}, where $h$ is the Hubble constant in units of $100\;\mathrm{km\ s^{-1}\ Mpc^{-1}}$. \cref{fig:mosaic_Cl0024} shows a mosaic of the cluster, taken from observations of the OSIRIS/GTC instrument.

\subsection{Ancillary data}\label{sec:ancillary}
In this study, we employ the recently calculated $\Halpha$ and \NII$\lambda6583$ fluxes and redshifts obtained through the inverse convolution technique, as detailed by \cite{Cedres2024}. In cases where the convolution results are classified as having high quality (tiers 1 to 3) we use the newest values of fluxes and redshifts, and in instances where the convolution quality falls below this criterion (tiers 4 to 6), we use the original flux estimates presented in \citetalias{SP15} to minimise uncertainties. Given the significant overlap in the results between the two datasets (see \citealt{Cedres2024} for a close analysis), we observe negligible differences on our results, except for uncertainties, which are based on stochastic instead of deterministic procedures for the latter.

In addition to those values, we also use the photometric data available and compiled in \citetalias{SP15}, coming from the public catalogues of the collaboration `A Wide Field Survey of Two $z = 0.5$ Galaxy Clusters' \citep{Treu2003, Moran2005}. These catalogues provide magnitudes in bands $B$, $V$, $R$, $I$, $J$, $K_{\rm s}$, and $F814W$ obtained from deep images. Matching the galaxies with spectroscopic redshift from these catalogues with the galaxies included in the footprint of the GLACE observations yields a total of 366 sources, including the 174 sources selected for this study. The additional 192 sources have been classified as non-ELGs during the creation of the sample \citepalias[see][]{SP15}.

\subsection{The GLACE multi-object spectroscopy data}\label{sec:glace_mos_data}
Our work is constrained to the study of the ELGs in the cluster reported by \citetalias{SP15}, composed of 174 sources detected using the $\Halpha$ emission line. The program was aimed to carry out MOS on the ELGs with the low-resolution R500B grism\footnote{A low-resolution grism was used because the spectral region planned to be observed (roughly from 4000 to 8000 \AA\ net), containing complementary strong emission lines in the optical, does not contain strong airglow emission.} of the OSIRIS instrument of the GTC. A total of 15 masks were prepared on the basis of the identification of sources made with TF, using multiple slits of $\sim12\arcsec\times1.2\arcsec$, each one aligned with a specific target, or to the empty sky in order to facilitate the sky subtraction. Each mask was acquired and exposed $3 \times 900$ seconds. Secondary spectro-photometric standard stars were also observed using a long slit of $4\arcsec$ wide (which contains $\sim99\%$ of the flux of the star for a median seeing of $0.7\arcsec$) to provide a general flux calibration and aperture correction of the MOS fluxes. The observations took place in service mode during 2-hour time slots over 8 nights in August and September of 2017, upon completing the awarded 15 hours.

The reduction of the data was performed using the IRAF-based pipeline GTCMOS \citep[see][]{pipeline_GTCMOS}. Details regarding the reduction of similar data can be found in Sect. 5 of \cite{lockman_mauro_reduction_MOS}. In the case of our MOS data, we have estimated that the flux measurements carry an inherent uncertainty of approximately $\sim$$10-20\%$ stemming from the procedure for relative flux calibration with standard stars. Additionally, there is an intrinsic uncertainty associated with the wavelength calibration, which is of the order of 0.01 \AA\ (RMS) for the grism used and the fit function adopted (R500B and cubic spline, respectively)\footnote{More information is available in the user manual of OSIRIS: \href{http://www.gtc.iac.es/instruments/osiris/media/OSIRIS-USER-MANUAL_v3_1.pdf}{http://www.gtc.iac.es/instruments/osiris/media/OSIRIS-USER-MANUAL\_v3\_1.pdf}.}, which results in an additional uncertainty of $\Delta z\sim 0.004$ in the estimation of the redshifts.

\begin{figure*}
    \centering
    \sidecaption
    \includegraphics[width=0.7\textwidth]{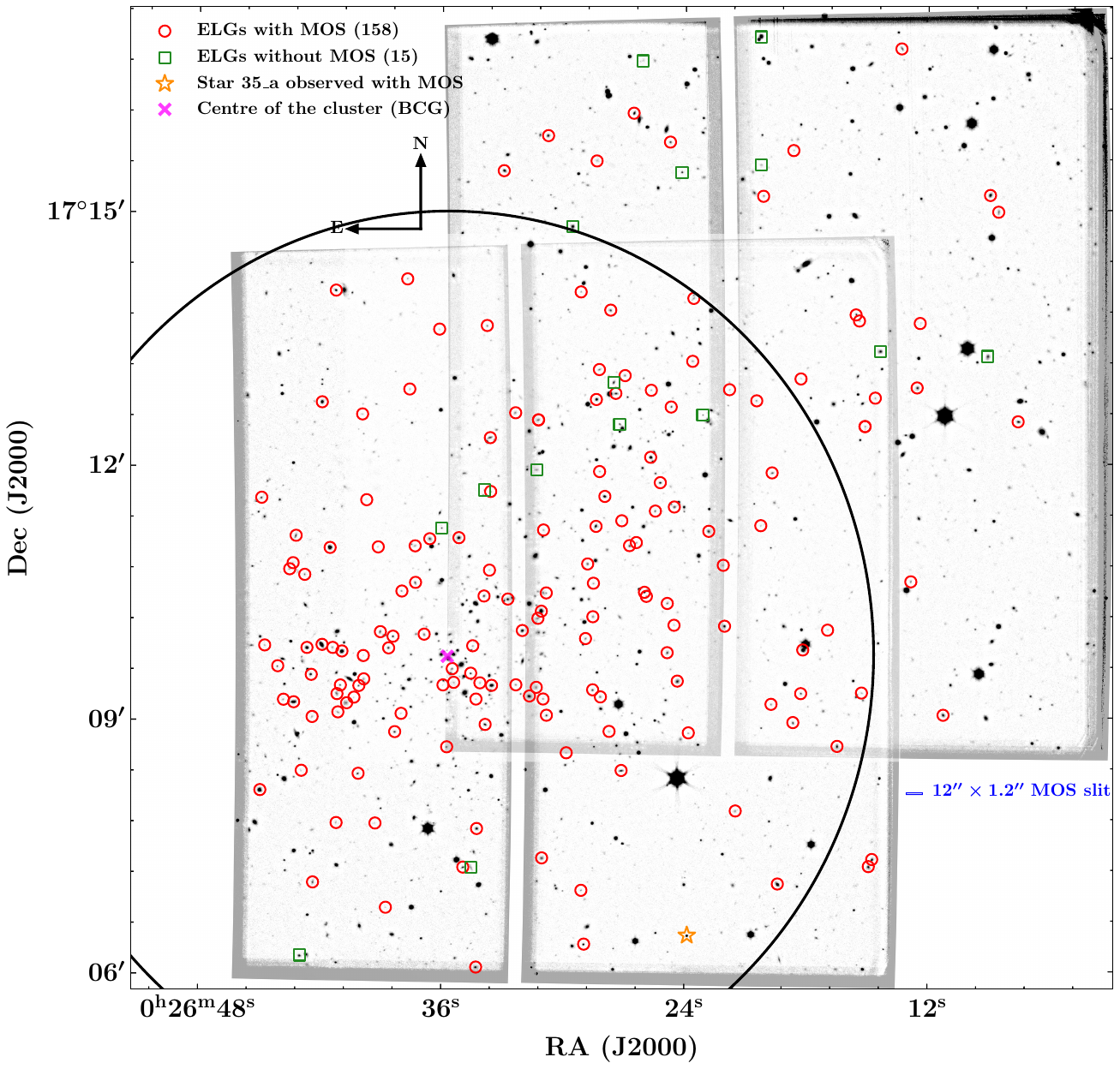}
    \caption{Two OSIRIS/GTC pointings towards Cl0024 with the ELGs catalogued by GLACE. Red circles show the 158 galaxies with MOS observations, the green squares show the 15 galaxies that were not observed with MOS, and the orange star shows the position of the source identified as an M-type star. The violet cross shows the centre of the cluster based on the brightest cluster galaxy (BCG) and the black circle denotes the $R_{200}$ of 1.7 Mpc \citep{Treu2003}. We also show the typical size of the MOS slits in the lower-right corner of the mosaic, as well as the north-east directions below the legend.}
    \label{fig:mosaic_Cl0024}
\end{figure*}

In total, out of 174 targets, 161 spectra were obtained for 159 sources\footnote{Please note that we refer in the article to some galaxies using their ID. The ID of the galaxies is formed by a number (from 4 to 41000, followed by `\_a' or `\_b'; the letter refers to the pointing and not to the kinematical structure in which the galaxy is found.}. (galaxies {\tt 647\_b} and {\tt 657\_b} have two spectra each). In addition, the source \texttt{35\_a} appears to be an M-type star. After comprehensive visual inspection, this object appears to be the only one with a star-like MOS spectrum. We removed it from all the ancillary data we use. We ended up with 160 MOS spectra for 158 galaxies, for a total of 173 ELGs, and 365 galaxies in total included in the footprint of the GLACE observations. We show the position of each source in \cref{fig:mosaic_Cl0024}.

In \cref{sec:examples_spectra}, we present a selection of MOS spectra with different signal-to-noise ratios (S/Ns) obtained from the reduction of data. The large range of S/Ns from the MOS spectra emission lines (between 1.3 up to 140) poses a challenge in the analysis because false detections can easily occur. Nevertheless, a majority of spectra have enough S/N to correctly and unequivocally identify the emission lines and the initial large sample size allows for a substantial number of spectra that can be used to study the SF in Cl0024.

\subsection{Multi-object spectroscopy data calibration}
The calibration of the MOS data was performed in several steps. The first task is to identify the emission lines for each spectrum. We have access to five main emission lines: \OII$\lambda\lambda3727,3729$, \OIII$\lambda\lambda4959,5007$ and $\Hbeta$. However, due to the spectral resolution of the data ($\Delta\lambda$$\approx$3.54 \AA), we cannot resolve the two lines of \OII, so we refer to the blended emission line as \OII, and we only use \OIII$\lambda5007$ in our analysis, the strongest of the two \OIII\ forbidden lines. In summary, for the purposes of this work we have attempted to identify \OII, $\Hbeta$ and \OIII$\lambda5007$.

For each visible emission line (some lines are not distinguishable from the stellar continuum), we applied a Gaussian fit to get the position, amplitude, and $\sigma$ of the line. After several tests, we decided to use a combination of manual and automatic fitting processes to obtain the most accurate results: we first used {\tt IRAF}\footnote{IRAF (Image Reduction and Analysis Facility) is a collection of software written at the National Optical Astronomy Observatory (NOAO) geared towards the reduction of astronomical images and spectra.} task \texttt{noao.onedspec.splot} to manually fit all the emission lines, using the redshift estimated with the $\Halpha$ line (from \citetalias{SP15} or \citealt{Cedres2024}); then, using Python packages \texttt{astropy.modelling}, \texttt{specutils.fitting} and \texttt{specutils.spectra} we made more accurate fittings, using \texttt{splot}'s results as prior. Once all the detected lines have been fitted, we obtained the following line properties: redshift, amplitude, full width at half maximum (FWHM), flux ($F$), continuum level, equivalent width (EW), and S/N of the line (estimated using eq. 1 from \citealt{Rodrigues2016}).

In addition, to obtain reliable galaxy fluxes, we adjust the fluxes in the MOS spectra and correct them for galactic reddening and extinction, stellar absorption, dust extinction, and the effect of metallicity on the SFR. We describe the main calculations for the corrections and calibrations in \cref{sec:data_corrections}.

\subsection{Star formation rate estimations}
One of the main products that the new MOS observations obtain are SFRs. Thus, the data now allow the SFR to be estimated using not only the $\Halpha$ emission line (using the TF observations) but also the $\Hbeta$ and \OII\ emission lines (from the MOS observations).

In the case of indicators based on the emission lines of the spectra of galaxies, the most common conversions between luminosity and SFR are those given by \cite{Kennicutt1998}. The conversion factors used in this study, initially presented in \cite{Hao_SFR_indic} and \cite{Murphy_SFR_indic}, rely on the outcomes derived from the {\tt STARBURST99} population synthesis code \citep{Leitherer_SFR_indic,Vazquez_SFR_indic}, but also on observations \citep{Osterbrock_1989}. These factors assume specific conditions, including a constant SF history, solar metallicity, a certain IMF, and a stellar population age of 100 Myr. We use this estimation, updated by \cite{KennicuttEvans2012} and adapted to the IMF of \cite{Chabrier_IMF}\footnote{The SFR using a specific IMF can be easily converted to another; here, we just multiplied by 0.63 to get the SFR for \citeauthor{Chabrier_IMF}'s IMF from the SFR calculated using the IMF of \cite{Salpeter_IMF}, as in \cite{Madau_Dickinson}.} to calculate the SFR of $\Halpha$ emission line. The equation for the $\Halpha$ SFR indicator is presented in \cref{eq:SFR_Ha}:

\begin{equation}\label{eq:SFR_Ha}
    {\rm SFR}\left({\rm M_\odot \; yr^{-1}}\right)_{\rm \Halpha} = 4.977 \times 10^{-42}\ \left[\frac{L\left(\Halpha\right)}{\rm erg\; s^{-1}}\right],
\end{equation}

\noindent where $L(\Halpha)$ is the luminosity of $\Halpha$ emission line, calculated using the corrected flux ($L = 4\pi d^2 F$, where $d$ is the luminosity distance and $F$ is the flux).

In our study, we primarily use the $\Halpha$ SFR since it tends to be more reliable due to the higher S/N of this emission line. However, we also use the $\Hbeta$ and \OII\ SFRs for several reasons. They are used in cases where the $\Halpha$ is unreliable (see \cref{sec:results_redshifts} for details on why $\Halpha$ might not be reliable and the cases when this happens) as well as to ensure the quality of the MOS data (see \cref{sec:data_corrections}) and provide these value-added products from the MOS observations.

In the case of $\Hbeta$ SFR, we converted the SFR of $\Halpha$ into the SFR of $\Hbeta$ using $(\Halpha/\Hbeta)_{\rm{int}} = 2.86$, as we did previously when calculating the colour excess. The equations for $\Halpha$ and $\Hbeta$ SFR indicators are presented in \cref{eq:SFR_Hbeta}:

\begin{equation}\label{eq:SFR_Hbeta}
    {\rm SFR}\left({\rm M_\odot \; yr^{-1}}\right)_{\rm \Hbeta} = 1.423 \times 10^{-41}\ \left[\frac{L\left(\Hbeta\right)}{\rm erg\; s^{-1}}\right],
\end{equation}

\noindent where $L(\Hbeta)$ is the luminosity of $\Hbeta$ emission line, calculated in a similar way to $L(\Halpha)$.

Finally, to estimate the \OII\ SFR, a correction needs to be applied based on the gas metallicity. Several studies \citep[e.g.][]{Hammer1997, Jansen_metal_SFR, Cardiel_SFR_met} showed that the luminosity of \OII\ is highly affected by metallicity. According to \cite{Jansen_metal_SFR}, not taking into account this effect may result in an overestimation of a factor of 3 in the SFR calculated with this emission line, which is why we choose to use the improved \cite{Kennicutt1998} calibration proposed by \cite{Kewley2004}.

\cite{Kewley2004} introduced an SFR indicator based on the ratio between \OII\ and $\Halpha$, incorporating considerations of metallicity. They presented the results for four indicators of metallicity: the \cite{KD02_met} \NII$\lambda6583$/\OII\ diagnostic, the \cite{M91_met} R23 diagnostic (hereafter M91), the \cite{Z94_met} R23 diagnostic (hereafter Z94), and the \cite{C01_met} `case F' diagnostic. The corrected SFR is

\begin{equation}\label{eq:SFR_OII}
    {\rm SFR}\left({\rm M_\odot \; yr^{-1}}\right)_{\rm \OII} = \frac{4.977\times10^{-42}}{a\; \left[{\log({\rm O/H})+12}\right]+b} {\left[\frac{L\left({\rm \OII}\right)}{\rm erg\; s^{-1}}\right]},
\end{equation}

\noindent where $\log(\rm{O/H})+12$ is the metallicity and $a$ and $b$ are the slope and y-intercept of the $\log(\rm{O/H})+12$ versus the \OII/$\Halpha$ ratio, respectively (Fig. 6 of \citealp{Kewley2004}).

We decided to use the Z94 \citep{Z94_met} and M91 \citep{M91_met} indicators to apply this correction for two reasons: first, they are the indicators with the best results in \cite{Kewley2004}, and second, they only rely on the R23 diagnostic \citep{Pagel1979}, and not directly on $\Halpha$ or \NII$\lambda6583$ fluxes (though we have access to those lines, we prefer to study the SFR from the \OII\ line with an semi-independent estimation of the metallicity). Nonetheless, the M91 diagnostic requires an initial guess of the oxygen abundance to determine which branch of the R23 curve to use, which is why this indicator is only semi-independent. Here we use the N2 \citep{pettini_oiiinii_2004} abundances estimated by \cite{Cedres2024}.

\subsection{Comparison samples}\label{sec:comparison_samples}
In our analysis, we make use of several samples to compare with our findings regarding Cl0024. In \cref{tab:comparison_samples} we display a summary of each sample, and a detailed description of the samples is provided in the subsequent sections.

\begin{table}
\centering
\caption{Properties and sizes of the comparison samples.}
\label{tab:comparison_samples}
\resizebox{\hsize}{!}{
\begin{tabular}{ccccc}
\hline\hline
Survey/Cluster & Environment & N & ${<z>}$ & Reference                         \\ \hline
Cl 0939+4713            & Cluster              & 62         & 0.41                  & \cite{Sobral2016}                 \\\\
RX J2248-443            & Cluster              & 58         & 0.348                 & \cite{Ciocan2020}                 \\\\
VVDS                    & Field                & 248         & 0.39                  & \cite{le_fevre_vimos_2005}           \\\\
\multirow{2}{*}{LCS}    & Cluster              & 497        & \multirow{2}{*}{0.03} & \multirow{2}{*}{\cite{Finn2018}}  \\
                        & Field                & 8220       &                       &                                   \\ \hline
\end{tabular}}
\tablefoot{The first column is the name of the survey or cluster used; the second column is the type of large-scale environment (cluster or field); the third column is the number of SFGs in each dataset; the fourth column is the mean redshift of each sample; and the fifth column is the reference where these data come from.}
\end{table}

\paragraph{\textit{Cl 0939+4713 and RX J2248-443}}
Cl 0939+4713 \citep{Sobral2016} and RX J2248-443 \citep{Ciocan2020} are two galaxy clusters at $z\sim0.41$ and $z\sim0.348$, respectively. \cite{Sobral2016} observed galaxies farther away from the cluster centre than GLACE, and \cite{Ciocan2020} observed a cluster ten times more massive than Cl0024. For Cl 0939+4713, we selected the SFGs confirmed by \cite{Sobral2016} which have a S/N $\geq5$ (flag < 4). For RX J2248-443, we used the SFGs based on the \cite{Kauffmann2003} criterion.

\paragraph{\textit{VVDS field sample}}
We selected 440 field galaxies from the VIMOS VLT Deep Survey (VVDS, \citealt{le_fevre_vimos_2005}), as a sample of field galaxies to compare with the cluster environment. This survey has a sensitivity close to that of the GLACE survey for Cl0024 and the required data to reproduce our AGN discrimination using the Baldwin-Phillips-Terlevich \citep[BPT;][]{BPT} diagnostic. The criteria of selection for the VVDS sample were the following: sources with a reliable redshift ({\tt z Flag}=[2, 3, 4, 9]) between 0.35 and 0.45, with $M_\star>10^{8}\ {\rm M_{\odot}}$, and a $R$ band magnitude brighter than 24 mags ({\tt MAG\_AUTO\_R\_CORR} < 24.21). To select the SFGs, we used the \cite{Kauffmann2003} criterion from the BPT analysis, which yields an initial number of 55 galaxies; however, because of the low resolution of the data ($R\sim230-300$ at $z\sim0.4$) and the poor spectral response of VIMOS at wavelengths of $\sim9200\ \AA$ (where \NII$\lambda6583$ emission lies at this redshift), 329 galaxies do not have any reliable flux for the \NII$\lambda6583$ emission line. For this reason, we cannot perform the classical BPT diagnostics on these galaxies and confidently select the SFGs. Therefore, to the initial sample of 55 SFGs we added 193 galaxies with confident estimates for the $\Halpha$, $\Hbeta$ and \OIII$\lambda5007$ emission lines and unknown flux for \NII$\lambda6583$, assuming they are SFGs. This yields a total of 248 SFGs in the field out of 440 in total.

\paragraph{\textit{Local Cluster Survey}}
The Local Cluster Survey (LCS) is a composite survey presented in \cite{Finn2018} and formed by nine groups and clusters at $0.02 < z < 0.04$. Details on the selection process can be found in Sect. 2 of \cite{Finn2018}. This survey has the great advantage of presenting both field and cluster galaxies (divided in infall, and core galaxies), which we can use to study in more detail the population of galaxies within their specific environment.

\subsection{{\tt LePhare} and {\tt CIGALE} spectral energy distribution fitting}\label{sec:CSP_fitting}
We used {\tt LePhare} to estimate general parameters of the galaxies without assuming any a priori AGN or SF contribution. Once the galaxies were classified using the BPT and WHAN \citep{Cid_Fernandes_LINERs} diagnostics (this classification is described in \cref{sec:diagnostics}), we employed {\tt CIGALE} to estimate parameters such as $E_{\rm s}(B-V)$ (used in \cref{sec:extinction_correction}), the 4000 \AA\ break ($D4000$ index, \citealt{D4000_index}) and $M_{\rm \star}$. In \cref{sec:sed_fittings_parameters} we show the parameters used to fit the spectral energy distribution (SED) of the galaxies in the cluster, with {\tt LePhare} and {\tt CIGALE}, respectively.

We ran {\tt CIGALE} on 363 of the full sample of 365 galaxies (see \ref{sec:ancillary} to understand how this number was obtained). Galaxies {\tt 358\_a} and {\tt 938\_a} lacked the minimum amount of photometric data to fit their SED correctly using CIGALE. We divided the sample of 363 galaxies in five distinct subsamples: SFGs with \cite{Moran2005} photometry (94), SFGs with Pan-STARRS photometry (6), AGN hosts/composite galaxies and unclassified galaxies initially classified as AGN or broad-line AGN (BLAGN)\footnote{The AGNs that display broad permitted and semi-forbidden emission lines in their spectra, with FWHMs of a few to several thousand kilometres per second.} by \citetalias{SP15} (respectively, 30 and 25), unclassified galaxies initially classified as SFG by \citetalias{SP15} (16), and non-ELGs (192).

\begin{figure}
     \centering
     \includegraphics[width=\hsize]{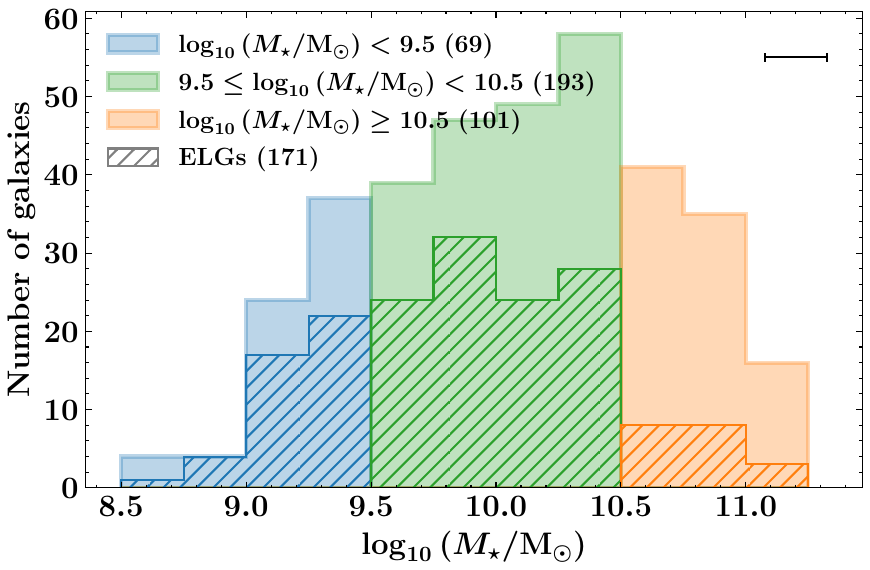}
     \caption{Distribution of $M_\star$ computed using {\tt CIGALE} for 363 galaxies of the cluster (171 ELGs and 192 non-ELGs). The distribution is divided into three segments: low-mass (in blue), intermediate-mass (in green), and high-mass galaxies (in orange). We show the distribution of ELGs with the hatched histograms. In the legend, between parentheses, we include the number of galaxies within each subsample, and in the upper-right corner we show the median of the uncertainty bars of $M_\star$.}
     \label{fig:Stellar_mass_histogram}
\end{figure}

In \cref{fig:Stellar_mass_histogram}, we show the distribution of $M_\star$ estimated using {\tt CIGALE}, successfully achieved for 363 galaxies, including 171 out of the 173 ELGs selected from the TF observations and 192 non-ELGs (later used in \cref{sec:local_environment} to estimate accurate fractions). We divided the histogram in three categories: the low-mass galaxies, with a mass $\log_{10}(M_\star/{\rm M_\odot})<9.5$, intermediate-mass galaxies with a mass $9.5\leq\log_{10}(M_\star/{\rm M_\odot})<10.5$ and high-mass galaxies with a mass $\log_{10}(M_\star/{\rm M_\odot})\geq10.5$. The majority of the galaxies lie in the middle segment (53\%) and at the high-mass end of the histogram (28\%), except for ELGs that have more galaxies at intermediate mass range.

\section{Results: MOS fittings, SFRs, and relations with cluster environment}\label{sec:results}
In this section we present our results based on the TF and MOS observations of the ELGs of Cl0024. We have created a multi-wavelength catalogue that summarises the data and the results obtained for the available SFGs (classification in \cref{sec:diagnostics}). In addition, an atlas of the results with MOS spectra, fits and additional information is also available on request. A full version of the catalogue, including all 173 ELGs, will be available in Sánchez-Portal et al. (in prep.).

\subsection{Redshifts}\label{sec:results_redshifts}
\begin{figure*}
        \centering
        \includegraphics[width=\textwidth]{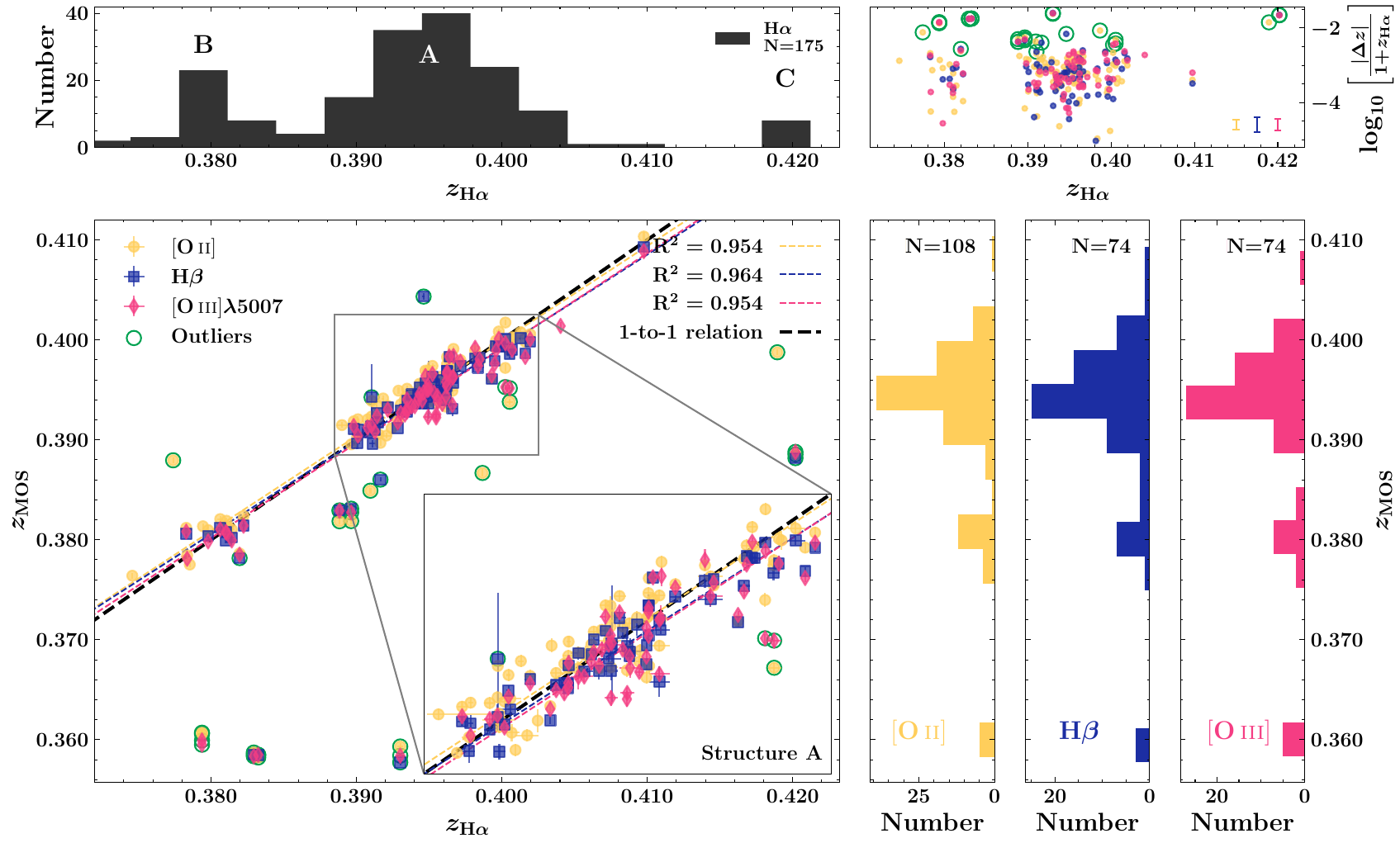}
        \caption{Results of the line fitting process applied on the 161 MOS spectra. \textit{(Lower-left panel)} The main plot of this figure shows the values of the MOS redshifts calculated using \OII\ (yellow), $\Hbeta$ (indigo), and \OIII$\lambda5007$ (magenta) emission lines with respect to $\Halpha$ redshifts (from \citetalias{SP15} or \citealt{Cedres2024}), in both cases with their respective uncertainties as error bars. The symbols of each emission line are shown in the upper-left region. The green circles show outliers, which are galaxies with MOS redshifts that are different from the $\Halpha$ redshifts. We linearly fitted each distribution (shown with the coloured dashed lines), taking care to exclude outliers. The dashed black line represents the one-to-one relation. In the lower-right corner we show a zoomed window on the structure A. \textit{(Upper-left panel)} The histogram shows the distribution of $z_{\Halpha}$ with the position of each structure (A, B and C) at their corresponding redshift and the number of detections in the upper-right corner (including duplicated $z_{\Halpha}$ for galaxies {\tt 647\_b} and {\tt 657\_b} that have two MOS spectra each). \textit{(Lower-right panel)} The histograms show the three distributions of MOS redshifts (\OII\ in yellow, $\Hbeta$ in indigo, and \OIII$\lambda5007$ in magenta) with the number of detections at the top. \textit{(Upper-right panel)} The scatter plot shows how distant are the MOS redshifts from the $\Halpha$ distribution, using $\log_{10}\left(|z_{\rm MOS}-z_{\rm \Halpha}|\right)/(1+z_{\rm \Halpha})$. These values were used to identify the outliers shown with green circles. We show in the lower-right corner the medians of the uncertainty bars of each distribution.}
        \label{fig:redshift_comparison}
    \end{figure*}

One of the direct parameters obtained after the reduction of the MOS data is the spectroscopic redshift of the galaxies. In \cref{fig:redshift_comparison} we compare between the redshifts obtained for each emission line of the MOS spectra ($z_{\rm MOS}$) and the redshift calculated using the $\Halpha$ emission line ($z_{\rm H\alpha}$, as explained in \cref{sec:ancillary}).

\begin{figure}
        \centering
        \includegraphics[width=\hsize]{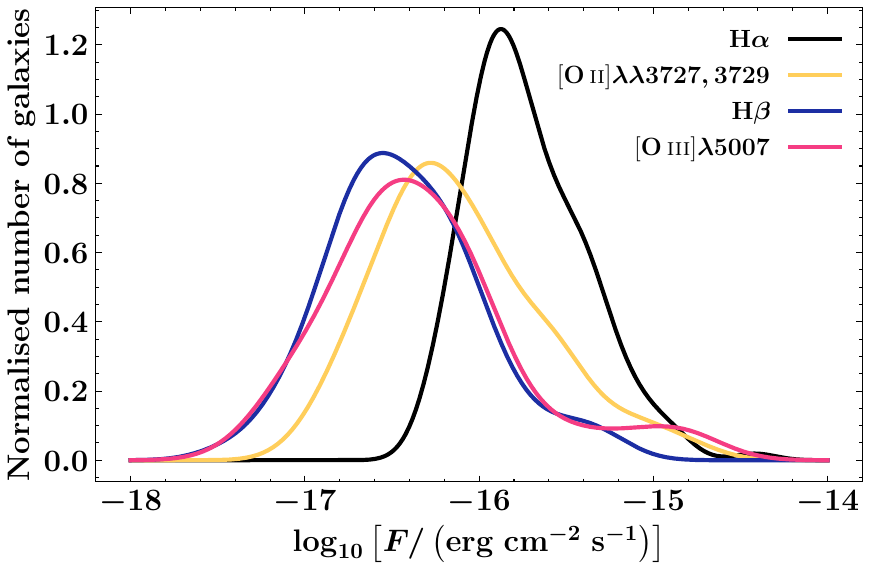}
        \caption{Distribution of the uncorrected fluxes of the emission lines. MOS emission lines generally have a lower flux than $\Halpha$. Among them, \OII\ fluxes are usually higher, while $\Hbeta$ and \OIII$\lambda5007$ fluxes display a similar distribution.}
        \label{fig:fluxes_hist}
    \end{figure}

Out of 160 spectra, we obtain spectroscopic redshifts for 108 \OII\ emission lines (67.50\%), 74 $\Hbeta$ (46.25\%) emission lines, and 74 \OIII$\lambda5007$ emission lines (46.25\%). The excess of detections of the \OII\ line can be explained by the higher fluxes in comparison to $\Hbeta$ and \OIII$\lambda5007$ emission lines, as shown in \cref{fig:fluxes_hist}. These fluxes are expected, as this emission line is the combination of two forbidden lines. Additionally, MOS observations targeted this particular emission line. The $\Hbeta$ and \OIII$\lambda5007$ percentages are similar due to their proximity on the spectrum, so the quality of the spectra will affect them similarly. Out of 109 spectra where at least one emission line was detected, we were able to measure the parameters of the three emission lines in 65 spectra (59.6\%), while we found only two emission lines in 17 spectra (15.6\%) and only one emission line in 27 spectra (24.8\%). This shows that in 75\% of the cases, the detection of emission lines leads to more than one detected emission line, and in 50.6\% of all the MOS spectra.

We used a Monte Carlo method similar to that used in \cite{Cedres2024} to linearly fit each distribution of redshifts in \cref{fig:redshift_comparison}. The values were uniformly randomised within the error bars and the fitting process was iterated 10\,000 times. The resulting fit is represented by the mean values of the parameters, specifically the slope and intercept. We can confirm the general agreement between the redshift estimated using TF data and MOS spectra, with linear correlation factors of 0.954, 0.964 and 0.954 for \OII$\lambda\lambda3727,3729$, $\Hbeta$ and \OIII$\lambda5007$, respectively. Overall, the precision of the redshifts, defined as ${|z_{\rm \mathrm{line}}-z_\Halpha|/(1+z_\Halpha)}$ in previous works \citep{precision_redshift_Hildebrandt,precision_redshift_Diego} and depicted in the upper-right panel of \cref{fig:redshift_comparison}, is varying between $10^{-2}$ and $10^{-4}$. This therefore confirms the spectroscopic quality of the MOS-derived redshifts.

Overall, MOS data result in several improvements with respect to the $\Halpha$ TF. We observe that structures A (centred at $z\simeq0.395$ with 133 objects) and B (centred at $z\simeq0.380$ with 40 objects) are clearly identified with the redshifts from the three emission lines. The structure C, detected behind the cluster central region at $z\sim0.42$, is missing. In the analysis made by \citetalias{SP15}, only 8 galaxies composed this region, and we only could detect emission lines in the spectra for one of them, {\tt 359\_a}, that can be attributed to the main structure A. This does not mean that the third structure does not exist, especially because \cite{Czoske2001,Czoske2002} also detected it, but with the data at our disposal we could not confirm its existence or characteristics.

We observe 33 `outliers': 13 redshifts calculated with \OII\ line (39.4\%), 10 calculated with $\Hbeta$ line (30.3\%) and 10 calculated with \OIII$\lambda5007$ line (30.3\%). We estimated the deviation of the 3 distributions of $|z_{\rm \mathrm{line}}-z_\Halpha|/(1+z_\Halpha)$ 
(one per line) using a Gaussian function and defined outliers as galaxies with a position in the histogram $|x|>\mu+3\sigma$. We think that these differences come from a false detection of the $\Halpha$ emission line. In the case of TF results, the derivation of the properties of the $\Halpha$ and \NII$\lambda6583$ lines was made under the assumption that the main visible feature in each pseudo-spectrum was $\Halpha$; but because only one line was visible, it could have been another line, such as [\ion{O}{I}]$\lambda6364$ or [\ion{S}{II}]$\lambda\lambda6716,6731$.

These variations in the spectroscopic redshifts of the galaxies can also change the structure in which the galaxies were originally observed. Two groups of redshifts (from two different galaxies) located at $z_{\rm \mathrm{MOS}}\sim0.382$ previously had $z_\Halpha\sim0.389$. In this case, instead of being part of structure A, they are probably located in the structure B. A similar case happens for a galaxy initially located in the structure C at $z_\Halpha\sim0.42$: the three emission lines of its MOS spectrum indicate a redshift of $z_{\rm \mathrm{MOS}}\sim0.382$, which would place this galaxy in the main structure A instead. Since we detected the three emission lines in the MOS spectrum of each galaxy, we favour that our estimation is correct and the $\Halpha$ emission line that was detected was probably another emission line. 

Out of the 33 outliers, 8 galaxies have a confirmed redshift calculated using MOS spectra (estimated with at least two emission lines) different from the $\Halpha$ redshift calculated with TF pseudo-spectra. After a thorough revision of the TF pseudo-spectra and MOS fittings, we found that the redshift of those galaxies ({\tt 96\_a}, {\tt 105\_a}, {\tt 359\_a}, {\tt 384\_b}, {\tt 433\_a}, {\tt 647\_b}, {\tt 657\_b}, {\tt 888\_a}) was poorly estimated using the pseudo-spectra, and the fitted line was probably [\ion{S}{II}]$\lambda\lambda6716,6731$. Galaxies {\tt 359\_a}, {\tt 433\_a}, {\tt 657\_b} and {\tt 888\_a} are included in this analysis using the revised and more accurate MOS redshift. Galaxies {\tt 96\_a}, {\tt 105\_a}, {\tt 384\_b} and {\tt 647\_b} happen to have a redshift $z\lesssim0.36$, which is much lower than the original TF estimate. They are possibly part of a new, foreground structure of the cluster (a new structure D), but for now we classify them as interlopers and will not include them in our sample. A future analysis of the galaxies at this redshift, focused on their dynamical properties, will determine if they fall within the outskirts of the cluster.

\subsection{Galaxy classification}\label{sec:diagnostics}
We need to distinguish the ionisation of SF from that produced by AGNs. One of the most common methods to do so is using the BPT analysis \citep{BPT}, but we also use the WHAN diagram, introduced by \cite{Cid_Fernandes_LINERs}, reproducing and following the method used in \citetalias{SP15}. An issue with the BPT diagnostic is that it requires at least four emission lines. However, similar results can be obtained using only two lines: \cite{Miller_EWan2}, \cite{Brinchmann2003} and \cite{S06} argued for a classification using the \NII$\lambda6583$/$\Halpha$ ratio only. Additionally, \cite{Cid_Fernandes_LINERs} presented the so-called WHAN diagram, that also uses the ratio between the \NII$\lambda6583$ and $\Halpha$ lines and the EW of $\Halpha$. We decided to use it for galaxies without $\Hbeta$ or \OIII$\lambda5007$ detections. Because \NII$\lambda6583$/$\Halpha$ is known for 156 galaxies of the sample,\footnote{The \NII$\lambda6583$ emission line flux from \citetalias{SP15} is not available for 19 spectra but for one of them the inverse convolution had a good enough quality to be used in our study, and another one was classified as BLAGN and was automatically removed. In total, the \NII$\lambda6583$/$\Halpha$ ratio is known for 156 galaxies.} we were able to characterise the ionisation source of the majority of the ELGs.

Prior to this, we used the classification of 25 BLAGN of \citetalias{SP15}, based on the FWHM of the $\Halpha$ line, to remove them from our sample of galaxies\footnote{\cite{Cedres2024} did not make a new classification based on the results of the inverse convolution because the software was not optimised yet to separate the components of these composed objects. A future update should allow for a better estimation of the fraction of BLAGN.} because when the $\Halpha$ line becomes broad, it is extremely difficult to separate $\Halpha$ and \NII$\lambda6583$ components, and they may be catalogued as SFGs. Hereafter, if we do not explicitly refer to them, BLAGN have been removed from our analysis, and we use the term `AGN' to refer to narrow-line AGN since the analysis of the population of AGN hosts did not meet the scope of this study. This study will be presented in a future article (Sánchez-Portal et al. in prep.).

\begin{figure}
        \centering
        \includegraphics[width=\hsize]{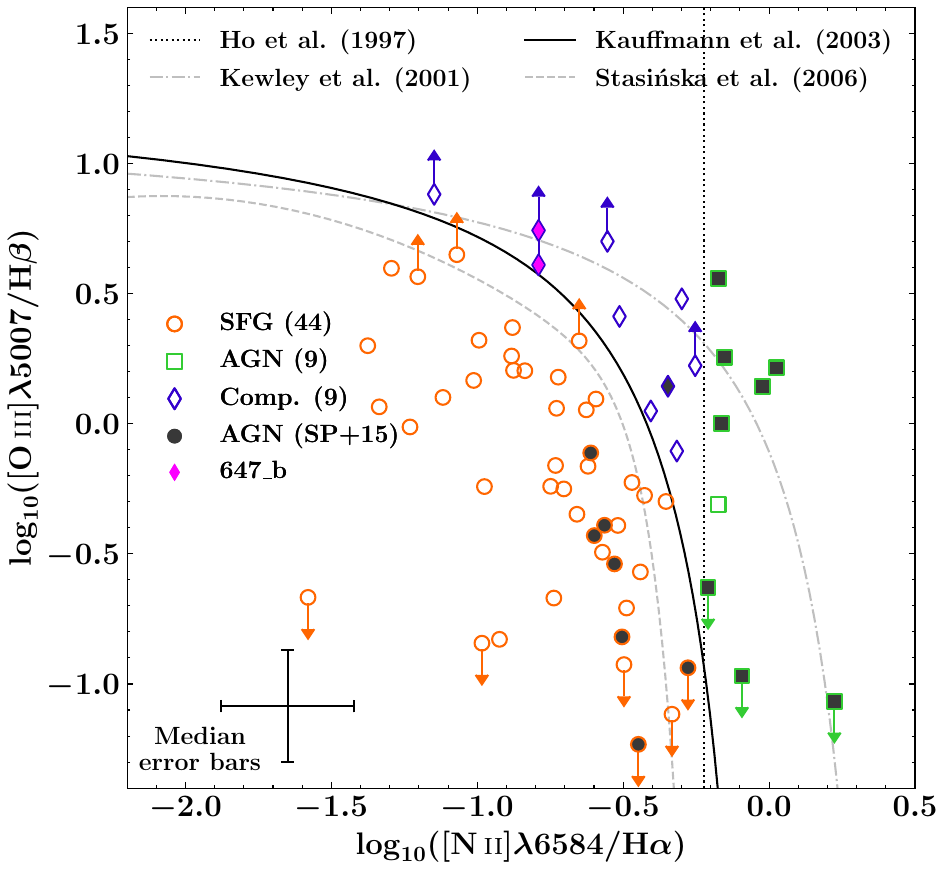}
        \caption{Baldwin-Phillips-Terlevich diagram of 62 galaxies of Cl0024. The position in the BPT diagram of each galaxy was calculated using \OIII$\lambda5007/\Hbeta$ and \NII$\lambda6583/\Halpha$ ratios when either \OIII$\lambda5007$ or $\Hbeta$ flux was available; in the case where only one of the two fluxes was available, an upper limit of the flux, estimated from the continuum level, was used (arrows on the diagram). The prescriptions of \cite{Ho1997}, \cite{Kewley2001}, \cite{S06}, and \cite{Schawinski2007} are shown in the diagram, as well as \cite{Kauffmann2003}'s, which we used as the separation between SFGs and AGN galaxies. Orange circles are SFGs, green squares are AGN galaxies and blue diamonds are composite galaxies. We also show the galaxies initially classified as AGN hosts by \citetalias{SP15} using grey-filled symbols and the two positions in the diagram of the galaxy {\tt 647\_b} with two pink diamonds since it has two MOS spectra with $\Hbeta$ detections. The median of the uncertainty bars are shown in the lower-left part of the figure.}
        \label{fig:BPT_diagram}
    \end{figure}

\cref{fig:BPT_diagram} shows the results of the BPT diagnostic. We note that in this diagram we correct for the stellar absorption of the $\Hbeta$ emission line using {\tt LePhare} SED fitting. Because we need both the \OIII$\lambda5007$ and $\Hbeta$ emission lines to use this diagnostic, we show the results for 62 galaxies with detections for either of the lines; when a line is undetected, we use an upper limit of its flux estimated with the continuum level in the region of the emission line, that we show with arrows. We show several classifications that can be used to separate AGN from SF-dominated galaxies \citep[i.e.][]{Ho1997, Kewley2001, Kauffmann2003, S06, Schawinski2007}. For the purpose of this work, we decided to use the prescription of \cite{Kauffmann2003}, which separates `pure SF' galaxies from any AGN or composite object, because it coincides with the prescription of \cite{Ho1997}, used for the WHAN diagram, as it is the same prescription used by \citetalias{SP15} and \cite{Zeleke2021}. The BPT diagram classifies the 62 galaxies into 44 SFGs, 9 AGN galaxies, and 9 composite galaxies (galaxies that are found between \cite{Kauffmann2003} and \cite{Ho1997} prescriptions).

\begin{figure}
        \centering
        \includegraphics[width=\hsize]{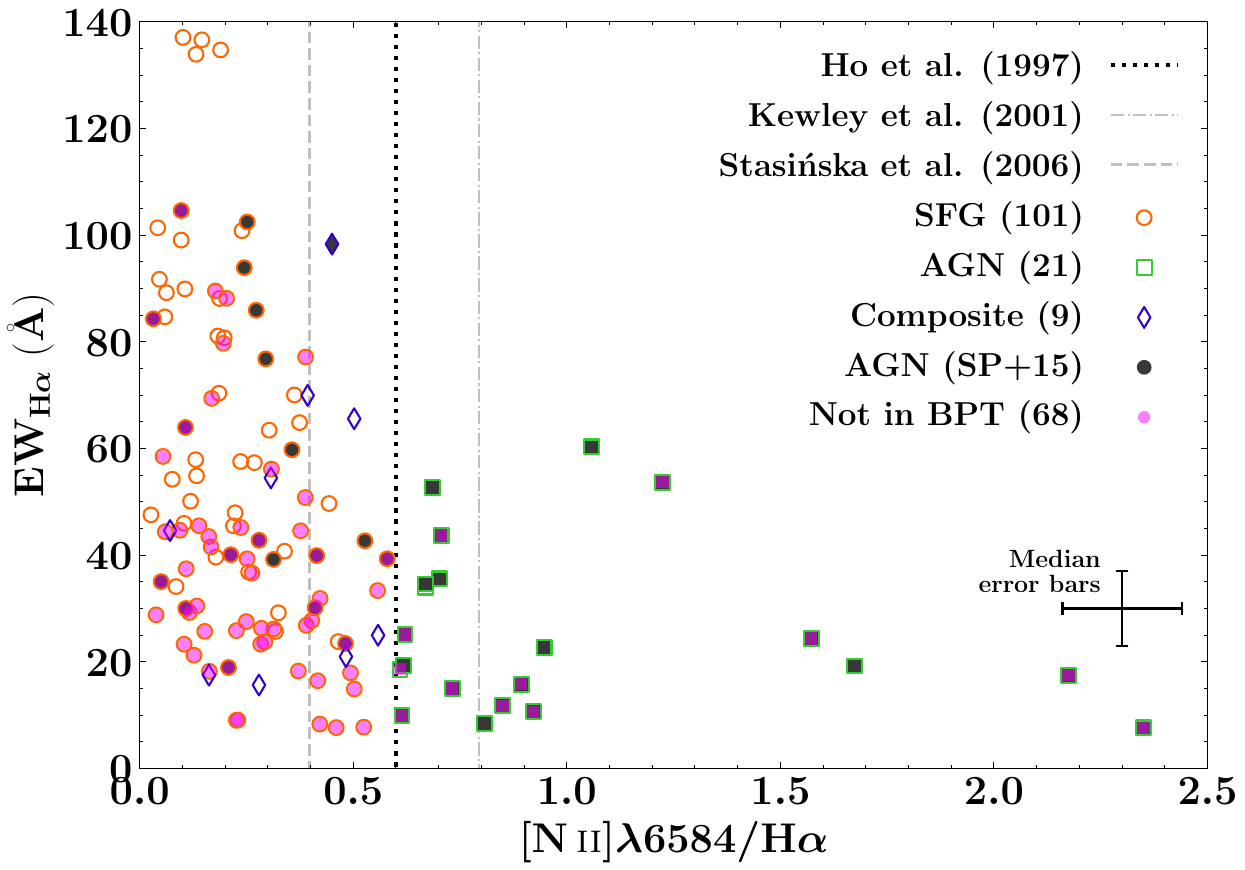}
        \caption{WHAN diagram of galaxies of Cl0024. 131 galaxies of the 173 ELGs are presented in this diagram. 42 galaxies are not present because they do not have a good estimation of \NII$\lambda6583$ flux or are BLAGN. The prescriptions of \cite{Ho1997}, \cite{Kewley2001} and \cite{S06} are shown in the diagram. The symbols are the same as in \cref{fig:BPT_diagram} and are described in the upper-right legend. We also show the galaxies that could not be classified in the BPT diagram with pink-filled symbols. The median of the uncertainty bars are shown in the lower-right part of the figure.}
        \label{fig:EWan2_diagram}
    \end{figure}

In \cref{fig:EWan2_diagram}, we show the WHAN diagram for 131 ELGs. We include the prescriptions of \cite{Ho1997}, \cite{Kewley2001} and \cite{S06}, which separate galaxies based on the \NII$\lambda6583$/$\Halpha$ ratio, and decided to use \cite{Ho1997} for its proximity with the \cite{Kauffmann2003} prescription in the BPT diagnostic. We obtain 101 SFGs, 21 AGN galaxies, and 9 composite galaxies. The 42 remaining galaxies in our sample are the 25 BLAGN and 17 galaxies without estimated \NII$\lambda6583$ flux. A summary of the results of the BPT and WHAN diagnostics is in \cref{tab:classification_BPT_EW}.

    \begin{table}
        \centering
        \caption{Summary of the final classification of ELGs into SFGs, composite galaxies, and AGNs.}
        \label{tab:classification_BPT_EW}
        \begin{tabularx}{\hsize}{X c}
        \hline\hline
            \rowcolor[HTML]{EFEFEF} Classification                  & Number (\%) \\ \hline
            \rowcolor[HTML]{FFFFFF} 
            Full sample                                             & 173 (100.0)          \\
            \rowcolor[HTML]{FFFFFF} 
            In BPT                                                  & 62 (35.8)            \\
            \rowcolor[HTML]{FFFFFF} 
            In WHAN                                                 & 131 (75.7)           \\ 
            \rowcolor[HTML]{EFEFEF} 
            SFGs (BPT - \citealt{Kauffmann2003})                    & 44 (71.0)            \\
            \rowcolor[HTML]{EFEFEF} 
            AGN (BPT - \citealt{Kauffmann2003})                     & 18 (29.0)            \\ 
            \rowcolor[HTML]{FFFFFF} 
            SFGs (WHAN - \citealt{Ho1997})                          & 110 (84.0)           \\
            \rowcolor[HTML]{FFFFFF} 
            AGN (WHAN - \citealt{Ho1997})                           & 21 (16.0)            \\ 
            \rowcolor[HTML]{EFEFEF} 
            SFGs (BPT + WHAN)                                       & 101 (58.4)           \\
            \rowcolor[HTML]{EFEFEF} 
            AGN (BPT + WHAN)                                        & 21 (12.1)            \\
            \rowcolor[HTML]{EFEFEF} 
            Composites (BPT + WHAN)                                 & 9 (5.2)              \\
            \rowcolor[HTML]{EFEFEF} 
            Unclassified                                            & 42 (24.3)            \\ \hline
        \end{tabularx}
    \tablefoot{Composite galaxies are classified as AGN galaxies in the BPT but SFGs in the WHAN. The unclassified sources are galaxies for which obtaining reliable measurements of the necessary emission lines for classification is not feasible or classified as BLAGN in \citetalias{SP15}.}
    \end{table}

This new classification of AGNs and SFGs using MOS emission lines also shows a substantial increment in the number of SFGs at the expense of the number of AGN galaxies, whose number was previously estimated to be 39 and is now 21. This `migration' of the former AGN population is clearly visible when we observe the two diagnostics: in the BPT diagram, seven galaxies that were previously classified as AGN hosts are now SFGs, and in the WHAN diagram, 20 misclassified AGN hosts are now SFGs. These new results are the consequence of more accurate estimations for the $\Halpha$ and \NII$\lambda6583$ emission lines since we partly used the results of the inverse convolution, but also because we now have access to more accurate fluxes for the \OIII$\lambda5007$ and $\Hbeta$ emission lines than the ones obtained by \citetalias{SP15} and \cite{Cedres2024}.

\subsection{Star formation rates}\label{sec:SFR_indicators}
\begin{figure}
        \centering
        \includegraphics[width=\hsize]{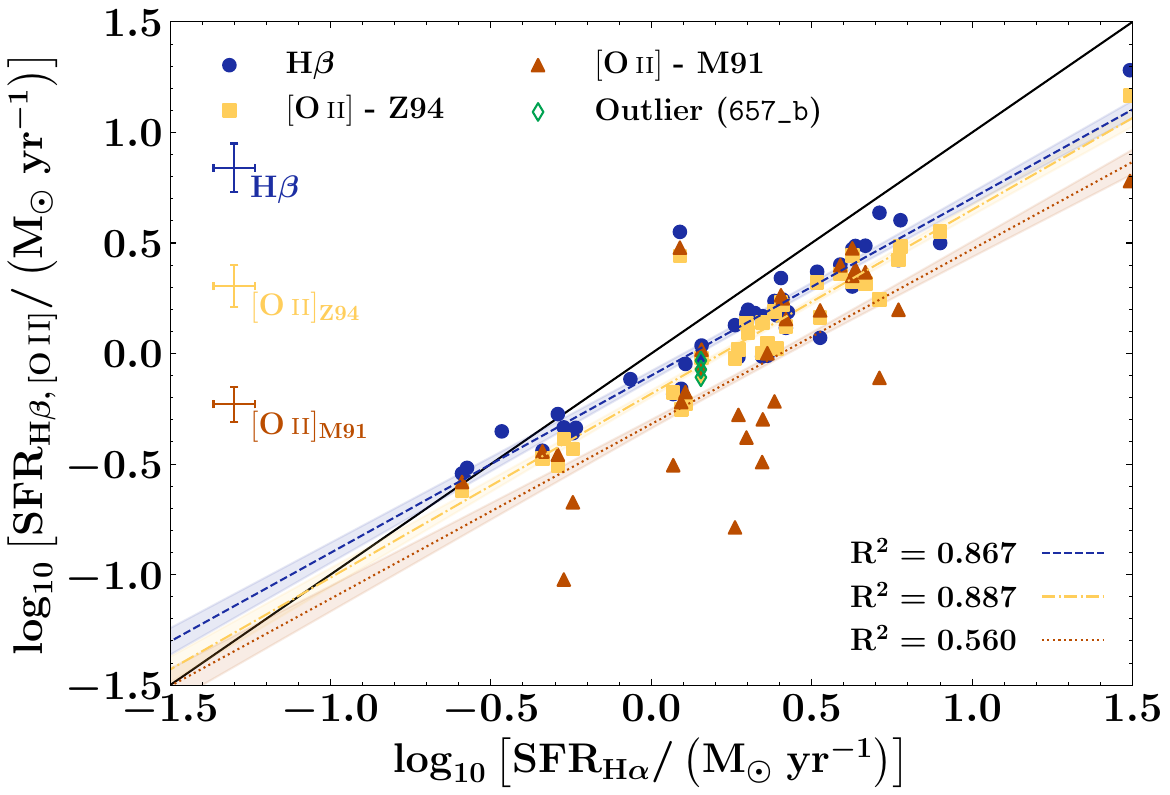}
        \caption{H$\beta$ and \OII\ SFRs with respect to $\Halpha$ SFR for the SFGs of our sample. We show the H$\beta$ (blue circles) and \OII\ SFR, with a correction based on the chemical abundance using Z94 (yellow squares) and M91 (brown triangles) gas metallicities. The solid black line is the one-to-one relation. The green diamonds  show the SFRs of one outlier, {\tt 657\_b}; the other seven outliers are non-SFGs, interlopers or do not have \OII\ or $\Hbeta$ SFRs). The coloured dashed lines represent the linear fittings of each distribution of SFRs, and the semi-transparent colour areas represent a $1\sigma$ deviation of each fit. On the left we show the medians of the uncertainty bars of each SFR distribution.}
        \label{fig:SFRs_comparison}
    \end{figure}

\cref{fig:SFRs_comparison} displays the SFRs calculated using $\Hbeta$ and \OII\ emission lines with respect to the $\Halpha$ SFRs. Using \cref{eq:SFR_Ha,eq:SFR_Hbeta,eq:SFR_OII}, from the sample of 101 SFGs, we are able to estimate the SFR of 100 galaxies using $\Halpha$ (99.0\%), 42 galaxies using H$\beta$ (41.6\% of the SFGs), 36 galaxies (35.6\%) using \OII\ corrected by Z94 metallicity and 31 galaxies (30.7\%) using \OII\ corrected by M91 metallicity. Overall, there is good agreement between the different tracers. The $\Halpha$ SFRs range from 0.25 to 31.05 $\mathrm{M_\odot\; yr^{-1}}$. However, when considering the \OII\ tracer corrected using M91 abundance, we observe a systematic underestimation and dispersion of the SFRs, with a correlation factor much lower than those of the other two estimates. This may be a consequence of the propagation of errors from the N2 estimator used to determine the branch of R23 to estimate M91 abundances. Consequently, we choose not to rely on the results from this tracer.

To obtain the results of this work, we primarily employed the $\Halpha$ tracer. This choice was driven by its superior S/N and its lesser susceptibility to potential effects involving \OII, H$\beta$, and \OIII$\lambda5007$ emission lines that would unavoidably introduce dispersion in analytical outputs. The MOS data have corroborated the findings obtained with the TF data, and any inconsistent observations between the two are excluded in future analyses. Out of the 8 outliers found initially (see \cref{sec:results_redshifts}), half of them are SFGs and we were able to estimate the MOS SFRs for only one of them ({\tt 657\_b}), depicted with the diamond symbols in \cref{fig:SFRs_comparison}. For this particular case, we cannot rely on the $\Halpha$ tracer since the emission line visible on the TF pseudo-spectrum may differ from $H\alpha$, and therefore we employed the SFR estimated using the \OII\ emission line, which has the highest S/N for this object.

In summary, we use the $\Halpha$ SFR for 97 galaxies and the \OII\ SFR for one outlier galaxy ({\tt 657\_b}), totalling 98 SFRs. The other three outlier SFGs are {\tt 105\_a}, an interloper galaxy, {\tt 433\_a}, an outlier galaxy without $\Hbeta$ or \OII\ SFR and {\tt 938\_a}, a galaxy without a reliable estimation of the corrected flux of its $\Halpha$ emission line (check \cref{sec:CSP_fitting} for more details on {\tt 938\_a}); these galaxies are removed from the final sample to avoid heterogeneity in the estimation of the SFRs.

\subsection{The $M_\star-{\rm SFR}$ relation}
\begin{figure*}
        \centering
        \includegraphics[width=\textwidth]{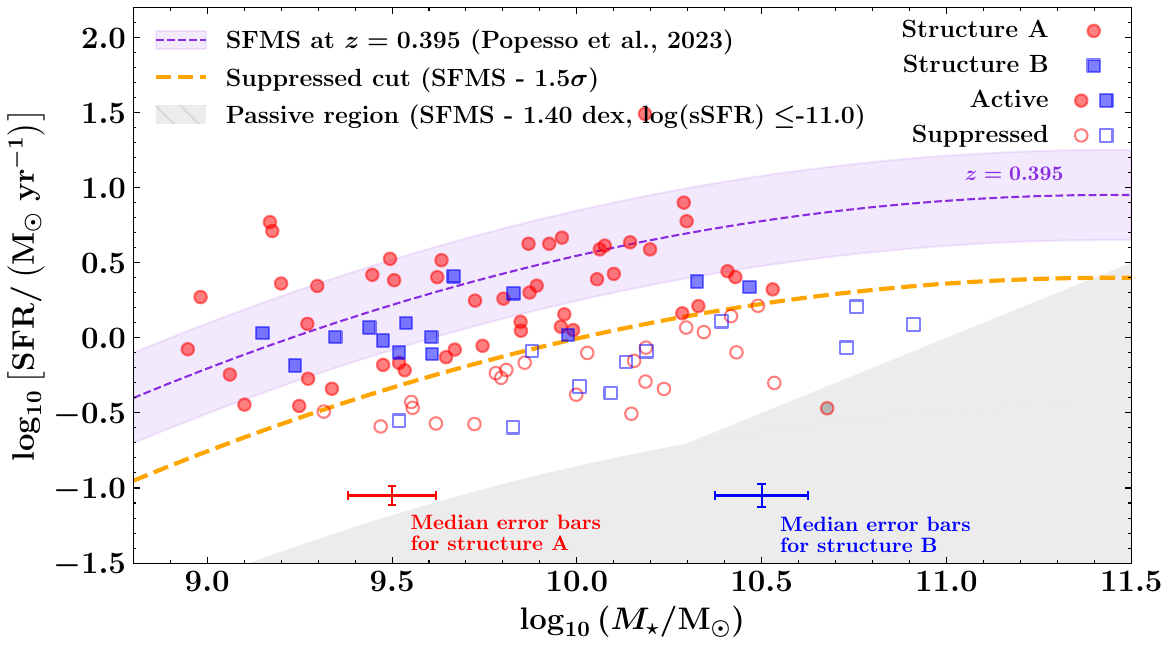}
        \caption{Relation of $M_\star-{\rm SFR}$ for the SFGs of Cl0024. Red circles are SFRs of galaxies from structure A and blue squares are SFRs of galaxies from structure B. The purple dashed curve is the SFMS of field galaxies at the redshift of Cl0024, based on \cite{Popesso2023}. The purple area around the curve represents a $\pm0.3$ dex uncertainty in the SFMS, which shows the active population of SFGs. The diagram includes a division between active and suppressed SFRs with the orange dashed curve and a passive region in the bottom right corner, in grey. Active galaxies, above the suppressed cut, have filled symbols and suppressed galaxies, below it, emptied symbols. The passive galaxy \texttt{882\_a} is shown with a grey-filled symbol in the passive region. The medians of the uncertainty bars for each distribution are shown at the bottom of the diagram.}
        \label{fig:SFR_mass}
    \end{figure*}

In \cref{fig:SFR_mass}, we present the relation between the $M_\star$ and SFRs of the SFGs. Among the 101 SFGs, we were able to confidently estimate 98 SFRs, and 99 $M_\star$, so we ended up with a total of 98 data points in the diagram. As stated in \cref{sec:Cl0024}, structures A and B display distinctive kinematic properties, leading us to showcase them separately in this figure using the separation $z_{\rm B} < 0.3876 < z_{\rm A}$. We also show the SFMS at $z=0.395$, the mean redshift of the cluster reported in \citetalias{SP15}, following the parametrisation of \cite{Popesso2023} shown in \cref{eq:popesso_MS}. This parametrisation is based on the analysis of 64 star-forming main sequences from 25 references covering a wide range of redshift ($0 < z < 6$) and $M_\star$ ($10^{8.5} - 10^{11.5}\ \mathrm{M_\odot}$):

    \begin{multline}\label{eq:popesso_MS}
        \log_{10}\left[{\rm SFR}/\left({\rm M_\odot\;yr^{-1}}\right)\right] = (-0.034 \;t + 4.722)\times\log_{10}(M_\star/{\rm M_\odot}) \\ - 0.1925\times\log_{10}^2(M_\star/{\rm M_\odot}) + (-26.1324 + 0.20 \;t),
    \end{multline}

\noindent where $t$ is the cosmic time. We observed that the majority of the SFGs lie below the SFMS. We also present a separation of the galaxies with active SFRs with the galaxies with suppressed SFRs and passive galaxies, based on the $M_\star-{\rm SFR}$. We describe the methodology followed to accurately separate each population in \cref{sec:SF_sup_pas}.

The SFMS at $z=0.395$ corresponds to the expected SFRs at the average redshift of Cl0024 when the Universe was 9.4 Gyr old, and what would be typically found for field galaxies. However, the cluster exhibits more advanced SF activity, with lower SFRs, and contains the majority of the galaxies whose SF has been suppressed below the SFMS. An Anderson-Darling test\footnote{We used {\tt scipy.stats.anderson\_ksamp} Python package. Populations are significantly different when the Anderson-Darling p-value<0.003.} also confirms that the distributions of SFRs for structures A and B are statistically similar (p-value=0.0429). Comparing the position of the galaxies in the $M_\star-{\rm SFR}$ diagram with the SFMS at the redshift of the cluster, we are able to determine that an important part of SFGs from the 173 ELGs are in a quenching phase, displaying lower SFRs than can be found in field galaxies at this redshift.

\subsection{Star-forming active and suppressed galaxies}\label{sec:SF_sup_pas}
Star-forming galaxies can exhibit varying levels of SF activity, which has led numerous studies to propose diverse techniques for separating them based on their activity and star formation history (SFH). For instance, \cite{Koyama2013} identified a redshift-dependent specific SFR (${\rm sSFR = SFR/}M_\star$) threshold, used for example by \cite{LaganaUlmer2018}, whereas \cite{Salim2014} and \cite{Guo2015} used sSFR thresholds for local galaxies, \cite{Davies2018} the colour of galaxies with a $u^{*}-r^{*}$ threshold and \cite{Schaefer2019} the EW of $\Halpha$.

We define three galaxy populations based on their position on the $M_\star-{\rm SFR}$ diagram: the `active' galaxies with SFRs similar to that of the SFMS, `passive' galaxies with the lowest SFRs in the diagram, and the `suppressed' galaxies, with their SF in an intermediate state between the active and passive galaxies.

    \begin{figure}
        \centering
        \includegraphics[width=\hsize]{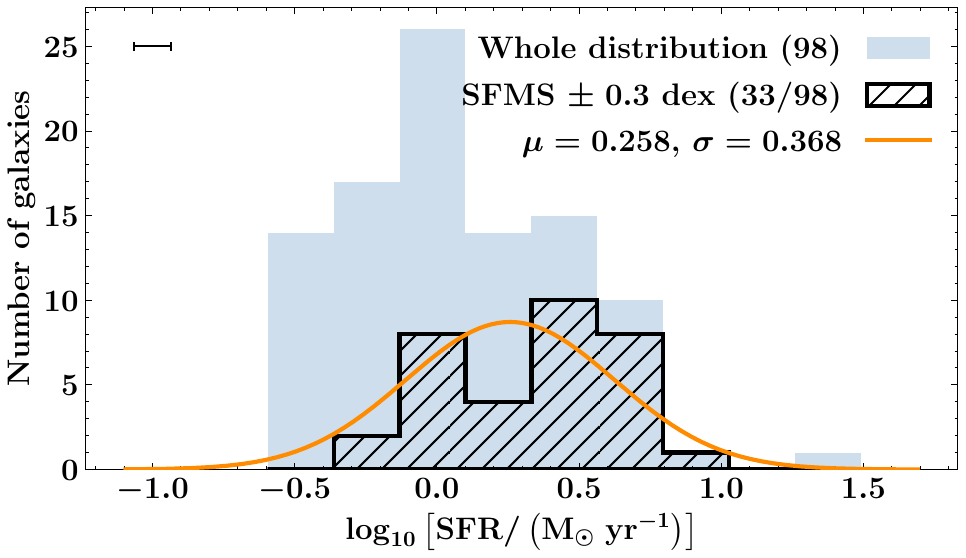}
        \caption{Histogram of SFRs for the Cl0024 cluster. The blue histogram represents the whole SFR distribution, while the hatched black histogram is the galaxies in the range of SFMS$\pm0.3$ dex. We indicate the number of galaxies between parentheses in the legend. We fit the black histogram with a single Gaussian fit, depicted by the orange curve. The value of $\sigma$ of the Gaussian fit is used to estimate the separation between active, suppressed and passive galaxies, following \cite{Finn2023} methodology. In the upper-left corner we show the median of the uncertainty bars of the SFRs.}
        \label{fig:hist_passive_suppressed_galaxies}
    \end{figure}

To accurately identify the level of SF activity of the galaxies within Cl0024, we first selected galaxies in a region of $\pm0.3$ dex around the SFMS at the redshift of the cluster in \cref{fig:SFR_mass}: it is commonly reported that the width of the SFMS is $\sim 0.3$ dex \citep[e.g.][]{janowiecki_xgass_2020,Popesso2023,Finn2023}. This approach led to the group of 33 active galaxies shown with the black dashed histogram in \cref{fig:hist_passive_suppressed_galaxies}.

Once the galaxies around the SFMS have been selected, we decided to follow the method depicted by \cite{Finn2023}: fit the distribution of SFRs of the 33 active galaxies to a Gaussian function and use the fitted parameter $\sigma$ to determine the separation between active, suppressed and passive galaxies.

To separate active galaxies from galaxies with suppressed SF, we define a cut at $1.5\sigma$ below the SFMS, where $\sigma$ is the standard deviation of the Gaussian fit of \cref{fig:hist_passive_suppressed_galaxies} (here, the cut is 0.55 dex below the SFMS). We define suppressed galaxies as those below this cut in the $M_\star-{\rm SFR}$ diagram. This factor of $1.5\sigma$ comes from \cite{Paccagnella2016} and $\mu\pm1.5\sigma$ represents 86.6\% of the Gaussian distribution area. According to this definition and assuming the galaxies are normally distributed within the best fit of the SFMS (i.e. their SFR is not affected by the environment), we should observe $\sim$$7\%$ of the galaxies in this category of suppressed galaxies. However, the total number of suppressed galaxies is 34 galaxies, constituting 35\% of the total number of galaxies shown in \cref{fig:SFR_mass}. Furthermore, a higher fraction of galaxies in structure B (44\% of the SFGs) are suppressed in comparison to structure A (only 32\%). This shows the complex dynamics of the cluster and the processes that galaxies in B are experiencing.

Finally, passive galaxies are defined as presented in \cite{Finn2023}: we set the passive cut using a curve that is parallel to the SFMS and intersects the sSFR limit of $\log_{10}({\rm sSFR/ yr^{-1}})<-11.0$ \citep{leethochawalit_evolution_2018}\footnote{\cite{Finn2023} used an sSFR limit of $10^{-11.5}\ {\rm yr^{-1}}$ from \cite{Salim2014}, but we decided to use the cut of $10^{-11.0}\ {\rm yr^{-1}}$ presented by \cite{leethochawalit_evolution_2018} that specifically select the passive galaxies for their analysis of Cl0024.} at the $\mathrm{95^{th}}$ percentile of the mass distribution ($\log_{10}(M_\star/{\rm M_\odot})=10.29$), resulting in a cut shifted by $-1.40$ dex below the SFMS fit. Subsequently, the passive cut is the strictest, which is the parallel cut for $\log(M_\star/{\rm M_\odot})\leq10.29$ and the $\rm \log(sSFR/yr^{-1})=-11.0$ cut for $\log(M_\star/{\rm M_\odot})>10.29$. \cref{fig:SFR_mass} shows that only one galaxy from our sample is considered passive, 882\_a, which has the lowest sSFR. This very low number of passive galaxies is expected since our observations focused on ELGs.

\cref{tab:types_SFG} shows the number of galaxies that are above the suppressed cut (active galaxies), below the suppressed cut (suppressed galaxies) and in the passive region (passive galaxies). We note that in the following sections, we use the adjectives `active' and `suppressed' to refer to the SFRs of galaxies or to the galaxies themselves.

    \begin{table}
       \centering
       \caption{Number of galaxies of each activity type estimated using the $M_\star-{\rm SFR}$ diagram.}
       \label{tab:types_SFG}
        \begin{tabularx}{\hsize}{X c}
           \hline\hline
           Type                                           & Number (\%)     \\ \hline
           Total                                          & 98 (100.0)    \\
           Active                                         & 63 (64.3)     \\
           Suppressed                                     & 34 (34.7)     \\
           Passive                                        &  1 (1.0)      \\ \hline
       \end{tabularx}
       \tablefoot{Next to each number we include in parentheses the respective percentage relative to the total number of SFGs shown in the diagram.}
    \end{table}

Finally, it is worth noting that more passive galaxies are present in the cluster, as part of the non-ELG population. To confirm this assumption, we crossmatched our sample of 365 sources with the catalogue of quiescent galaxies assembled by \cite{leethochawalit_evolution_2018}. This catalogue contains 62 sources that have been classified as quiescent for having EWs of \OII$\lambda3727$ smaller than 5 \AA\ and rest-frame $\rm FUV - V$ colours larger than 3, which is equivalent to having a cut in sSFR at approximately $\rm 10^{-11}\ yr^{-1}$. Out of those 62 sources, 45 fall within the footprint of GLACE observations and 93\% of them (42) are found in our non-ELGs sample. The other 3 galaxies have been classified in our work as AGN (\texttt{193\_a}), suppressed (\texttt{453\_a}) and passive (\texttt{882\_a}). Nonetheless, this only accounts for 23\% of non-ELGs, so we cannot assume that all the non-emitters are passive galaxies.

In the following sections, we study the population of SFGs drawn from the ELGs classified from GLACE observations. The study of passive galaxies is beyond the scope of this work, for several reasons. First of all, the observations carried out in GLACE were specifically designed to study ELGs, and our objective is to continue this analysis. Furthermore, we saw that the preparation of a reliable sample of passive galaxies in the explored field is not easy, and using a possible incomplete sample would bias the following analysis. However, in analyses involving fractions (\cref{sec:local_environment}) we compare our sample of ELGs with our sample of 192 non-ELGs and one passive galaxy in order to draw conclusions about the evolutionary picture of the galaxies.

\subsection{The $M_\star-{\rm sSFR}$ relation}
\begin{figure*}
        \centering
        \includegraphics[width=\textwidth]{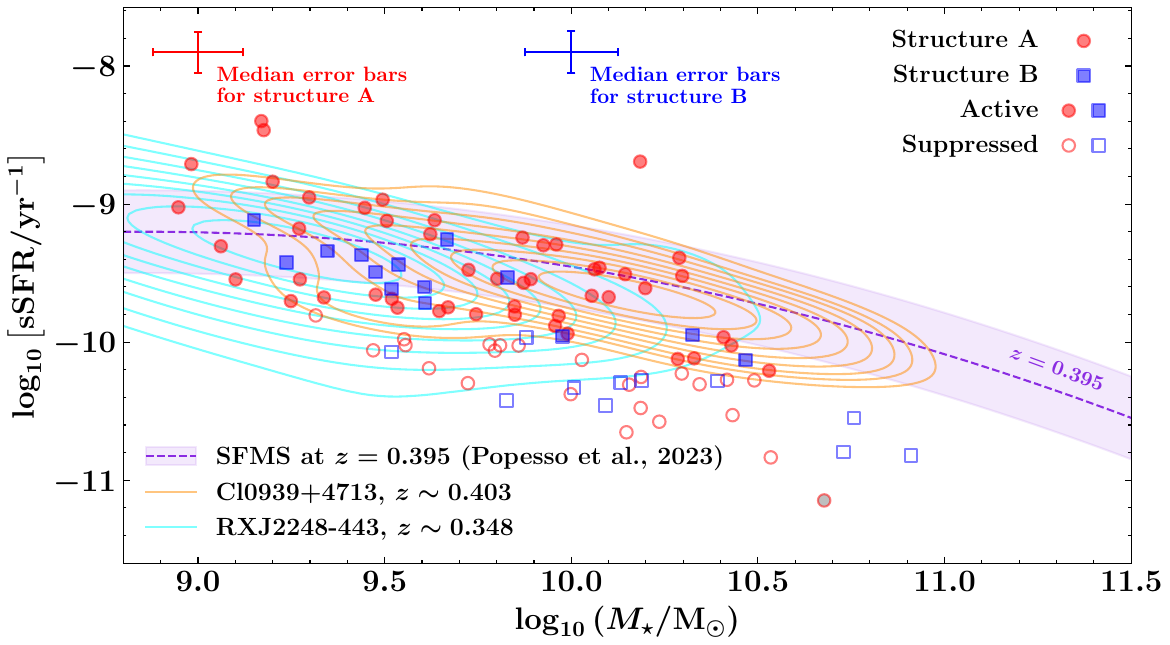}
        \caption{Relation of ${\rm sSFR}-M_\star$. The symbols and colours used in this figure are identical to those in \cref{fig:SFR_mass}. The orange and blue contours represent the distributions of ${\rm sSFR}-M_\star$ of the clusters Cl0939+4713 and RX J2248-443, respectively. The medians of the uncertainty bars for the data of the structures A and B are shown at the top of the diagram.}
        \label{fig:sSFR_mass}
    \end{figure*}

An additional aspect to explore involves analysing the sSFR in relation to the $M_\star$ of galaxies. From \cref{fig:sSFR_mass} we notice that the trend of the sSFR is a net anticorrelation with respect to $M_\star$. The suppressed galaxies display a trend almost parallel to $-0.5$ dex below the theoretical SFMS at $z=0.395$. Furthermore, there are more massive objects in the suppressed sample, which can have been formed by mergers of galaxies. Once again, the results are the same for structures A and B.

We also include the sSFRs from \cite{Sobral2016} and \cite{Ciocan2020} of the clusters Cl 0939+4713 ($z\sim0.41$) and RX J2248-443 ($z\sim0.348$), respectively. RX J2248-443 shows a similar anti-correlation trend, but with galaxies with a lower $M_\star$, and Cl 0939+4713 sSFRs are closer to the SFMS, displaying active SF,\footnote{In \cref{fig:sSFR_mass} we only show the SFMS at the redshift of Cl0024. The SFMS of Cl 0939+4713 and RX J2248-443 are not displayed since they fall within the uncertainty range of $\pm0.3$ dex depicted by the purple area.} in the range of $9.25<\log_{10}\left(M_\star/{\rm M_\odot}\right)<11.0$.

Cl0024 and RX J2248 have similar sSFRs, although RX J2248 is ten times more massive than Cl0024. Based on these results, the total mass of a cluster does not appear to be relevant in accelerating the evolution of SFGs. Furthermore, the galaxies of Cl 0939 appear to be more active, possibly because they were observed at a greater distance from the cluster centre than Cl0024,\footnote{In the Fig. 7 of \cite{Sobral2016} we observe that the cluster-centric distances get to $\sim6\ \rm Mpc$, which is approximately $2.81\ R_{200}$, while GLACE covers observations up to $1.8\ R_{200}$.} where the local density is generally lower.

\subsection{Field-cluster environment comparison}
To study the effect of the environment on the evolution of galaxies, we use the sample of 248 field galaxies from the VVDS to compare their SFRs with the ones we obtained for Cl0024. To compare both samples, we looked at the offset of each galaxy’s SFR with respect to their SFMS at a specific $M_\star$, $\Delta\log_{10}({\rm SFR})$ as defined in \cref{eq:delta_SFR}:

    \begin{equation}\label{eq:delta_SFR}
        \Delta\log_{10}({\rm SFR}) = \log_{10}({\rm SFR_{obs}}) - \log_{10}({\rm SFR_{SFMS}}),
    \end{equation}

\noindent where $\log_{10}({\rm SFR_{obs}})$ is the observed SFR and $\log_{10}({\rm SFR_{SFMS}})$ is the value predicted by the SFMS at each $M_\star$ for the specific redshift of the sample ($z=0.395$).

\begin{figure}
        \centering
        \includegraphics[width=\hsize]{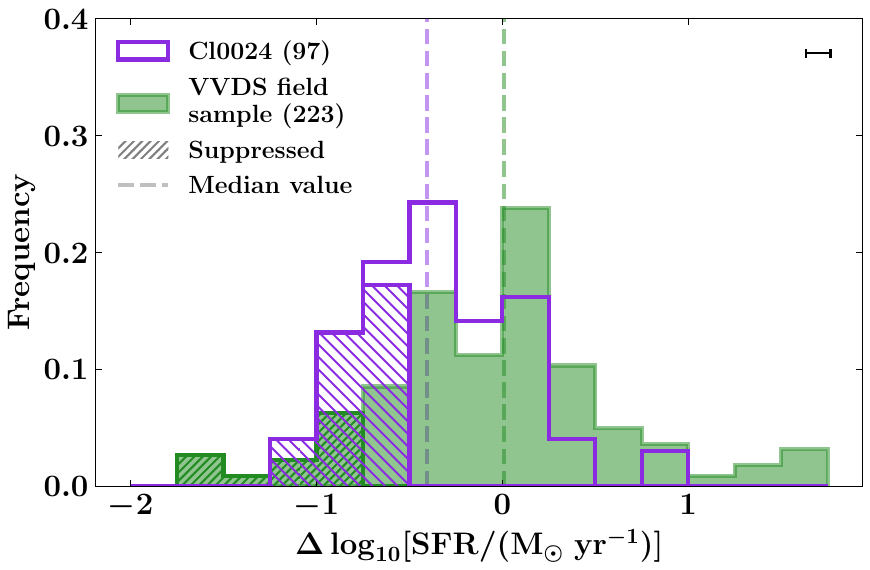}
        \caption{Histograms of $\Delta\log_{10}({\rm SFR})$ for Cl0024 and VVDS field galaxies. We plot the frequency of active and suppressed galaxies in each sample. The purple histogram is the distribution from Cl0024, while the green histogram is from the VVDS field galaxies. We also show the distribution of the suppressed galaxies with the hatched histograms of the same colour. In addition, we show the median of each distribution with the vertical dashed lines. Next to each sample legend, we show the total number of galaxies in parentheses. In the upper-right corner we show the median of the uncertainty bars of the SFRs.}
        \label{fig:Delta_SFR_histograms}
    \end{figure}

\cref{fig:Delta_SFR_histograms} shows the distributions of $\Delta\log_{10}({\rm SFR})$ of the cluster and the field samples, after removing the passive galaxies. We also show the suppressed galaxies for each environment, where the suppressed sample of field galaxies was estimated using the same criteria as for the cluster, and does not include AGN nor passive galaxies. We observe a negative shift of the cluster distribution with respect to the field sample, which indicates a quenching of the SF within the cluster. The median of $\Delta\log_{10}({\rm SFR})$ is $0.008$ dex for the field galaxies, and $-0.41$ dex for the cluster galaxies, so a difference of 0.42 dex. The field distribution is also composed of fewer suppressed galaxies, with only 11.0\% of the sample (including passive galaxies) against 34.7\% for cluster galaxies (3.25 times more). Additionally, we performed an Anderson-Darling test between both samples and obtained that they are statistically different (p-value of 0.0010).

\subsection{Relation between star formation and local environment within the cluster}\label{sec:local_environment}
\begin{figure}
        \centering
        \includegraphics[width=\hsize]{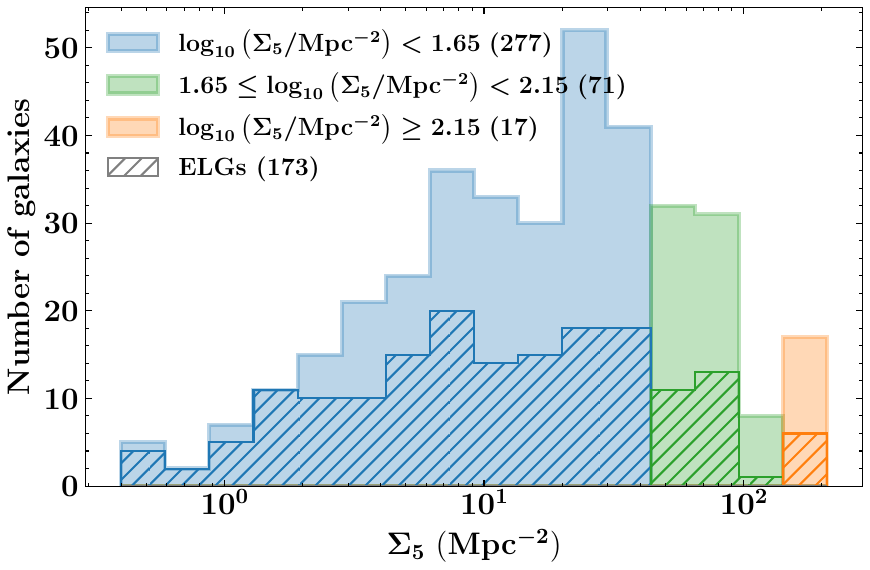}
        \caption{Distributions of $\Sigma_5$ for 365 galaxies from Pérez-Martínez et al. (in prep.). Based on \cite{Koyama2008}, we divided the distribution in low-density (in blue), intermediate-density (in green) and high-density (in orange) samples. We show the distribution for 365 galaxies of Cl0024 with plain histograms and, in hatched histograms, of 173 ELGs selected from GLACE observations.}
        \label{fig:Sigma5_histogram}
    \end{figure}

We are interested in understanding how the local environment of the cluster affects the evolution of galaxies. \cite{Cedres2024} presented new parameters to characterise the local environment of the cluster based on the work of Pérez-Martínez et al. (in prep.). One of them is $\Sigma_5$, the projected surface density of the galaxies, estimated as the source density in the area encircled between the object and the $\mathrm{5^{th}}$ nearest galaxy above a magnitude I $\sim$ 21.5 mag. We show the distribution of $\Sigma_5$ for the galaxies of Cl0024 and particularly the ELGs in \cref{fig:Sigma5_histogram}. We divided the distribution in three subsamples, as defined by \cite{Koyama2008}: low-density, with $-0.5\leq\log_{10}(\Sigma_5/{\rm Mpc^{-2}})<1.65$, intermediate-density, with $1.65\leq\log_{10}(\Sigma_5/{\rm Mpc^{-2}})<2.15$ and high-density with $2.15\leq\log_{10}(\Sigma_5/{\rm Mpc^{-2}})<2.50$. A majority of the galaxies (76\% of the total and 82\% of the ELGs) are found in the low-density range, possibly due to the truncation of SF activity in the densest regions of the cluster.

\begin{figure}
        \begin{subfigure}[c]{\hsize}
         \centering
         \includegraphics[width=\textwidth]{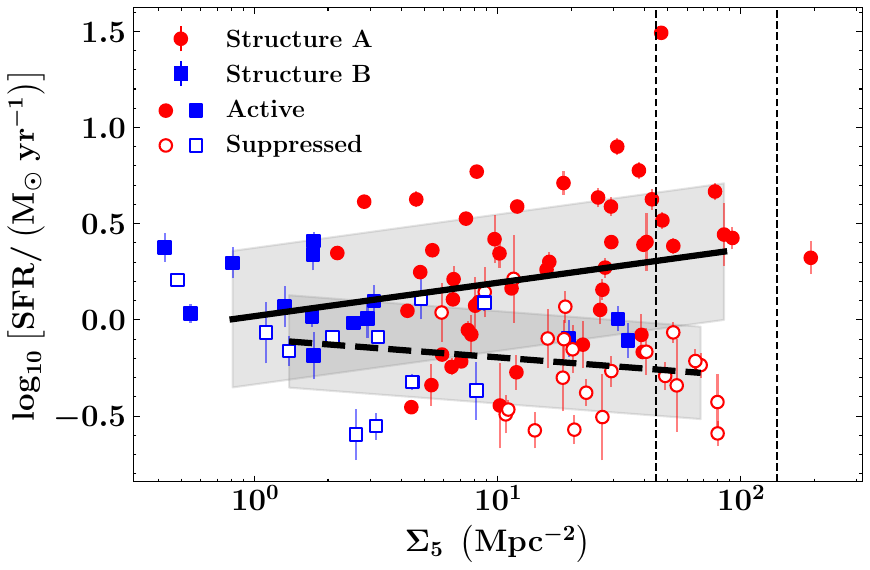}
         \caption{SFR vs $\Sigma_5$}
         \label{fig:SFR_Sigma5}
     \end{subfigure}
     
     \begin{subfigure}[c]{\hsize}
         \centering
         \includegraphics[width=\textwidth]{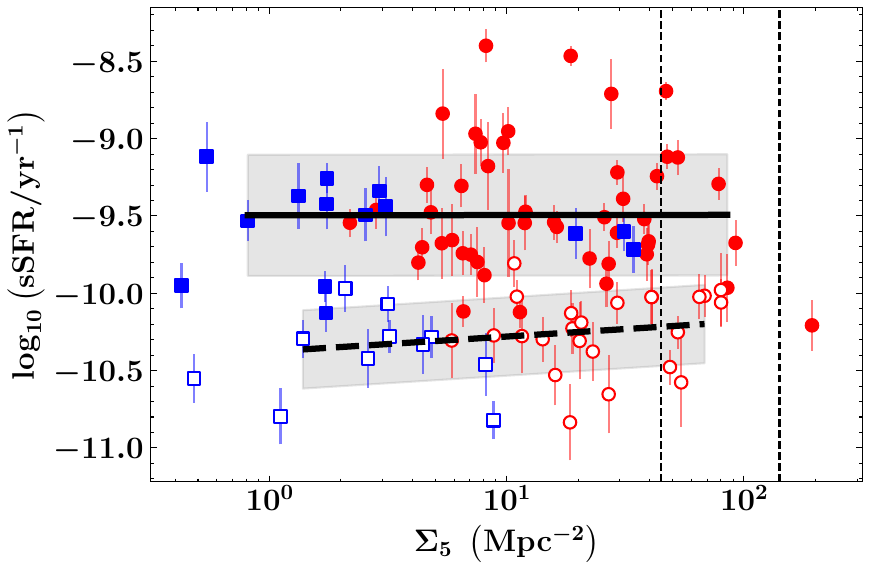}
         \caption{sSFR vs $\Sigma_5$}
         \label{fig:sSFR_Sigma5}
     \end{subfigure}
    \caption{\textit{(Top)} Star formation rate and \textit{(bottom)} sSFR with respect to $\Sigma_5$. We show the relation for structure A with red circles and for the structure B with blue squares. Filled symbols represent active SFRs while the emptied symbols are suppressed SFRs. Bins of $\Sigma_5$ are indicated by the vertical dashed lines and show the separation in low, intermediate and high density subsamples. The black lines represent the fits to the running median of the SFRs/sSFRs of the galaxies of both structures, calculated in bins of 5 points, and the grey-filled area represent the $1\sigma$ deviation of the SFR and sSFRs. The solid line is for the active galaxies and the dashed one for the suppressed galaxies.}
    \label{fig:SFR_sSFR_Sigma5}
    \end{figure}

In \cref{fig:SFR_sSFR_Sigma5} we show the relation between the SFR and sSFR, and the local density, $\Sigma_5$, for active and suppressed galaxies separately. In each case, we linearly fitted the values to a running median calculated in bins of 5 points, to show trends without considering the outliers. We clearly see the differentiation between each population and structures, with galaxies from structure A being in denser regions and suppressed galaxies having lower SFRs and sSFRs. The SFR appears to be increasing with $\Sigma_5$ for active galaxies, and decreasing for suppressed galaxies; however, the scatter of the data is important, and it is difficult to state that there is a correlation between both axes. Furthermore, this trend disappears when the SFR is normalised by the stellar mass and we represent sSFR as a function of $\Sigma_5$.
    
\begin{figure}
    \centering
    \includegraphics[width=\hsize]{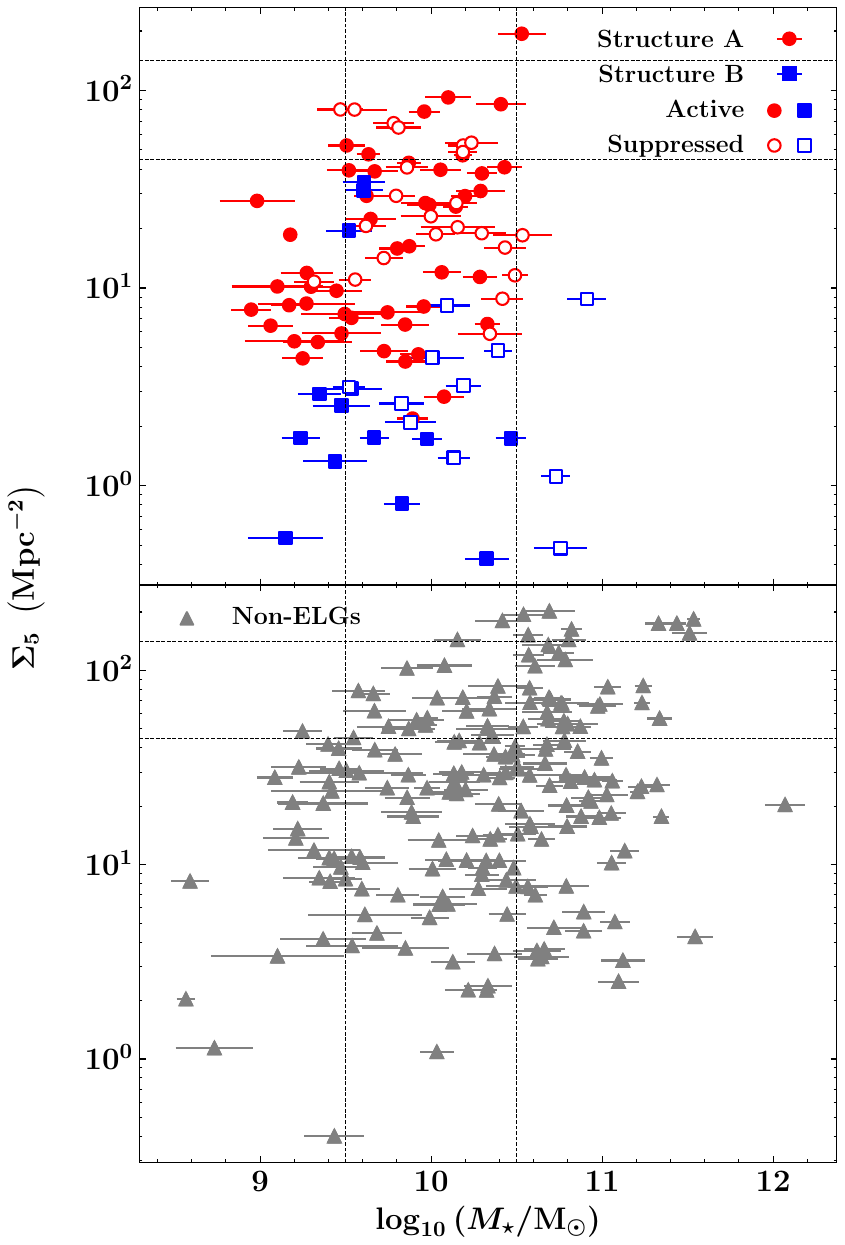}
    \caption{Relation of $\Sigma_5-M_\star$ for \textit{(upper panel)} active, suppressed, and \textit{(lower panel)} non-emitter galaxies. Symbols in the upper panel are the same as in \cref{fig:SFR_sSFR_Sigma5}. Bins of $\Sigma_5$ and $M_\star$ are indicated by the horizontal and vertical dashed lines, respectively, and show the separation in low, intermediate and high density or mass subsamples.}
    \label{fig:Mstar_Sigma5}
\end{figure}

Since SFRs are normalised by $M_\star$ in \cref{fig:sSFR_Sigma5}, this could indicate an underlying relation between $M_\star$ and $\Sigma_5$. In \cref{fig:Mstar_Sigma5} we show the values of $\Sigma_5$ with respect to $M_\star$ for the populations of 97 active and suppressed galaxies and 193 passive and non-ELGs. We do not observe any trend in either property for active or suppressed galaxies; for non-emitters, more massive galaxies tend to be in denser regions, but the correlation presents a large scatter.

An additional investigation shedding light on the correlation between SF in galaxies and cluster environmental properties is the analysis of the fractions of SFGs \citep[e.g.][]{Zeleke2021}. This process involves binning the property of interest, such as $\Sigma_5$ or $M_\star$, calculating the number of SFGs in each bin, and then dividing this count by the total number of galaxies in the bin. It is crucial to accurately estimate the total number of galaxies in each bin, especially when comparing different samples.

\begin{figure*}
        \begin{subfigure}[c]{0.45\textwidth}
         \centering
         \includegraphics[width=\textwidth]{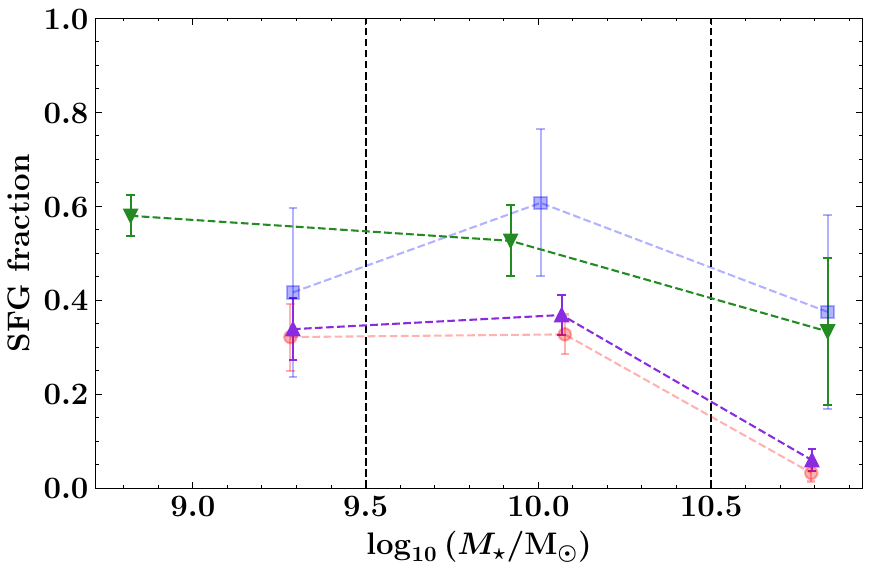}
     \end{subfigure}
     \hfill
     \begin{subfigure}[c]{0.45\textwidth}
         \centering
         \includegraphics[width=\textwidth]{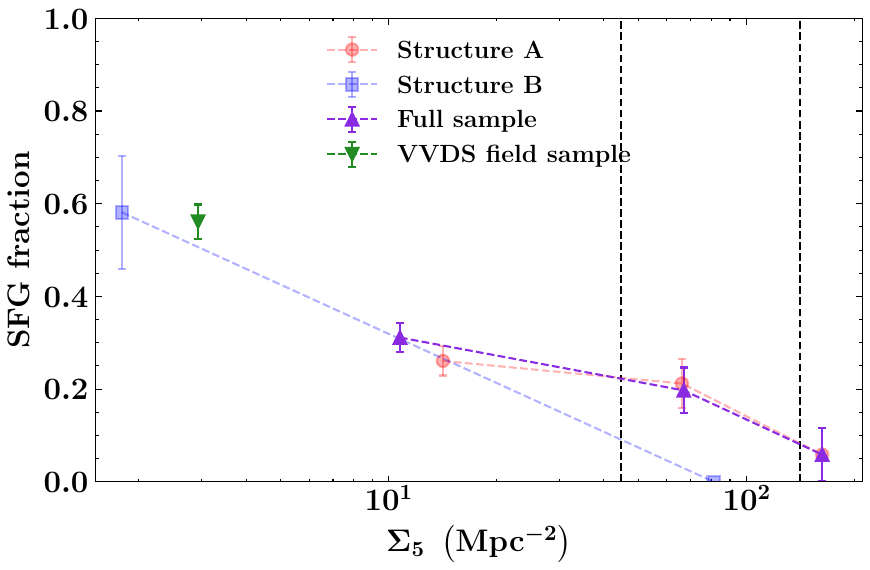}
     \end{subfigure}
    \caption{Fractions of SFGs with respect to \textit{(left)} $M_\star$ and \textit{(right)} $\Sigma_5$. Bins are indicated by the vertical dashed lines and show the separation in low, intermediate, and high mass or density subsamples. We show fractions from structure A with red circles, from structure B with blue squares and from the full cluster sample (sum of structures A and B) with purple triangles; the fractions from the VVDS field sample are depicted by the green inverted triangles. For $M_\star$, a total of 359 sources were employed, while for $\Sigma_5$ a total of 361 sources were employed. Error bars represent the Poisson statistics errors associated with the fractions in their specific bin.}
    \label{fig:SF_fractions}
    \end{figure*}

In \cref{fig:SF_fractions}, we present the fractions of SFGs as a function of $M_\star$ and $\Sigma_5$. For both parameters, the analysis is based on the total number of galaxies with calculated properties: 359 galaxies for $M_\star$ and 361 galaxies for $\Sigma_5$ (in both cases we removed the 4 interlopers). We selected the three bins of $\log_{10}\left(M_\star\right)$ and $\log_{10}\left(\Sigma_5\right)$ displayed in \cref{fig:Stellar_mass_histogram,fig:Sigma5_histogram}, respectively, so we can more easily compare with results from \cite{Zeleke2021}. Vertical error bars were computed using Poisson distribution errors, proportional to $\sqrt{N}$, where $N$ represents the absolute number of SFGs in each bin (additional details can be found in \citealt{Martinez2001}).

The fractions of SFGs for the full sample of Cl0024 are $34\pm7\ \%$ in the low-mass bin, $37\pm4\ \%$ in the intermediate-mass bin and $5.9\pm2.4\ \%$ in the high-mass bin (left panel in \cref{fig:SF_fractions}). This trend is similar to what was found by \cite{Zeleke2021} and indicates that more massive galaxies tend to have less SF activity than low-mass galaxies. Furthermore, the fraction of SFGs for the field sample is constant across all the mass bins within the error bars, with fractions around 50\%. This is similar to the fractions found in structure B of the cluster, but higher than the fractions found in the full sample of cluster environment; this shows the higher suppression of the SF affecting the galaxies at all mass bins in the cluster. In the field, the general trend seems to indicate a decrement of the fractions at higher masses, but the error bars does not allow for a conclusive analysis. Fractions in structure B are higher than what we observe in the total sample of the cluster, which shows that galaxies in this group evolve differently than in the rest of the cluster.

The fraction of SFGs are $31\pm3\ \%$ in the low-density bin, $20\pm5\ \%$ in the intermediate-density bin and $6\pm6\ \%$ in the high-density bin (right panel in \cref{fig:SF_fractions}). The fractions in the low-density bin are 1.55 times higher than in the intermediate-density bin, i.e. towards the centre of the cluster. These results confirm that the cluster environment suppresses the SF activity of SFGs where the local density is higher, but due to the low number of galaxies in the densest bin, better study of these regions is required. For the field sample, only a fraction of $56\pm4\ \%$ is available in the low-density bin since no galaxies populate the other two denser bins.

\subsection{The $D4000-{\rm sSFR}$ relation}
We already mentioned several methods in \cref{sec:SF_sup_pas} to study the quenching of galaxies. An additional test could be the analysis of the $D4000$ index \citep{D4000_index}. The spectral index $D4000$ is a measurement used to quantify the strength of the Balmer break\footnote{The Balmer break refers to a discontinuity in the spectrum caused by the absorption lines of hydrogen in the Balmer series.} in the spectrum of a galaxy or a stellar population. A high $D4000$ value indicates a strong Balmer break, which usually implies old stellar populations or galaxies with a higher proportion of old, redder stars.

\begin{figure}
        \centering
        \includegraphics[width=\hsize]{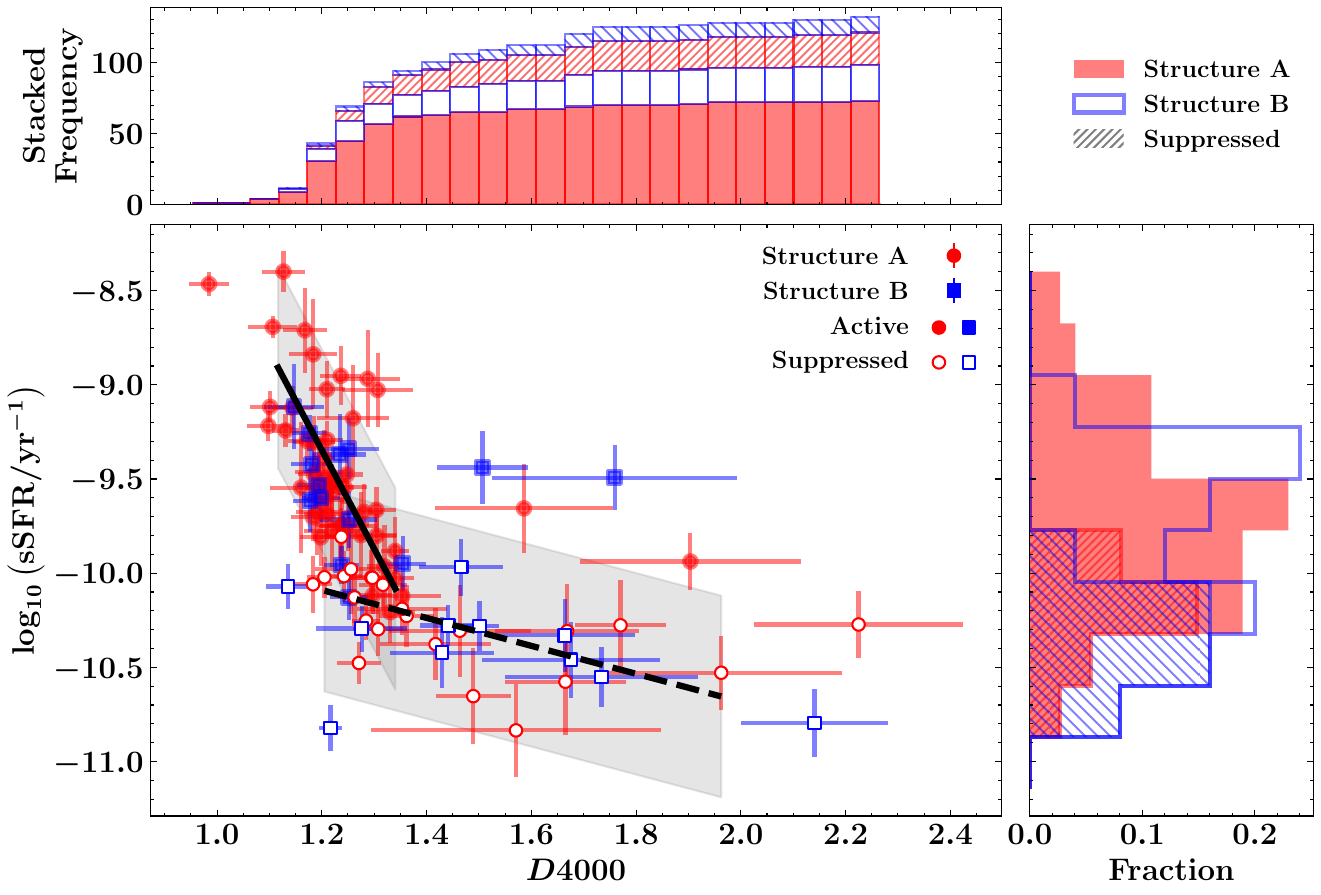}
        \caption{Relation of $\log_{10}({\rm sSFR})-D4000$. Symbols are the same as in \cref{fig:SFR_sSFR_Sigma5}. The black lines represent the fits to the running median of the sSFRs of the galaxies of both structures A and B, calculated in bins of 5 points, and the grey-filled areas represent the $1\sigma$ deviation of the sSFRs. For active sSFRs (solid line), we sorted the points vertically, and horizontally for the suppressed sSFRs (dashed line). Upper and right histograms show the distributions of $D4000$ (stacked cumulative histogram) and $\log_{10}({\rm sSFR})$, respectively, for active and suppressed populations, divided in structures A and B.}
        \label{fig:sSFR_D4000}
    \end{figure}

In the case of active and suppressed galaxies, it would be then logical to find several relations between the variables we previously studied and the $D4000$ index. To do so, we used the values of the $D4000$ index calculated by {\tt CIGALE}, and in \cref{fig:sSFR_D4000} we represent the sSFRs of the galaxies of the cluster as a function of this index. We used a linear fit to the running median of the sSFRs of the galaxies of both structures A and B, calculated in bins of 5 points, and the grey-filled area shows the $1\sigma$ scatter of the sSFRs. We observed different trends for active (solid line) and suppressed (dashed line) galaxies: $\log_{10}({\rm sSFR/yr^{-1}})$ for active galaxies strongly decreases monotonically as $D4000$ increases--it is especially visible with the values of $D4000<1.4$, that have smaller uncertainties. In contrast, in the case of the suppressed galaxies, they seem to be much less dependent on the $D4000$ index. Furthermore, the upper and right histograms show the individual distributions of $D4000$ and $\log_{10}({\rm sSFR/yr^{-1}})$. Suppressed galaxies are found at lower values of $\log_{10}({\rm sSFR/yr^{-1}})$, while they are evenly scattered along the $D4000$ axis. However, they represent the dominant population of galaxies at higher $D4000$ values. This is due to the continuous quenching suffered by these galaxies, which causes them to populate lower values of $D4000$ but dominate the higher ones.

Nonetheless, we observe a larger scatter of the suppressed sSFRs, which also tends towards the idea that the relation between sSFRs and $D4000$ vanishes for this population. These results support the theory that the suppressed galaxies within this cluster suffered an environmental quenching, causing $D4000$ to no longer represent the secular evolution of their stellar populations.

\subsection{Suppression of star formation and positions in the cluster}\label{sec:ppsd_results}
\begin{figure*}
         \centering
         \sidecaption
         \includegraphics[width=0.7\textwidth]{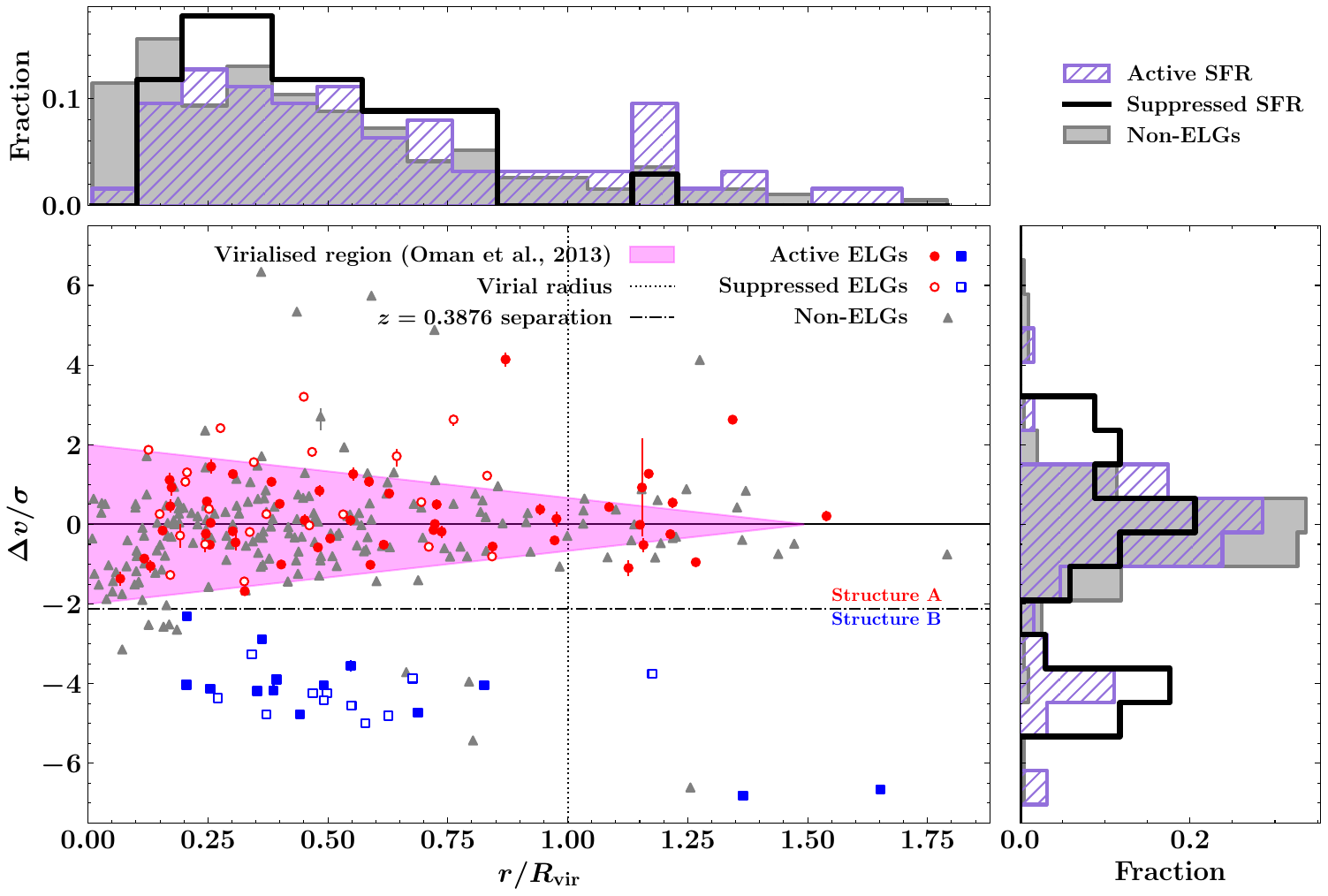}
         \caption{Phase-space diagram for 290 galaxies of Cl0024 at $z=0.395$. The red and blue symbols show the SFGs, divided in structure A and B, respectively. Open symbols represent the suppressed SFGs and filled ones the active SFGs, and grey triangles are non-ELGs and the passive ELG \texttt{882\_a}. The pink region corresponds to the virialised region delimited by \cref{eq:Oman_limit} from \cite{Oman2013a}. The black vertical dotted line marks the distance to $R_{\rm vir}$ and the black horizontal dot-dashed line shows the separation of $z=0.3876$ used to separate structures A (above) and B (below). In the top and right panels, histograms show the distributions of $r/R_{\rm vir}$ and $\Delta v/\sigma$, respectively. The active values are shown with the purple histogram, the suppressed values with the stepped black histogram and the non-ELGs values with the grey histogram.}
        \label{fig:phase_space}
    \end{figure*}

We can study the population of cluster galaxies based on their position and kinematics within the cluster, through a phase-space diagram. In \cref{fig:phase_space} we show the phase-space diagram of the galaxies of Cl0024.

This phase space is built using the distance $r$ to the centre of the cluster (also called cluster-centric distance) normalised by the virial radius of the cluster ($R_{\rm vir}=1.21\pm0.1\ \mathrm{Mpc}\ h^{-1}$) in the horizontal axis, while the vertical axis displays the relative velocity of the galaxies with respect to the cluster recessional velocity, $\Delta v$, normalised by the velocity dispersion of the cluster ($\sigma$). We calculated it following

\begin{equation}
    \frac{\Delta v}{\sigma} = \frac{c(z-z_{\rm{cl}})}{(1+z_{\rm{cl}})\sigma},
\end{equation}

\noindent where $c$ is the speed of light, $z$ is the redshift of a given galaxy, $z_{\rm{cl}}=0.395$ \citepalias{SP15} the redshift of the cluster, and $\sigma=1050\ {\rm km\ s^{-1}}$ \citep{Czoske2002} its velocity dispersion. We have included in this diagram 290 galaxies: 63 active SFGs divided in structures A and B in red and blue, respectively, 34 galaxies with suppressed SFR with open symbols, and in grey the 192 non-emitters and the passive ELG (\texttt{882\_a}). Additionally, we show in the upper panel the histogram for the distributions of $r/R_{\rm vir}$ and in the right panel the histogram for the distributions of $\Delta v/\sigma$.
    
Our objective is to separate galaxies in the virialised region of the cluster from galaxies in the infalling region, similarly to \cite{Finn2023}. We use a phase-space cut from \cite{Oman2013a}, that divides the galaxies into galaxies in the cluster core and the infalling region. This cut is defined as follows:

    \begin{equation}\label{eq:Oman_limit}
        \left|\frac{\Delta v}{\sigma}\right| = -\frac{4}{3}\frac{r}{R_{200}} + 2.
    \end{equation}

We note that we are only using it to separate galaxies in the core and infall regions. The phase-space diagram is a powerful tool that can be used to trace the assembly history of clusters \citep[e.g.][]{Taranu_PPSD}, and the terms `core' and `infall' can have different meanings. In our analysis, they refer to the regions within (core) and outside (infall) the virialised zone traced by \cref{eq:Oman_limit}. They are not a strict definition of the kinematical phase of the galaxies, especially because we are not studying the orbits of the galaxies, if they are totally virialised or if they are gravitationally bound to the cluster. Other studies \citep{Czoske2002,Moran2005,Costa2024} have already presented some results on the kinematics of Cl0024 and we recommend readers who wish to learn more about this topic to consult these papers.

In \cref{fig:phase_space}, a majority of non-ELGs and passive galaxies are present in the core of the cluster (78.8\%), even more than active and suppressed galaxies (54.6\%). Following the upper histogram showing the distribution of $r/R_{200}$, the projected distances to the centre, we can see that active and suppressed galaxies, and non-ELGs follow the same trend, with more of them being closer to the centre. The right histogram shows that the active and suppressed galaxies occupy both structures A and B (we show with the dashed black line the separation between both structures), but the non-emitters are mainly concentrated in the structure A. Structure B, below the dashed black line, contains a 57.9\% of the suppressed galaxies that lie in the infall region (11 out of 19 galaxies). Furthermore, Anderson-Darling tests conclude that suppressed and active galaxies' distributions of $r/R_{200}$ and absolute velocities, $|\Delta v|/\sigma$, are statistically similar, with p-values of 0.0866 and 0.0388, respectively, similarly to what is found at local redshifts \citep{Finn2023}.
    
\begin{figure}
        \centering
        \includegraphics[width=\hsize]{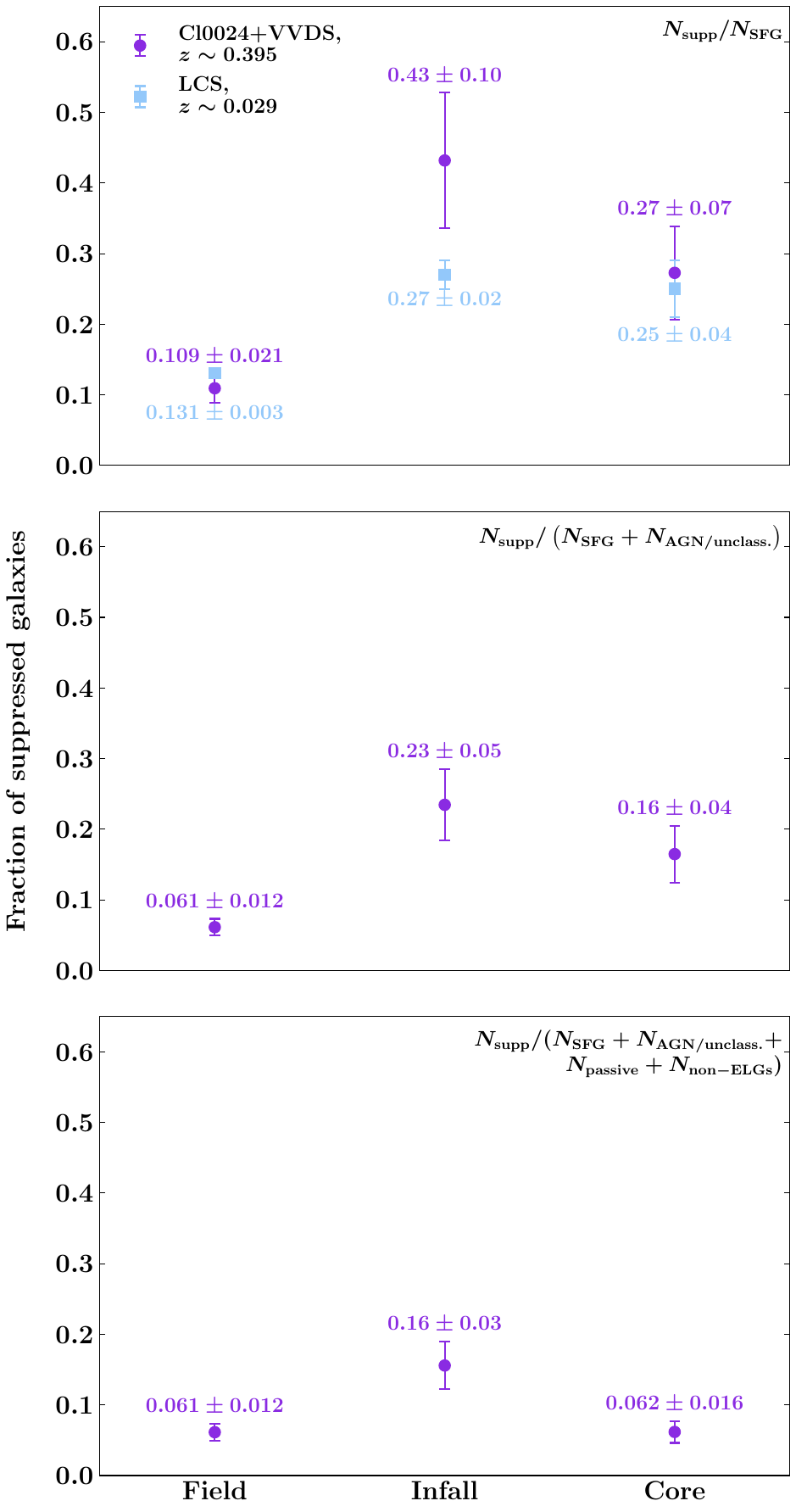}
        \caption{Fraction of suppressed galaxies within the SF population of galaxies within the field, infall and core environments, including in the comparison sample (\textit{left}) SFGs, (\textit{middle}) SFGs, AGNs, and unclassified galaxies (the sample of non-passive ELGs); and (\textit{right}) the whole sample of ELGs, non-ELGs, and passive galaxies. We display the results for Cl0024/VVDS samples at $z\sim0.395$ with the purple circles and the results of the LCS at $z\sim0.029$ with the blue squares, with the values of the fractions above and below the data points. The error bars are estimated following a Poissonian distribution statistics.}
        \label{fig:suppressed_galaxies_environment}
    \end{figure}

\cref{fig:suppressed_galaxies_environment} presents the fraction of suppressed galaxies in the field, infall and core regions, for both Cl0024+VVDS samples and the LCS \citep{Finn2023}. In the left panel, the fractions were estimated using only SFGs in the denominator, which allows us to make comparisons with the LCS values. In the middle panel, we added in the denominator the AGN and unclassified population of galaxies, which results in the whole sample of ELGs, except one passive galaxy (882\_A). In the right panel, we show the results including the passive galaxy and the non-ELGs (which also contain passive galaxies; see \cref{sec:SF_sup_pas}). The field sample at $z\sim0.395$ comes from the VVDS survey, with a total of 440 galaxies, while the infall region contains 118 galaxies and the core region 238 galaxies. The suppressed galaxies were estimated following the same methodology as for the galaxies of Cl0024.

Including only the SFGs in the denominator (left panel of \cref{fig:suppressed_galaxies_environment}), the fractions are $10.9\pm2.1\ \%$ in the field, $43\pm10\ \%$ in the infall region and $27\pm7\ \%$ in the core region. The trend is similar to the one obtained by \cite{Finn2023}, with close values in the field and core regions. However the fraction of suppressed galaxies of Cl0024 in the infall region is much higher than what is found in the LCS.

When adding the AGN and unclassified galaxies to include all the ELGs (except the passive galaxy \texttt{882\_a}) in the comparison sample (middle panel of \cref{fig:suppressed_galaxies_environment}), the fractions generally decrease, with $6.1\pm1.2\ \%$ in the field, $23\pm5\ \%$ in the infall region, and $16\pm4\ \%$ in the core region. The fraction in the core and the field did not change significantly within the error bars. However, the fraction in the infall region is $47\%$ lower when adding the AGN and unclassified populations; this shows that the infall region is dominated by the AGN and unclassified galaxies. Furthermore, we observe almost no statistical difference between the core and infall regions, similarly to what was found by \cite{Finn2023}.

Nevertheless, when we include the non-ELGs and passive galaxies in the denominator (right panel of \cref{fig:suppressed_galaxies_environment}), we obtain a similar trend, more pronounced and with smaller error bars: fractions of suppressed galaxies are $6.1\pm1.2\ \%$ in the field, $16\pm3\ \%$ in the infall region and $6.2\pm1.6\ \%$ in the core region. In proportion, the fraction of suppressed galaxies in the infall region is 2.6 times higher than in the core region, while the field has a similar fraction of suppressed galaxies to that of the core region, within the error bars. The difference between the infall and core regions indicate that the cluster's local environment suppresses the SF activity of galaxies differently depending on the region, and this suppression is apparently more efficient during their infall than when they are in the core.

\begin{figure}
        \centering
        \includegraphics[width=\hsize]{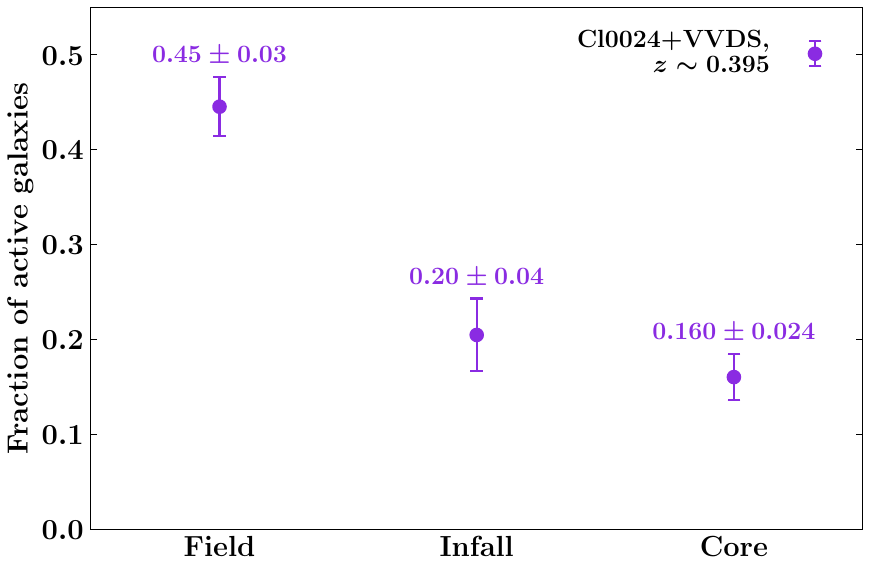}
        \caption{Fraction of active galaxies within the SF population of galaxies within the field, infall, and core environments, including the passive galaxies in the comparison sample. On top of each data point we express the values in percentage. The error bars are estimated following a Poissonian distribution statistics.}
        \label{fig:active_galaxies_environment}
    \end{figure}

In \cref{fig:active_galaxies_environment} we take a complementary view and analyse the fractions of active galaxies with respect to their position in the phase-space diagram. Fractions decrease as we enter the cluster and the local density increases: the fractions are $45\pm3\ \%$ in the field, $20\pm4\ \%$ in the infall region and $16\pm3\ \%$ in the core. This result is not surprising and confirms that the cluster environment suppresses the SF of galaxies more than what is found in the field, similarly to what we found with our previous results. The fraction of active SFGs is similar in the infall and core regions, which shows that changes we observe between both environment in the fractions of suppressed SFGs are mainly due to the presence of passive (in the core) and AGN/unclassified galaxies (in the infall region).

\subsection{Effect of the mass of the cluster}
\cite{Cedres2024} hypothesised that the mass of the cluster may have an effect on the evolutionary path of galaxies, especially on their metallicity. We compare the $M_\star-{\rm SFR}$ relation obtained for Cl0024 with other clusters to test this hypothesis. We use the same clusters as in \cite{Cedres2024} to maintain the comparison samples.

In \cref{tab:clusters} we present the main characteristics of these clusters and the information available for each sample of SFGs. We also calculated the results presented in this subsection by restricting the range of $M_\star$ of the galaxies to the lowest common mass range, to ensure that our comparison between clusters is not biased. The lowest common mass range for all samples is $\log_{10}\left(M_{\rm 
\star}/{\rm M_\odot}\right)=[9.22,\ 10.84]$, but there were no significant differences when applying this restriction, so we present the results without this mass constraint to increase their statistical significance.
    
    \begin{table*}[]
        \centering
        \caption{Main characteristics of the galaxy clusters employed in the study.}
        {\renewcommand{\arraystretch}{1.3}
        \begin{tabular}{lcccc}
        \hline
        \hline
         Cluster Name   & Redshift  & $M_{200}$ [$M_\odot$] & $R_{200}$ [Mpc] & $<\!Z/Z_{\rm \odot}\!>$\\
         \hline
         RX J2248--443 & 0.348 & $2.81 \times 10^{15}$ (1) & 2.6 (2) & 0.26 (3)\\
         Cl0024 (this work) & 0.395 & $5.85 \times 10^{14}$ (4) & 1.73 (4) & 0.22 (5)\\
         MACS J0416.1--2403 & 0.397 & $1 \times 10^{15}$ (6) & 1.8 (6) & 0.24 (7)\\
         Cl 0939+4713 & 0.41 & $1.7 \times 10^{15}$ (2) & 2.13 (2) & 0.2 (8)\\
         XMM--LSS J02182--05102 & 1.62 & $7.7 \times 10^{13}$
         (9) & 0.49 (9) & N/A\\
         PKS 1138--262 & 2.16 & $1.71 \times 10^{14}$ (10) & 0.53 (10) & N/A\\
         \hline
        \end{tabular}}
        \tablefoot{First column is the cluster name, second column is the redshift of the cluster, third column is the $M_{200}$ in solar masses, fourth column is the $R_{200}$ in Mpc, and fifth column is the mean value of the iron abundance in the intracluster gas in units of solar metallicity. In parenthesis we show the reference of each value.}
        \tablebib{(1) \citet{Kesebonye2023}; (2) \citet{Koyama2011}; (3) \citet{DeFilippis2003}; (4) \citetalias{SP15}; (5) \citet{Zhang2005}; (6) \citet{Bonamigo2018}; (7) \citet{Bonamigo2017}; (8) \citet{Rahaman2021}; (9) \citet{Pierre2012};  (10) \citet{Shimakawa2014}.}
        \label{tab:clusters}
    \end{table*}

\begin{figure*}
        \centering
        \sidecaption
        \includegraphics[width=0.7\textwidth]{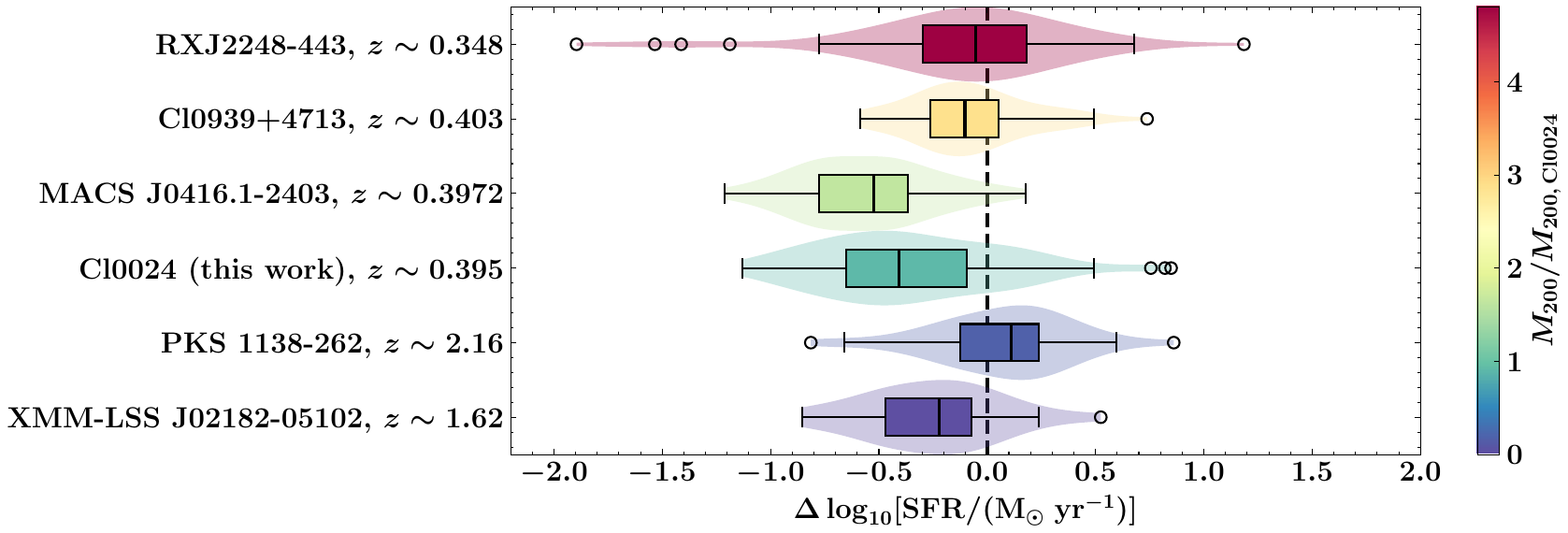}
        \caption{Distributions of $\Delta\log_{10}({\rm SFR})$ for several clusters, shown as box and violin plots sorted by virial mass. Each box extends from the first quartile to the third quartile of the data, with a vertical line at the median. The whiskers extend from the box to the furthest data point within 1.5 times the interquartile range of the box. Lonely points are those beyond the end of the whiskers. The dashed vertical line marks the value where the distance to the SFMS is zero.}
        \label{fig:DeltaSFRs_Mass_several_samples}
    \end{figure*}

To study the effect of the cluster mass on the quenching suffered by the SFGs of each cluster, we estimate $\Delta\log_{10}({\rm SFR})$ for the galaxies of each sample, based on \cref{eq:popesso_MS} for the particular redshifts of each cluster. We then represent the distribution of $\Delta\log_{10}({\rm SFR})$ for each cluster in \cref{fig:DeltaSFRs_Mass_several_samples}, where the colour indicates the mass of the cluster. We have sorted the clusters by their $M_{200}$, with the most massive at the top and the least massive at the bottom. If the virial mass of the clusters had an effect on $\Delta\log_{10}({\rm SFR})$ we would expect to see the most massive clusters to the left and the least massive to the right, but this is not the case and their distributions do not seem to depend on their virial mass.
    
\section{Discussion}\label{sec:discussion}
This section presents a discussion of the results. In particular, we discuss the implications of the trends we observed in terms of the suppression of the SF and the causes behind this phenomenon.

\subsection{Improvements using multi-object spectroscopy}
The use of MOS spectra improves largely the results obtained in the past by GLACE collaboration. We now have a better estimate of the spectroscopic redshift of the galaxies (see \cref{fig:redshift_comparison}), using not only one but four emission lines ($\Halpha$, $\Hbeta$, \OII\ and \OIII$\lambda5007$). We removed one false detection of an M-type star and found that at least 8 galaxies had a false detection of the $\Halpha$ emission line.

We also calculated integrated fluxes corrected for galactic reddening, dust extinction, and stellar atmosphere absorption, offering more robust fluxes for the community to study the galaxies of Cl0024 (see \cref{sec:data_corrections} for an extensive description of the corrections). This also allowed us to perform CSP fittings and obtain high-quality products that will be studied in more detail in the future (especially the SFHs and parameters that trace the suppression of the SFR, such as the $D4000$ index). Finally, the combination of the BPT and WHAN diagnostics give better estimates of the source of ionisation for the galaxies. We only applied it here to separate AGN from SFGs, but a future study of the specific population of AGNs is already in preparation (Sánchez-Portal et al., in prep).

\subsection{Suppression of the star formation in the galaxies of Cl0024}
Our results provide evidence that the galaxies in Cl0024 are experiencing a suppression of their SFR with respect to the galaxies in the field environment. In \cref{fig:Delta_SFR_histograms}, we observe 3.25 times more galaxies with suppressed SF in Cl0024 than in the field sample, which clearly demonstrates an effect of the dense environment of the cluster on the capacity of galaxies to form stars. We also observe that galaxies in structure B are, in proportion, more affected by this suppression than the galaxies in structure A. In \cref{sec:suppression_position_discussion} we discuss this result in relation with the dynamical state of the cluster in more detail.

\subsection{Effect of the environment in Cl0024 on the suppression of the star formation}
In \cref{fig:SFR_sSFR_Sigma5}, we observe that the SFRs of galaxies with active SF tend to increase with $\Sigma_5$, while they decrease for galaxies with suppressed SF. However, this trend seems to only be a result of the selection of each population, as it disappears when using sSFRs. In \cref{fig:Mstar_Sigma5} we tried to determine a possible trend between $\Sigma_5$ and $M_\star$, but massive galaxies do not appear to be located in the densest regions of the cluster. A relation appears when comparing both properties for passive galaxies and non-emitters, where we observe more massive galaxies in denser environments.

The absence of significant trends between the density of galaxies within the cluster and their SFRs and sSFRs aligns with previous observations. For instance, \cite{LaganaUlmer2018} derived the SFR of 17 galaxy clusters between $0.4<z<0.9$ using stellar population synthesis models. In their analysis, they confirmed the dependence of SFR with $M_\star$ but they did not find any evidence of a correlation between SFR and projected distance to the centre of the cluster (which is equivalent to $\Sigma_5$), for both SF and quenched galaxies. When studying the TF data of Cl0024, \cite{Zeleke2021} found no clear correlation between SFRs and $\Sigma_5$, similarly to us. However, when they represented sSFRs against $\Sigma_5$, they observed a positive correlation for active galaxies and a negative correlation for quenched galaxies. Our results, that are obtained using more precise estimations of SFRs, $M_\star$ and redshifts (that are used to estimate $\Sigma_5$), invalidate this previous finding.

Furthermore, the absence of relation between $\Sigma_5$ and $M_\star$ in \cref{fig:Mstar_Sigma5} is surprising, as we would expect more massive galaxies to be localised in the densest regions of the cluster. This can be explained by the fact that this assumption relies on the fact that, with time, galaxies fall towards the centre of the cluster and suffer cannibalism and galaxy-galaxy interaction, increasing the mass of the galaxies at the centre \citep[e.g.][]{OstrikerTremaine1975}; we can then imagine that ELGs are galaxies that are still orbiting the cluster and undergoing the infall process. This means that their $M_\star$ is not greatly affected by the presence of neighbours. However, passive galaxies, that have already experienced SF quenching and fallen to the centre of the cluster, have a positive correlation between their mass and the local density. The large scatter can be attributed to the fact that Cl0024 is a non-relaxed cluster \citep[e.g.][and references therein]{Costa2024}.

Nonetheless, we observe significant trends when we study the fractions of SFGs in \cref{fig:SF_fractions}. In the left panel, there are in proportion fewer cluster SFGs with a high $M_\star$, while the fraction of SFGs in the field does not vary significantly with $M_\star$. In addition, the cluster sample has lower SFG fractions than the field for all mass bins. This demonstrates that the quenching of the galaxies in the cluster is a superposition of internal quenching (also called `mass quenching') and environmental quenching, especially for the most massive galaxies. Furthermore, galaxies in structure B are less affected by the environmental quenching since their SFG fraction is similar to that of field galaxies.

In the right panel of \cref{fig:SF_fractions}, the inverse relation between the fraction of SFGs and $\Sigma_5$ indicates that galaxies that form stars are statistically more common in the less dense regions of the environment. In the particular case of Cl0024, this corresponds to the outskirts of the cluster (see Fig. 5 of \citealt{Cedres2024}), which agrees with the models of galaxy cluster dynamics. This result aligns with previous findings, and the trend we observe appears even more conclusive than what had been found by \cite{Zeleke2021}, with a larger difference between the low-density and intermediate-density bins.

Finally, we also observe a trend when the sSFRs are plotted against the $D4000$ index in \cref{fig:sSFR_D4000}. The turnover of the relation between these two quantities has already been studied with larger samples of galaxies \citep[e.g.][]{duarte_puertas_mass-metallicity_2022}, but in our work we can clearly see how it coincides with the separation between galaxies with active and suppressed SFRs, although the scatter in sSFR is important, especially for the suppressed population. Additionally, this trend can also be found in other relations; for example, \cite{McNab2021} studied the SF quenching in galaxies at $z\gtrsim 1.0$ and in their Fig. 3 they show that their quiescent (Q) population is spanning over a large range of the $D4000$ index, while not varying their (U-V) colour. A future study of the colours of the galaxies of Cl0024 might confirm these findings.

The fact that we observe a significant trend between the fractions of SFGs and $\Sigma_5$, and the sSFRs and $D4000$, but not between SFRs/sSFRs and $\Sigma_5$, tells us that the suppression of SF is a slow process. On the one hand, the SF activity of the ELGs we are tracing through the SFR is only of the recent SF, so the absence of a trend leads us to conclude that the local density of the cluster does not affect the galaxies in the short term (the SFRs we are using in this work trace the SFH of the past 100 Myr). On the other hand, the fractions of SFGs are tracing a longer SFH than the SFRs, so the confirmation that the fraction of SFGs is indeed decreasing in the regions where $\Sigma_5$ is higher shows that it still has an effect, but in the medium or long term.

\cite{Rhee2020} estimate that quenching remains constant for $\sim2\ {\rm Gyr}$ during the first infall of galaxies in the cluster (they call this the `delay time'). \cite{Rhee2020} and several studies such as \cite{Oman2016}, that used $N$-body simulations and observations from the Sloan Digital Sky Survey (low $z$), \cite{muzzin_phase_2014} that studied the phase-space diagram of galaxies from nine clusters at $z\sim1.0$ with simulations, find that after the delay time, galaxies become passive in $\sim 1\ {\rm Gyr}$, possibly due to the environmental effects that are more intense near the centre of the cluster.

\subsection{Spatial distribution of the galaxies with suppressed star formation in Cl0024}\label{sec:suppression_position_discussion}
Focusing on the fractions of suppressed galaxies in the left panel of \cref{fig:suppressed_galaxies_environment}, we observe several results. First of all, we see similar fractions in Cl0024 and the LCS, except in the case of the infall region. In the case of Cl0024, the fraction of suppressed galaxies is significantly higher in the infall region. We think this increment is not due to the difference in redshift of the two samples, because we observe similar values in the field and the core regions (especially in the core, where we would think a difference in redshift would be more visible, as other effects such as the Butcher-Oemler effect were seen in this environment). Instead, we think the origin of this increment in the fraction of suppressed galaxies in the infall region is due to the particular dynamics of the cluster. We know the galaxies in structure B, located in the infall region, are experiencing a suppression of their SF in larger proportion than structure A (44\% of SFGs in structure B are suppressed, and only 32\% are in structure A). Additionally, structure B is falling onto the core of the cluster \citep{Czoske2001,Czoske2002}, where structure A is located; for these reasons, we think galaxies in structure B are particularly affected by the environmental quenching in the cluster, and especially due to a possible ram pressure stripping that would affect the galaxies falling towards the centre of the cluster.

On the other side, one would expect to observe a higher fraction towards the core, where the galaxies tend to be redder and more passive. \cite{Finn2023} was unable to infer a difference between the infall and core regions of the clusters of the LCS from this diagram, as the error bars of both values were too large to distinguish between the two populations. Nevertheless, considering only the values, a lower fraction in the core appears to have been estimated at low $z$. We observe the same differences in the galaxies of Cl0024 (left panel of \cref{fig:suppressed_galaxies_environment}), with a higher fraction of suppressed galaxies in the infall region than in the core. We decided to compute the fractions including the AGN and unclassified ELGs (middle panel) and passive and non-ELGs (right panel), to confirm this initial result. In this case, the higher number of galaxies in each subsample allows us to differentiate the values between the environments, and we can confirm that the fraction of suppressed galaxies is lower in the core than in the infall region of the cluster.

Additionally, we observe a significant decrement of the fractions in the core, not when comparing with the whole sample of ELGs (middle panel) but when adding the non-ELGs/passive galaxies in the comparison sample (right panel); this demonstrates that the core is in majority populated by galaxies that were suppressed and have now become passive. The fact that the fraction in the infall region stays the same when adding the non-ELGs/passive galaxies means that they do not populate this region, although we know the suppressed galaxies are found in majority in the infall region, where the fraction is higher with respect to other environments. This confirms that galaxies suffer a suppression of their SF during their infall, and become passive when reaching the core of the cluster, as suggested by \cite{Rhee2020}

Finally, combining \cref{fig:suppressed_galaxies_environment,fig:active_galaxies_environment} together, we observe that the highest fraction of suppressed galaxies is found in the infall region, while the fraction of active galaxies is lowest in the core--and not in the infall region. This can be explained by referring to the passive galaxies, which are mainly found in the core (as seen in the right histogram of \cref{fig:phase_space}) and generally decrease the fractions in this region. However, when we compare the fractions in the infall region, we observe that the fraction of suppressed galaxies ($16\pm3\ \%$) is statistically similar but slightly lower than that of active galaxies ($20\pm4\ \%$), demonstrating that the suppressed population is particularly prevalent in the infall region. This finding, once again, relates to the presence of structure B, where the majority of the infalling suppressed galaxies are located.

This result is surprising since one would assume that the majority of the suppressed galaxies are in the denser and more virialised regions of the cluster. It could be explained by the presence of an important quenching phenomenon that would mainly affect galaxies as they fall through the gravitational potential of the cluster, especially in structure B, but would have a dimmer effect as they reach the cluster core, where the galaxies fall once they are fully quenched. This result is consistent with the `delayed-then-rapid quenching' scenario, where galaxies are slowly quenching in the infall region after their initial accretion by the cluster. Nevertheless, as we explained in \cref{sec:ppsd_results}, our analysis using the phase-space diagram is simplistic since we infer the falling stage of the galaxies only from the separation of the virialised region; future analyses combining a more complete use of the phase-space diagram may confirm our initial conclusions.

\subsection{Effect of the mass of the cluster on the suppression of the star formation}
Finally, we wanted to assess the hypothesis that galaxies in clusters with higher $M_{\rm vir}$ tend to be more quenched. \cite{Cedres2024} showed that bigger was the mass of the cluster, larger were the zones of higher density, which may favour galaxy-galaxy encounters as well as galaxy-cluster interactions and, ultimately, galaxy quenching.

We observe in \cref{fig:DeltaSFRs_Mass_several_samples} the deviation of the galaxies' SFRs of several clusters with respect to the SFMS. The lack of differences between the clusters tends to refute this hypothesis. It is also important to take the effect of the redshift into account, as explained by \cite{Cedres2024}. For example, the cluster PKS 1138-262 has a median of $\Delta\log_{10}({\rm SFR})>0$, which can be expected since it is a proto-cluster at redshift $z>2$, but even when we compared the four clusters at $z\sim0.4$, RX J2248-443, Cl0939+4713, MACS J0416.1-2403, and Cl0024, we did not see any significant difference between the $\Delta\log_{10}({\rm SFR})$ distributions.

A more precise study of the phenomena that are causing the quenching in clusters could more confidently confirm or reject this hypothesis. It is possible that gas abundances reflect a phenomenon not visible through SFRs. A future work for GLACE will be to study the fundamental relations between properties, now that more precise SFRs, gas abundances and $M_\star$ were estimated. Furthermore, we also plan to study the neutral cold gas within the cluster, to help understand how the pristine gas is affected by the cluster's environment in both active and suppressed galaxies.

\section{Summary}\label{sec:summary}
In this work, we have used the new MOS data and previous TF observations from the GLACE survey to study the $M_\star-{\rm SFR}$ relation of the SFGs in the cluster Cl0024. This led us to analyse the population of galaxies with suppressed SF based on the local environment and in the context of the dynamical state of the cluster. The results we obtained are resumed as follow:

    \begin{itemize}
        \item We now have better estimates of redshifts for 158 of the 173 ELGs of Cl0024 based on the MOS observations of the $\Hbeta$, \OII, and \OIII$\lambda5007$ emission lines made by the GLACE collaboration. These data confirm the existence of the two kinematical structures of the cluster, A and B \citepalias{SP15}. Structure C could not be detected using MOS data.
        
        \item The MOS spectrum of the source {\tt 35\_a} helped identify it as an M-type star and allowed it to be removed from the GLACE catalogues. The $\Halpha$ redshifts of eight galaxies of the GLACE sample were poorly estimated in previous works because the emission line identified was not $\Halpha$. Notably, {\tt 359\_a}, {\tt 433\_a}, {\tt 657\_b}, and {\tt 888\_a} now have a revised MOS redshift. Galaxies {\tt 96\_a}, {\tt 105\_a}, {\tt 384\_b}, and {\tt 647\_b} have an MOS redshift $z\lesssim0.36$ and are now catalogued as interlopers; they may be part of a fourth foreground structure. However, further analysis is needed to confirm this.
        
        \item The BPT and WHAN diagnostics separate the 173 ELGs into 101 SFGs, 21 AGN galaxies, and 9 composites. We note that 25 galaxies display BLAGN characteristics in their pseudo-spectra, and 17 do not have a confident estimation of \NII$\lambda6583$ emission line flux to confidently categorise them.
        
        \item The SFRs estimated using $\Hbeta$ and \OII\ agree well with $\Halpha$ SFRs, except in the case of \OII\ with a M91 metallicity correction, for which the scatter with respect to the one-to-one relation is large. We used the $\Halpha$ SFR estimator in our analysis due to its higher signal-to-noise ratio, and in the case of the outlier \texttt{657\_b}, we used the \OII\ SFR estimator corrected by Z94 metallicity.
        
        \item The $M_\star-{\rm SFR}$ of 98 SFGs confirms the quenching of galaxies in Cl0024. A majority of the galaxies lie below the SFMS at $z=0.395$, and the SFRs of 34 of them are considered suppressed. Only one galaxy is considered passive (\texttt{882\_a}), which is expected since GLACE observations focused on ELGs.
        
        \item The ${\rm sSFR}-M_\star$ diagram shows a distribution similar to that of two other clusters at the same redshift: RX J2248-443 ($z\sim0.348$) for a lower $M_\star$ and Cl0939+4713 ($z\sim0.403$) for a higher $M_\star$. This correspondence with respect to Cl0024 takes place despite the fact that their virial mass is different.
        
        \item The comparison with field galaxies revealed that cluster galaxies exhibit lower SF activity, with a median difference in their $\Delta\log_{10}({\rm SFR})$ of 0.42 dex. There is also a higher percentage of galaxies with a suppressed SFR in cluster galaxies (34.7\%) compared to galaxies with a suppressed SFR in the field (11.0\%). In addition, based on an Anderson-Darling test, the distribution of SFRs in the field and cluster are statistically different.
        
        \item We observed a correlation between SFRs and $\Sigma_5$ when separating galaxies with active and suppressed SFRs; however, it disappeared when normalising the SFRs by the $M_\star$, which is similar to the findings of previous studies \citep[e.g.][]{LaganaUlmer2018}. This indicates that the environment does not directly change the SFR of active and suppressed galaxies.
        
        \item We observed a gradient of the fractions of SFGs with respect to $M_\star$ and $\Sigma_5$. On one hand, $34\pm7\ \%$ of the SFGs are in the low-mass regime, but only $5.9\pm2.4\ \%$ are in the high-mass regime. On the other hand, the fractions are invariant (within the error bars) and higher for field galaxies ($\sim50 \%$) due to the higher proportion of SFGs with respect to AGNs and passive galaxies. All galaxies suffer mass quenching, but the cluster galaxies additionally suffer environmental quenching, especially the most massive ones.
        
        \item Similarly, low-density regions exhibit fractions 1.55 times higher than intermediate-density regions, and the fraction is close to zero ($6\pm6\ \%$) for high-density regions, indicating that more ELGs tend to suffer a suppression of their SF activity in the densest regions of the cluster.

        \item The $D4000$ index shows that the quenching experienced by the galaxies in the cluster has a lasting effect on the stellar populations in the long term. The sSFR decreases as a function of $D4000$ for active galaxies, while it is constant for suppressed galaxies.

        \item The suppression of SF in the ELGs of Cl0024 occurs at a low rate, since it is only visible through the study of fractions of SFGs and parameters that trace the longer SFH, such as the $D4000$ index.

        \item The environmental quenching especially affects galaxies in structure B (44\% of its galaxies are suppressed, against 32\% in structure A), possibly due to the infall of this group onto the core of the cluster.
        
        \item The fraction of suppressed galaxies at intermediate redshift follows a similar trend to what is observed at local redshift \citep{Finn2023}, namely, it is higher in the infall region than in the field but within the error bars. When passive galaxies are included, this trend becomes statistically significant, with similar fractions in the field and the infall region, and it is 2.6 times higher in the infall region than the core. We also observed that the fractions in the infall region decrease when including AGNs and unclassified galaxies in the comparison sample but not when adding passive and non-ELGs, while the contrary happens in the core. This indicates that the process of quenching of the SFRs of galaxies is more efficient--or at least, visible--in the infall region of the cluster, while galaxies mainly become passive in the core, supporting the `delayed-then-rapid quenching' scenario.
        
        \item The virial mass of the cluster appears to have no direct effect on the quenching of the cluster galaxies when studying the deviation of the SFRs of several clusters from their respective SFMS.
    \end{itemize}

This study has provided new and valuable data for the community and important results that push forward the analysis of the quenching of galaxies within Cl0024. The analysis of the SF activity in the context of the internal parameters of the galaxies, such as their $M_\star$ and stellar populations (through the $D4000$ index), and the local environment of the cluster through the $\Sigma_5$ parameter and phase-space diagram allowed us to better characterise the causes of quenching and how it affects galaxies. Nevertheless, there are still some questions remaining, such as why the quenching affects mainly galaxies during their infall, how much of this quenching is due to the infall of structure B onto the core of the cluster, whether this explains why we observe that the suppression of SF occurs on the long term, how much of the quenching can be attributed to the cluster environment and the internal mechanisms of the galaxies. To answer these questions, studies involving the analysis of the cold neutral gas and the hot ICM will be key as well as advanced simulations to further understand the complex kinematics within the cluster.

\section*{Data availability}\label{sec:data_availability}
The multiwavelength catalogue presented in this work is only available in electronic form at the CDS via anonymous ftp to \href{cdsarc.u-strasbg.fr}{cdsarc.u-strasbg.fr} (130.79.128.5) or via \href{http://cdsweb.u-strasbg.fr/cgi-bin/qcat?J/A+A/}{http://cdsweb.u-strasbg.fr/cgi-bin/qcat?J/A+A/}.

\begin{acknowledgements}
    This article is based on observations made with the Gran Telescopio Canarias (GTC) at Roque de los Muchachos Observatory on the island of La Palma.
    This research uses data from the VIMOS VLT Deep Survey, obtained from the VVDS database operated by Cesam, Laboratoire d'Astrophysique de Marseille, France.

    \\
    We acknowledge the support of the Spanish Ministry of Science, Innovation and Universities through the projects PID-2021-122544NB-C41 and PID-2021-122544NB-C43.

    \\
    SBD acknowledges financial support from the grant AST22.4.4, funded by Consejería de Universidad, Investigación e Innovación and Gobierno de España and Unión Europea –- NextGenerationEU, and projects PID2020-113689GB-I00 and PID2023-149578NB-I00, financed by MICIU/AEI/10.13039/501100011033.
    
    Part of this work was supported by the \emph{University of Granada (UGR)}, through the FisyMat master extracurricular internship program (reference number 380837), funded by the Institut de Radioastronomie Millimétrique (IRAM).

    DE acknowledges support from a Beatriz Galindo senior fellowship (BG20/00224) from the Spanish Ministry of Science and Innovation; projects PID2020-114414GB-100 and PID2020-113689GB-I00 financed by MCIN/AEI/10.13039/501100011033, projects PID2023-150178NB-I00 and PID2023-149578NB-I00, financed by MICIU/AEI/10.13039/501100011033, and by FEDER, UE; project P20-00334 financed by the Junta de Andalucía; and project A-FQM-510-UGR20 of the FEDER/Junta de Andalucía-Consejería de Transformación Económica, Industria, Conocimiento y Universidades.
    
    CCC acknowledges financial support from the grant CEX2021-001131-S funded by MICIU/AEI/10.13039/501100011033, from the grant PID2021-123930OB-C21 funded by MICIU/AEI/10.13039/501100011033, by ERDF/EU and from the grant TED2021-130231B-I00 funded by MICIU/AEI/10.13039/501100011033 and by the European Union NextGenerationEU/PRTR.

    GTR acknowledges financial support from the research project PRE2021-098736, funded by MCIN/AEI/10.13039/501100011033 and FSE+.

    MGO acknowledges financial support from the State Agency for Research of the Spanish MCIU through Center of Excellence Severo Ochoa award to the Instituto de Astrofísica de Andalucía CEX2021-001131-S funded by MCIN/AEI/10.13039/501100011033, and from the grant PID2022-136598NB-C32 ``Estallidos8''. MGO acknowledges support by the project ref. AST22\_00001\_Subp\_11 funded from the EU – NextGenerationEU.

    JAD acknowledges support from UNAM-PAPIIT IN116325.

    MC acknowledges support by grant PID2022-136598NB-C33 funded by MCIN/AEI/10.13039/501100011033 and by “ERDF A way of making Europe”.

    MALL acknowledges support from the Ramón y Cajal program RYC2020-029354-I funded by MICIU/AEI/10.13039/ 501100011033 by ESF+, and the Spanish grant PID2021-123417OB-I00, funded by MCIN/AEI/10.13039/501100011033/FEDER, EU.

    JIGS acknowledges the support of the Spanish Ministry of Science, Innovation and Universities through the project PID-2021-122544NB-C44.

    ICG acknowledges financial support from DGAPA-UNAM grant IN-119123 and CONAHCYT grant CF-2023-G-100.

    CAN acknowledge the support from projects SECIHTI CBF2023-2024-1418, PAPIIT IA104325 and IN119123.

    \\
    This work made use of the following software packages: \texttt{astropy} \citep{astropy:2013, astropy:2018, astropy:2022}, \texttt{matplotlib} \citep{Hunter:2007}, \texttt{numpy} \citep{numpy}, \texttt{pandas} \citep{mckinney-proc-scipy-2010, pandas_10304236}, \texttt{python} \citep{python}, \texttt{IRAF} \citep{1993ASPC...52..173T, 1986SPIE..627..733T}, \texttt{seaborn} \citep{Waskom2021}, \texttt{Lmfit} \citep{Lmfit}, \texttt{SciencePlots} \citep{SciencePlots}, \texttt{Petrofit} \citep{Petrofit}, \texttt{SExtractor} \citep{SExtractor}, \href{https://www.cfht.hawaii.edu/~arnouts/LEPHARE/lephare.html}{\texttt{LePhare}} \citep{Arnouts_LePHARE,Ilbert_LePHARE}, \texttt{CIGALE} \citep{CIGALE}, \texttt{TOPCAT} \citep{TOPCAT}, and \texttt{MARZ} \citep{MARZ}.\\
    
    Software citation information aggregated using \texttt{\href{https://www.tomwagg.com/software-citation-station/}{The Software Citation Station}} \citep{software-citation-station-paper, software-citation-station-zenodo}.
    
    \\
    SBD would like to thank Robel Geda for his help and advice with the \texttt{Petrofit} software, Dr. Sean Moran for providing the B, V, R and I images from the CFHT CFH12k camera used to extract the photometric apertures, and Dr. Rose Finn for providing some of the data presented in \cite{Finn2023} from the LCS.\\
    Specific results of our analysis, as well as the multi-wavelength atlas, are available upon reasonable request.
\end{acknowledgements}

\bibliographystyle{aa.bst}
\bibliography{references}

\begin{appendix}

\onecolumn
\section{Examples of multi-object spectroscopy spectra}\label{sec:examples_spectra}
\begin{figure*}[h!]
     \centering
     \begin{subfigure}[c]{0.85\textwidth}
         \centering
         \includegraphics[width=\textwidth]{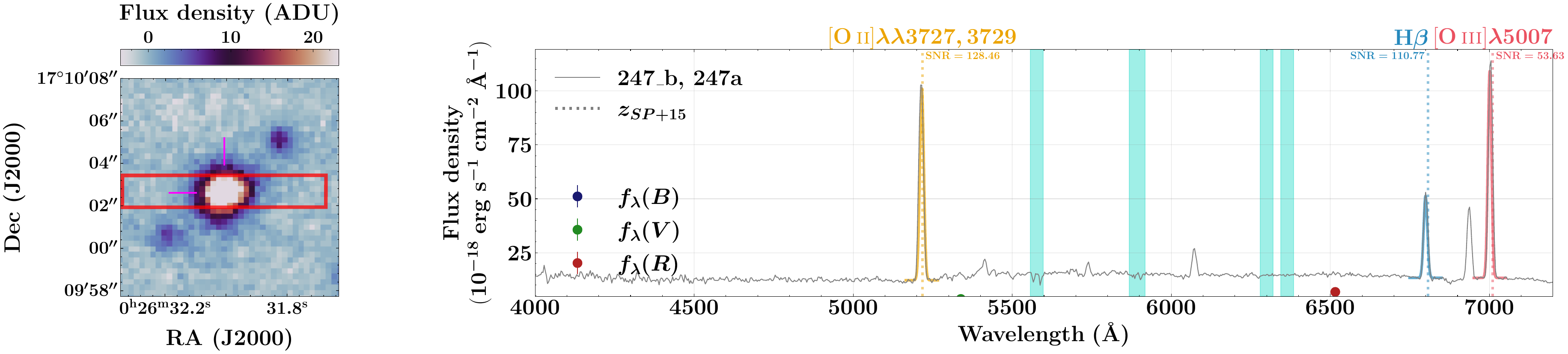}
     \end{subfigure}
     \\
     \begin{subfigure}[c]{0.85\textwidth}
         \centering
         \includegraphics[width=\textwidth]{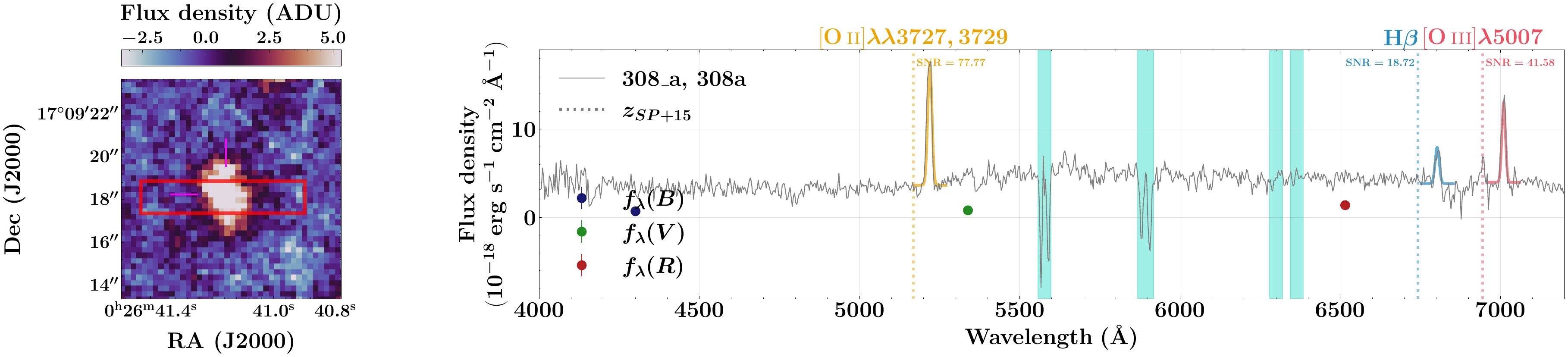}
     \end{subfigure}
     \\
     \begin{subfigure}[c]{0.85\textwidth}
         \centering
         \includegraphics[width=\textwidth]{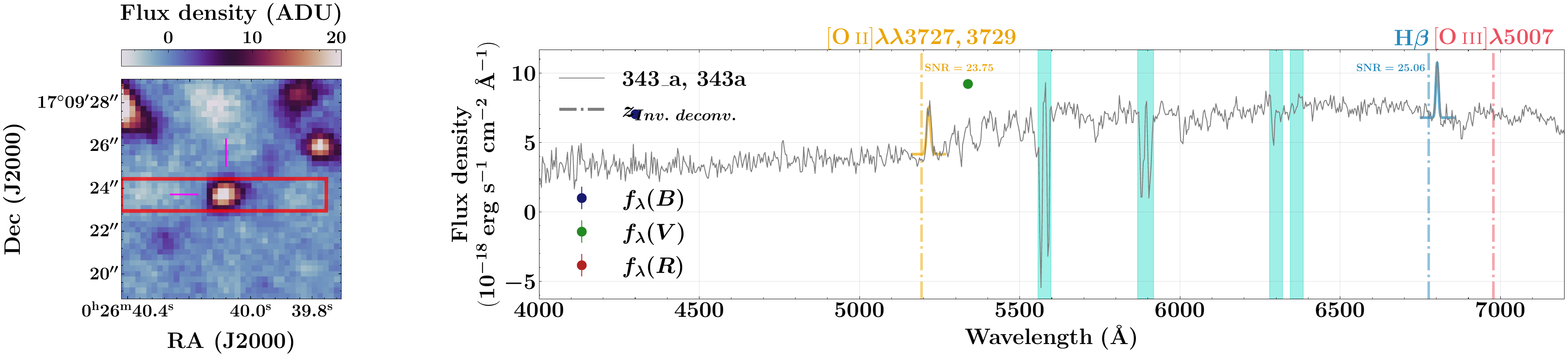}
     \end{subfigure}
     \\
     \begin{subfigure}[c]{0.85\textwidth}
         \centering
         \includegraphics[width=\textwidth]{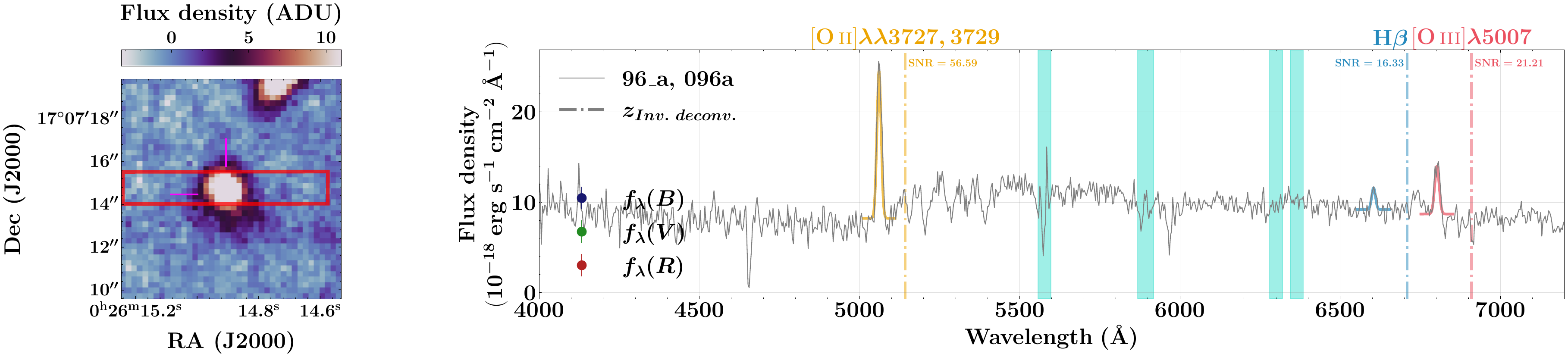}
     \end{subfigure}
     \\
     \begin{subfigure}[c]{0.85\textwidth}
         \centering
         \includegraphics[width=\textwidth]{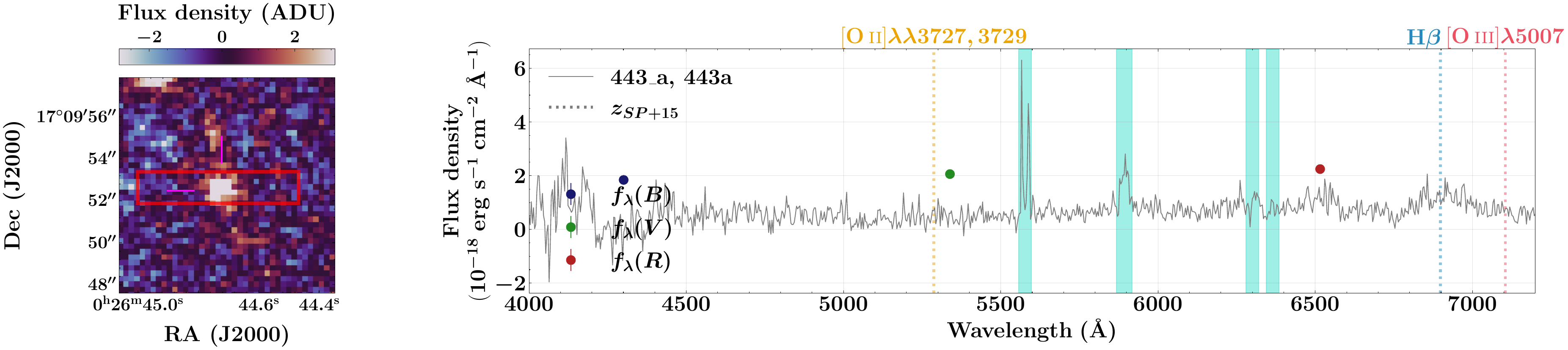}
     \end{subfigure}
     \caption{Five examples of the spectra and line identification from the MOS sample. In the left panel of each subfigure, we show a snapshot of the deep $\Halpha$ image centred on the galaxy, as well as the individual MOS aperture for each source. The right panel shows the spectrum of the galaxy obtained after reduction, with the IDs and redshift source in the upper-left legend. Using the $\Halpha$ emission line, we indicate the estimated position of \OII, $\Hbeta$ and \OIII$\lambda5007$ emission lines as well as four telluric absorption bands (in cyan), where some residuals of the sky subtraction are visible. The photometric fluxes from \cite{Moran2005} are depicted with blue, green and red circles for bands B, V and R, respectively. We also show the S/N of each fitted emission line below its name.}
    \label{fig:spectra_example_1}
\end{figure*}

In \cref{fig:spectra_example_1} we show the MOS spectra of five galaxies in the range of 4000 \AA\ to 7200 \AA. We also show a snapshot of the galaxy from the $\Halpha$ deep images, where the MOS slit is depicted as a red rectangle.

The spectra of galaxies {\tt 247\_b} and {\tt 308\_a} exhibit a high S/N for both the emission lines and the continuum, with minimal residuals from sky subtraction (shown with blue bands). In contrast, the spectrum of {\tt 343\_a} maintains a good S/N for \OII\ and $\Hbeta$ emission lines but \OIII$\lambda5007$ is not distinguishable from the continuum, that appears noisier while significant sky residuals are noticeable. In the spectrum of {\tt 96\_a}, \OII, $\Hbeta$ and \OIII$\lambda5007$ are detected, although with a noticeable deviation from their expected positions based on $\Halpha$ redshift, $z_{\rm{SP+15}}$. Finally, the spectrum of {\tt 443\_a} does not exhibit any emission line.

\FloatBarrier
\twocolumn

\begin{figure}[h!]
    \centering
    \includegraphics[width=\hsize]{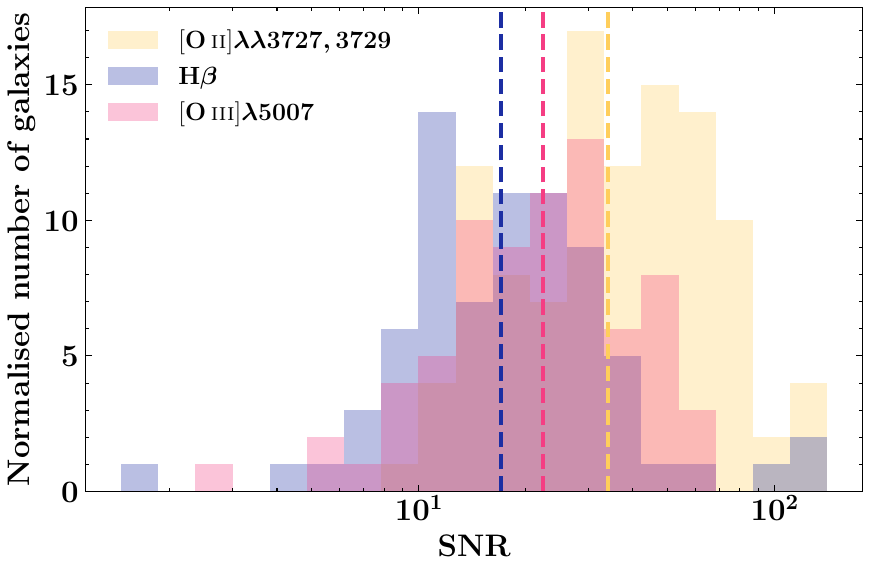}
    \caption{Multi-object spectroscopy S/N distributions for \OII, $\Hbeta$ and \OIII$\lambda5007$ emission lines.}
    \label{fig:SNRs_MOS}
\end{figure}

In \cref{fig:SNRs_MOS} we show the distributions of the S/N for each MOS line as well as the median value. As expected, the \OII\ line tends to have a higher S/N. Nonetheless, all the median values are higher than $\rm S/N=10$, which shows the high quality of the MOS observations. Although some spectra, such as {\tt 443\_a}, show a very low S/N, the majority of the MOS spectra (at least two thirds, based on the minimal number of MOS redshifts estimated) display enough S/N to correctly and unequivocally identify the emission lines.

\section{Corrections and calibrations of the data}\label{sec:data_corrections}
Several corrections and calibrations were required for the MOS data. This subsection describes the steps taken to apply these corrections and calibrations.

\subsection{Flux adjustments of MOS spectra}\label{sec:flux_adjustments}
An aperture correction was implicitly applied to the MOS spectra during the reduction of the data, but additional adjustments due to the use of MOS slits are required. Apart from the uncertainties in the flux calibration inherent to the MOS observations (see \cref{sec:glace_mos_data}), the observations of the secondary photometric standards used for the flux calibration were not made immediately before or after the science target observations, but at the beginning or end of the observing nights, with seeing and atmospheric extinction conditions not necessarily similar. In addition, the flux calibration applies to the whole mask indifferently, but diffraction effects inherent to the use of short slits appear depending on the position of each MOS slit. Taking all these effects together, there may be small differences between the fluxes in the continuum of the MOS spectra and the one corresponding to the photometric band for a given object and wavelength.

The mean effective radius of the galaxies in band $I$ is $0.53\arcsec\pm0.19\arcsec$, which is less than half the width of the MOS slits used in the observations. Additionally, the typical difference between the flux in the $R$-band and MOS fluxes is smaller than $2.2\times10^{-18}\ {\rm erg\ s^{-1}\ cm^{-2}\ \AA^{-1}}$. Therefore, we can assume that not using them does not introduce much additional uncertainty in the estimated properties.

We compared several methods to adjust the MOS fluxes due to the size of the galaxies relative to the size of the slits:

\begin{itemize}
    \item \underline{`Optical photometry'}: we divided the continuum of MOS spectra by the $V$ and $R$ `AUTO' photometric fluxes estimated by \cite{Moran2005} in these same photometric bands; when we did not have access to one of these photometric fluxes, we used the $g$, $r$ and $i$ Pan-STARRS photometric bands \citep{panstarrs} instead, for the \OII, $\Hbeta$ and \OIII$\lambda5007$ emission lines, respectively.
    \item \underline{`Direct method'}: we divided the area of each galaxy covered by the slits by the total area of the galaxy, estimated using deep $\Halpha$ images from HST and \texttt{Petrofit}.
    \item \underline{`\OIII\ cont / TF cont.'}: we divided the continuum next to the \OIII$\lambda5007$ emission line by the continuum extracted from the TF pseudo-spectra.
    \item \underline{`Band I MOS / TF cont.'}: we divided the continuum in the band I by the continuum extracted from the TF pseudo-spectra.
    \item \underline{`Band I MOS / Band I cont.'}: we divided the continuum in the band I by the flux estimated using \cite{Moran2005} photometric catalogue in this same band.
\end{itemize}

\begin{figure*}
         \centering
         \sidecaption
         \includegraphics[width=0.6\textwidth]{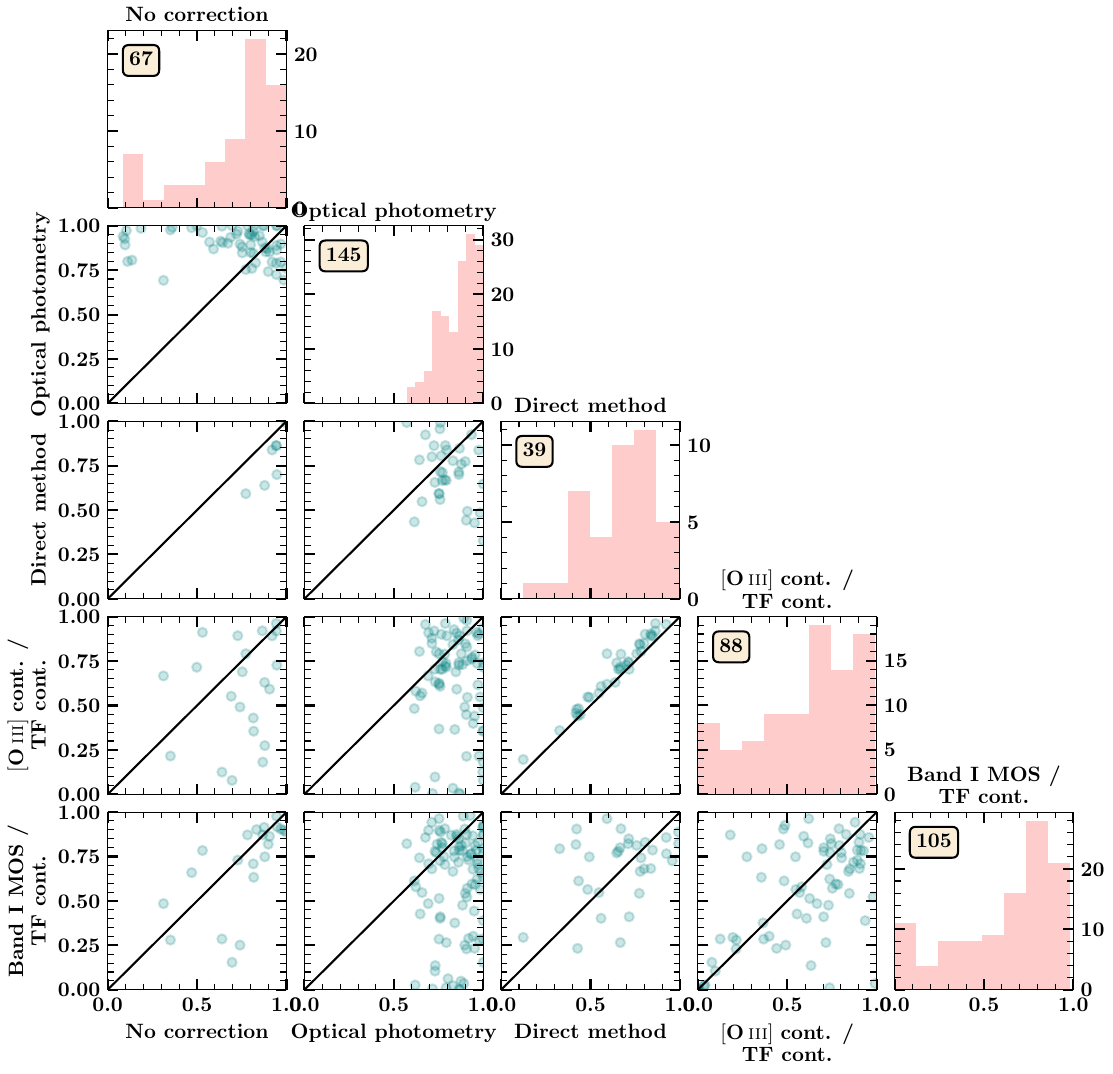}
         \caption{Histograms and comparisons of the flux adjustment factors obtained using several methods.}
         \label{fig:aperture_correction_methods_histograms}
    \end{figure*}

The correction factors of each method are shown in \cref{fig:aperture_correction_methods_histograms}. As explained in \cref{sec:flux_adjustments}, not all the galaxies are corrected for the effect of the MOS apertures, and the number of galaxies corrected also changes depending on the method used.

\begin{figure*}
        \centering
        \sidecaption
        \includegraphics[width=0.6\textwidth]{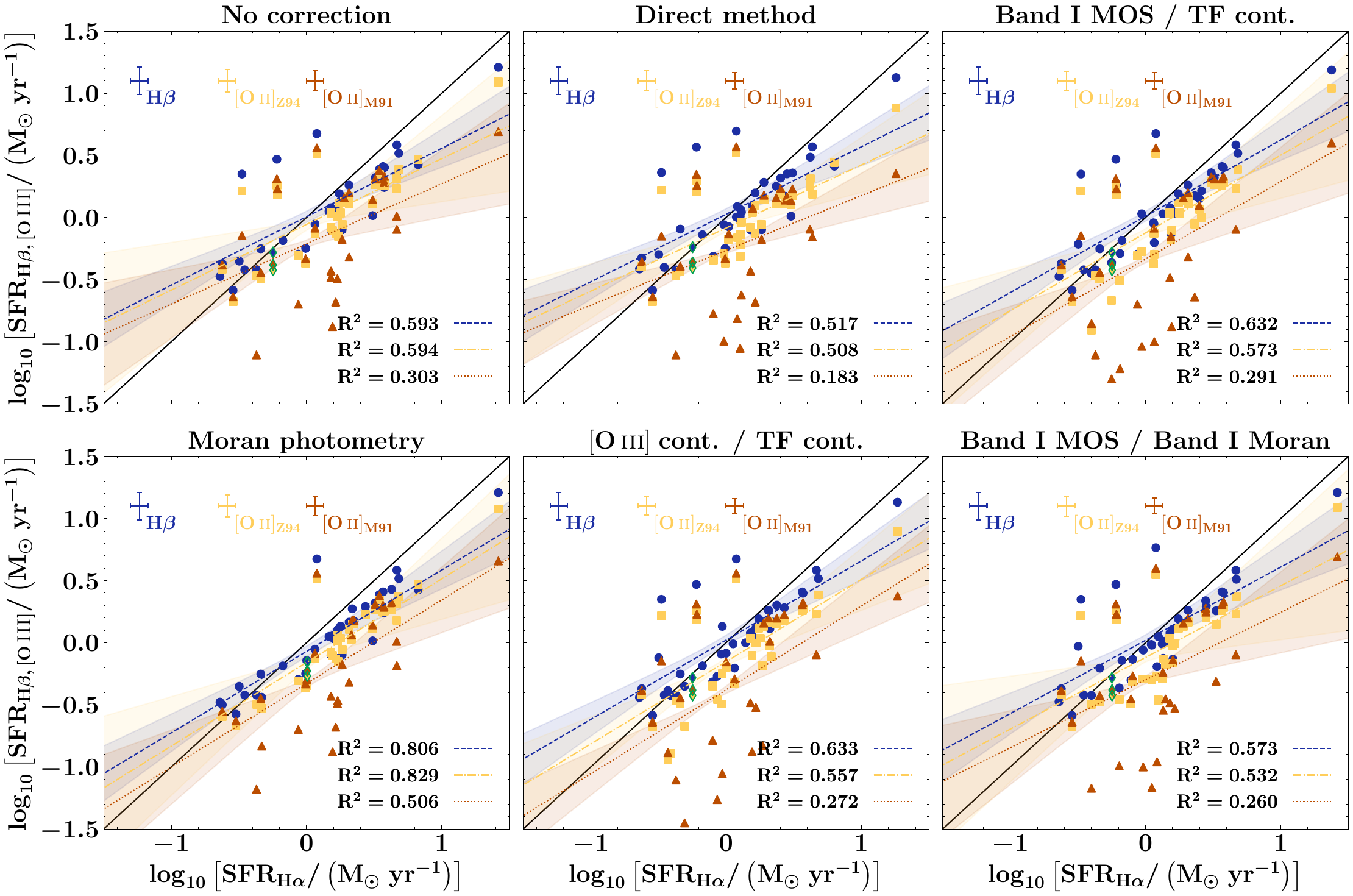}
        \caption{H$\beta$ and \OII\ SFR with respect to $\Halpha$ SFR for the 101 SFGs, using several methods for the flux adjustment. Symbols and colours are the same as in \cref{fig:SFRs_comparison}.}
        \label{fig:aperture_correction_methods_SFRs_comparison}
    \end{figure*}

To determine which correction factor was the most effective, we decided to recreate \cref{fig:SFRs_comparison} applying each method. In \cref{fig:aperture_correction_methods_SFRs_comparison} we show the results for each method and also applying no correction (first panel). The SFRs are very similar, but the method with the highest correlation factors ($R^2$) and for which $\Hbeta$ and \OII corrected using Z94 metallicities are closer to the $\Halpha$ SFRs is `Optical photometry'. We also looked for a dependence of the factors with respect to $M_\star$ and $r/R_{200}$, which would conclude to a systematic bias, but we did not find any, concluding that this method is robust.

We then decided to use the photometric data available and compiled in \citetalias{SP15}, coming from the public catalogues of \cite{Treu2003} and \cite{Moran2005}, except for 6 galaxies that have unreliable optical fluxes from these catalogues: {\tt 87\_a}, {\tt 97\_a}, {\tt 700\_b}, {\tt 783\_a}, {\tt 929\_a}, and {\tt 1157\_b}, for which we used Pan-STARRS fluxes \citep{panstarrs}. To reconcile the flux observed in the continuum with the one estimated from the photometric data, we calculated the total flux of the galaxies in each photometric band and scaled up the observed flux close to the three emission lines. Because we only want to correct the flux of three emission lines, \OII, $\Hbeta$, and \OIII$\lambda5007$, we used the $V$-band to correct the first one and the $R$-band to correct the other two since these are the bands closer to each emission line (with respect to their effective wavelength). For galaxies \texttt{358\_a} and \texttt{938\_a}, the $V$ and $R$ magnitudes are unknown, so we use the magnitude from the $F814W$-band, available for all the galaxies. In the cases where we used Pan-STARRS data, we employed fluxes in bands $g$, $r$ and $i$ to correct the aperture factor of lines \OII, $\Hbeta$ and \OIII$\lambda5007$, respectively. For 46 cases (e.g. galaxy {\tt 443\_a}, \cref{fig:spectra_example_1}), the total photometric flux seems lower than the one estimated from MOS observations: this would mean that MOS observation received more light than the deep observations, and we cannot explain this phenomenon without strong assumptions, which is why we decide not to apply any flux adjustment when this occurs, to avoid inducing further uncertainty.

\subsection{Galactic reddening and extinction}
A common and important effect concerning the analysis of spectra is interstellar reddening and extinction. The interstellar medium (ISM) contains dust that absorbs the light emitted by stars and re-emits it at longer wavelengths (mainly far infrared), which implies that we need to correct this extinction to obtain the flux emitted by the stars. Because we are within the Milky Way, we first have to correct the extinction produced by the ISM of the Galactic plane. Following the study by \cite{Treu2003} and \cite{Moran2005}, we apply the same galactic reddening correction, assuming a Galactic extinction of $E(B-V)=0.056$, and using the empirical selective extinction function of Cardelli, Clayton, and Mathis \citep{CCM_gal_extinc}\footnote{$E(B-V)$ represents the colour excess in the $(B-V)$ colour due to dust attenuation. It can be calculated as the difference between the observed colour $(B-V)_{\rm{obs}}$ and the intrinsic colour $(B-V)_{\rm{int}}$.}, based on the maps of \cite{Schlegel_gal_extinc}.

\subsection{Stellar atmosphere absorption}
In the case of helium and hydrogen recombination lines, such as the $\Halpha$ and $\Hbeta$ Balmer lines, the absorption of the existing stellar populations should be considered. The atmosphere of stars produces absorption lines of the same transition, which results in a variation in the flux when fitting the emission line. We measured the EW of the Balmer lines in absorption in the best-fitted composite stellar population (CSP) templates to correct this effect. This method compares a synthetic stellar population spectrum to the observed absorption features, making it possible to correct the hydrogen emission lines of the effect of stellar atmospheres. More details about the CSP fits can be found in \cref{sec:CSP_fitting}.

This correction is directly applied on the EW, following \cite{Kong_II}:

\begin{equation}
    F^{\rm{cor}}_{\rm{line}} = F^{\rm{obs}}_{\rm{line}} \left[1+{\rm EW^{abs}_{line}}(1+z_{\rm cl})/{\rm EW^{obs}_{line}}\right],
\end{equation}

\noindent where $F^{\rm{cor}}_{\rm{line}}$ and $F^{\rm{obs}}_{\rm{line}}$ are the absorption corrected and the observed emission line fluxes in $\rm erg\ s^{-1}\ cm^{-2}$, respectively, ${\rm EW^{abs}_{line}}$ and ${\rm EW^{obs}_{line}}$ are the EWs, in \AA, of the underlying stellar absorption line (calculated using CSP fits) and of the observed emission line, respectively, and $(1+z_{\rm cl})$ is accounting for the redshift of the stellar continuum. When ${\rm EW^{abs}_{line}}$ is unknown we use $<{\rm EW^{abs}_{line}}>=4.0$ \AA\ derived by \cite{Zeleke2021} and when ${\rm EW^{obs}_{line}}$ is unknown because the emission line was not estimated, we assume ${\rm EW^{abs}_{line}}=0$ to avoid further uncertainties. 

\subsection{Dust extinction}\label{sec:extinction_correction}
In addition to correcting the Galactic extinction and after taking into account the absorption produced by the underlying stellar population of the galaxies, we have to correct for dust extinction of the galaxies. For this purpose, we apply a correction factor to the observed fluxes of the emission lines using $E(B-V)$, as it is shown in \cref{eq:extinction_correction}:

\begin{equation}\label{eq:extinction_correction}
    F_{\rm{int}}(\lambda) = F_{\rm{obs}}(\lambda)10^{0.4E_{\rm s}(B-V)k(\lambda)}.
\end{equation}

In this equation, $F_{\rm{int}}(\lambda)$ and $F_{\rm{obs}}(\lambda)$ represent the intrinsic and observed fluxes, respectively, measured in $\rm erg\ s^{-1}\ cm^{-2}$; $E_{\rm s}(B-V)$ represents the colour excess of the stellar continuum, and $k(\lambda)$ is the \cite{Calzetti_2000} extinction law we use to describe the wavelength-dependent nature of interstellar reddening.

While it is easier to calculate $E(B-V)$ using the colour excess derived from the nebular gas emission lines, we need $E_{\rm s}(B-V)$ to apply \cref{eq:extinction_correction} since $k(\lambda)$ is derived using the stellar continuum. These two quantities are proportional \citep{Calzetti_2000}: $E_{\rm s}(B-V) = (0.44\pm0.03) \times E(B-V)$.

It is common to use the Balmer decrement, which is the ratio between the $\Halpha$ and $\Hbeta$ emission lines, to estimate $E(B-V)$:

    \begin{equation}
        E(B-V) = \frac{E(\Hbeta - \Halpha)}{k(\lambda_{\rm \Hbeta}) - k(\lambda_{\rm \Halpha})} =\frac{2.5}{k(\lambda_{\rm \Hbeta}) - k(\lambda_{\rm \Halpha})}\log_{10}\left[\frac{(\Halpha/\Hbeta)_{\rm{obs}}}{(\Halpha/\Hbeta)_{\rm{int}}}\right].
    \end{equation}

The MOS observations allow us to precisely estimate the Balmer decrement for a majority of the galaxies, using the $\Hbeta$ emission line. Following \cite{Dominguez_extinc}, we assume the value of $A_{\rm V} = (\Halpha/\Hbeta)_{\rm{int}} = 2.86$ for SFGs, corresponding to a temperature $T=10^4$ K and an electron density $n_{\rm e} = 10^2\ {\rm cm^{-3}}$ for Case B recombination \citep{Osterbrock_1989}. These considerations lead to the following expression:

\begin{equation}\label{eq:E_B_V}
    E(B-V) = 1.97 \log_{10}\left[\frac{F_\Halpha/F_\Hbeta}{2.86}\right],
\end{equation}

\noindent where $F_\Halpha$ is the flux of the $\Halpha$ emission line, and $F_\Hbeta$ of the $\Hbeta$ emission line, corrected for Galactic extinction, aperture and underlying absorption on stellar population. Combining \cref{eq:extinction_correction,eq:E_B_V}, we can then correct the dust extinction for the four emission lines we are interested in: $\Halpha$, \OII, $\Hbeta$ and \OIII$\lambda5007$. In cases where the colour excess estimated using the Balmer decrement is negative, the extinction is assumed to be zero and in cases where it is not possible to estimate it (i.e. the $\Hbeta$ line is not available), we use the colour excess provided by CSP fits.

\section{Spectral energy distribution fittings}\label{sec:sed_fittings_parameters}
In \cref{tab:sed_fit_params_lephare,tab:sed_fit_params_cigale} we show the parameters used to fit the SED of the 363 galaxies used in this work, using \texttt{LePhare} and \texttt{CIGALE}, respectively. 

\begin{table}[h]
        \centering
        \caption{Parameters used for the SED fitting of the galaxies of Cl0024 using {\tt LePhare} software.}
        \label{tab:sed_fit_params_lephare}
        \begin{tabular}{l c}
        \hline\hline
            \rowcolor[HTML]{FFFFFF} 
            Module parameters                  & Values \\ \hline
            \rowcolor[HTML]{FFFFFF} 
            Templates                   & \cite{BC03}         \\
            \rowcolor[HTML]{FFFFFF} 
            IMF                         & \cite{Chabrier_IMF}           \\
            \rowcolor[HTML]{FFFFFF} 
            Metallicities               & 0.4, 0.2 and 0.04 $\times\ Z_\odot$    \\ 
            \rowcolor[HTML]{FFFFFF} 
            $E_{\rm s}(B-V)$            & [0.0, 0.5] mag, $\Delta=0.05$ mag	    \\
            \rowcolor[HTML]{FFFFFF} 
            SFH                         & ${\rm SFR} \propto \tau^{-2}\ t\ e^{-t/\tau}$	 ($t<10$ Gyr)         \\ 
            \rowcolor[HTML]{FFFFFF} 
            $\tau$                      & [0.1, 30.0] Gyr                        \\ \hline
        \end{tabular}
    \end{table}

\begin{table}[h]
       \centering
       \caption{Parameters used for the SED fitting of the galaxies of Cl0024 using {\tt CIGALE} software.}
       \label{tab:sed_fit_params_cigale}
       \resizebox{\hsize}{!}{
       \begin{tabular}{l c}
       \hline\hline
           Module & \\
           parameters        &  Values \\ \hline
           {\tt sfhdelayedbq}        &         \\ 
           {\tt tau\_main}           & 500, 1000, 3000, 5000, 7000 [Myr] \\
           {\tt age\_main}           & 3000, 5000, 7000, 9000 [Myr] \\
           {\tt age\_bq}             & 0, 100, 300, 500, 1000 [Myr] \\
           {\tt r\_sfr}              & 0.01, 0.1, 0.5, 1, 2 \\
           {\tt sfr\_A}              & 1 \\ \hline
           {\tt bc03}                &         \\
           {\tt IMF}                 & \cite{Chabrier_IMF} \\
           {\tt metallicity}         & 0.004, 0.02, 0.05 [Z] \\
           {\tt separation\_age}      & 10 [Myr] \\ \hline
           {\tt nebular}             &          \\
           {\tt logU}                & -4, -2.5, -1 \\
           {\tt zgas}                & 0.004, 0.02 [Z] \\
           {\tt lines\_width}         & 250 [km/s] \\
           {\tt emission}            & True \\ \hline
           {\tt dustatt\_modified\_starburst} &    \\
           {\tt E\_BV\_lines}        & 0.01, 0.05, 0.1, 0.2, 0.3, 0.4, \\
                                     & 0.5, 0.6, 0.7, 0.9, 1.0, 1.1 [mag] \\
           {\tt E\_BV\_factor}       & 0.44 \\
           {\tt uv\_bump\_wavelength} & 217.5 [nm] \\
           {\tt uv\_bump\_width}       & 35 [nm] \\
           {\tt uv\_bump\_width}       & 0, 3 \\
           {\tt power\_law\_slope}   & -1.5, -0.5, 0, 0.5 \\
           {\tt Ext\_law\_emission\_lines} & 1 (Milky Way) \\
           {\tt Rv}                  & 3.1 [$A_{\rm V}/E(B-V)$] \\ \hline
           {\tt dale2004}            &         \\
           {\tt fracAGN}             & 0 \\
           {\tt alpha}               & 0.5, 2, 3 \\ \hline
           {\tt rest\_frame\_parameters} &     \\
           {\tt beta\_calz94}        & True \\
           {\tt D4000}               & True \\
           {\tt IRX}                 & True \\ \hline
           {\tt redshifting}         & (spectroscopic redshifts from this work) \\ \hline
       \end{tabular}}
   \end{table}

\end{appendix}

\end{document}